\documentclass[10pt,twocolumns]{emulateapj}
\usepackage{lscape}

\slugcomment{ApJ, in press \today}

\shorttitle{CMR}
\shortauthors{Mei et al.}

\begin{document}


\title{Evolution of the Color--Magnitude Relation in Galaxy Clusters at $z \sim 1$ from the ACS Intermediate Redshift Cluster Survey}

\author{Simona Mei\altaffilmark{1,2,3},
Brad P.~Holden \altaffilmark{4},  
John P. Blakeslee\altaffilmark{5},
Holland~C.~Ford\altaffilmark{1}, 
Marijn Franx\altaffilmark{6},
Nicole L. Homeier\altaffilmark{1},
Garth D. Illingworth \altaffilmark{4},
Myungkook J. Jee\altaffilmark{7},
Roderik Overzier  \altaffilmark{1,8},
Marc~Postman\altaffilmark{9}, 
Piero Rosati\altaffilmark{10}, \\
Arjen Van der Wel\altaffilmark{1},
James G. Bartlett \altaffilmark{11}
}
\altaffiltext{1}{Department of Physics and Astronomy, Johns Hopkins University, Baltimore, MD 21218}
\altaffiltext{2}{GEPI, Observatoire de Paris, CNRS, Univ. Paris Diderot; Place Jules Janssen, 92190 Meudon, France; simona.mei@obspm.fr}
\altaffiltext{3}{University of California, Berkeley, CA 94720}
\altaffiltext{4}{Lick Observatory, University of California, Santa Cruz, CA 95064}
\altaffiltext{5}{Herzberg Institute of Astrophysics, National Research Council of Canada, Victoria, BC  V9E2E7, Canada}
\altaffiltext{6}{Leiden Observatory, Postbus 9513, 2300 RA Leiden,
Netherlands.}
\altaffiltext{7}{University of California, Davis, CA 95616}
\altaffiltext{8}{Max-Planck-Institut fur Astrophysik, Karl-Schwarzschild-Strasse 1, D-85748 Garching bei Munchen, Germany}
\altaffiltext{9}{Space Telescope Science Institute, 3700 San Martin Drive, Baltimore, MD 21218}
\altaffiltext{10}{European Southern Observatory, Karl-Schwarzschild-Str. 2, D-85748 Garching, Germany}
\altaffiltext{11}{APC, 10 rue Alice Domon et L\'eonie Duquet, 75205 Paris Cedex 13, France (UMR 7164, CEA, CNRS, Obs. de Paris, Univ. Paris Diderot)}
\begin{abstract}


We apply detailed observations of the Color--Magnitude Relation (CMR) with the ACS/HST  
to study galaxy evolution in eight clusters at $z\approx$~1. 

 The early--type red sequence is well defined and elliptical and lenticular galaxies lie on similar CMRs. We analyze CMR parameters -- scatter, slope and zero--point -- as a function of redshift, galaxy properties and cluster mass. 
For bright galaxies ($M_B < -21$~mag), the CMR scatter of the elliptical population in cluster cores is smaller than that of the S0 population, although the two become similar at faint magnitudes  ($M_B > -21$~mag). While the bright S0 population consistently shows larger scatter than the ellipticals, the scatter of the latter increases in the peripheral cluster regions. If we interpret these results as due to age differences, bright elliptical galaxies in cluster cores are on average older than S0 galaxies and peripheral elliptical galaxies (by about $0.5$~Gyr, using a simple, single burst solar metallicity stellar population model). 

CMR zero point, slope, and scatter in the $(U-B)_{z=0}$ rest--frame show no significant evolution out to redshift $z \approx$~1.3 nor significant dependence on cluster mass.  Two of our clusters display CMR zero points that are redder (by $\approx 2 \sigma$) than the average $(U-B)_{z=0}$ of our sample. 

We also analyze the fraction of morphological early--type and late--type galaxies on the red sequence.
We find that, while in the majority of the clusters most (80\% to 90\%) of the CMR population is composed of early--type galaxies, in the highest redshift, low mass cluster of our sample, the CMR late--type/early--type fractions are similar ($\approx 50\%$), with  most of the late--type population composed of galaxies classified as S0/a. This trend is not correlated with the cluster's X--ray luminosity, nor with its velocity dispersion, and could be a real evolution with redshift.

\end{abstract}


\def\hst{{\it HST}}

\section{Introduction}

Observing the evolution of galaxy properties in clusters allows us to probe galaxy formation on the peaks of the dark matter distribution. In particular, cluster cores harbor most of the early--type galaxy population in the universe and are therefore ideal environments to constrain the formation epoch of these galaxies and their assembly history, a key issue for galaxy 
formation theories.  

In the local universe, galaxies follow well--defined relations, such as the ubiquitous relation between galaxy color and magnitude (color--magnitude relation, hereafter CMR.  Bower et al. 1992; van Dokkum et al. 1998, Hogg et al. 2004; L\'opez--Cruz et al. 2004; Baldry et al. 2004; Bell et al. 2004, Bernardi et al. 2005; McIntosh et al. 2005; De Lucia et al. 2006; Gallazzi et al. 2006). The CMR displays a bimodal galaxy distribution with a tight red concentration defining what is called the red sequence and more diffuse blue distribution known as the blue cloud.   The origin of this segregation is central
to understanding the processes driving galaxy formation.  Notably, most of the early--type galaxy population 
lies on the red sequence, while the majority of star--forming galaxies fall within the blue cloud.  
The existence of the red sequence indicates that star formation has been reduced, or quenched, for most early--type
galaxies, an important clue to their evolution.  Unless stated otherwise, we hereafter use the term CMR in this paper to refer to the early--type galaxy CMR, i.e., the red sequence.

Two main processes appear responsible for building the red sequence:  quenching of star formation in galaxies in the blue cloud, and merging of less luminous, already red galaxies (see also e.g. Bell et al. 2004, van Dokkum 2005).  The relative importance of the two is not clear, and the mechanisms that cause quenching are not yet well understood. 
In this light, it is significant that a decrease in the S0 population is observed in high redshift clusters (e.g. Dressler et al. 1999; Smith et al. 2005; Postman et al. 2005; Desai et al. 2007); one explanation is that late--type galaxies falling into the cluster potential  have undergone quenching and a morphological transformation, thereby ``migrating''  onto the 
red sequence as early--type galaxies (probably S0 galaxies.  See, e.g. Poggianti et al. 2006; Moran et al. 2007; 
Tran et al. 2007).  

Faber et al. (2007) give a good overview of our current understanding of the way this migration occurs.
The results of Poggianti et al. (2006) suggest that only part of the current early--type population experienced infall and quenching, while another part  constitutes a pristine, older galaxy population established during the early moments of cluster formation.  For these authors, the latter population would correspond to the cluster elliptical population, while the S0 population results from quenched galaxies.

A wealth of ground--based and space--based observations have shown that the CMR exists out to redshift $ z \sim$1 (Ellis et al.\ 1997; Stanford, Eisenhardt, \& Dickinson 1998; van Dokkum et al. 2000, 2001; Blakeslee et al. 2003a; Bell et al. 2004; De Lucia et al. 2004, 2007; Holden et al. 2004; Lidman et al. 2004; Tanaka et al. 2005;  Blakeslee et al. 2006;  Cucciati et al. 2006; Franzetti et al. 2007; Homeier et al. 2006; Mei et al. 2006a,b; Stanford et al. 2006; Willmer et al. 2006; Cooper et al. 2007; Tanaka et al. 2007; Arnouts et al. 2008).   CMR cluster studies have also been extended to redshifts as high as $2<z<4$ by targeting known proto-clusters believed to be the progenitors of $10^{14-15}$ $M_\odot$ galaxy clusters that
later virialize between $z\sim1$ and $z=0$ (Steidel et al. 1998, 2005; Pentericci et al. 2000; Venemans et al. 2002, 2007; Kurk et al. 2004; Overzier et al. 2008).  For example,
Kodama et al. (2007) and Zirm et al. (2008) recently pushed CMR studies to $z>2$ and find an excess of red galaxies  around radio galaxies, suggesting that the bright end of the CMR (galaxies with masses $M_*>10^{11}$ $M_\odot$) may already be in place at $z\approx2$ (but
not at $z\approx3$). The significance of these results, however, must await spectroscopic confirmation.

Steidel et al. (2005) find that protocluster galaxies are on average older and more massive than similar galaxies in the field, although there is no evidence for a correlation of morphology with environment at these redshifts (Peter et al. 2007; Overzier et al. 2008). 
If these structures are representative of massive clusters, this would suggest that their high density environments accelerate galaxy evolution compared to more average environments, so that their assembly epoch can be considered as an upper limit to that of the cluster CMR. 

Cassata et al. (2008) studied the rest--frame CMR between redshifts $z$=1.4 and 3, by combining spectroscopy from the Galaxy Mass Assembly ultradeep Spectroscopic Survey (GMASS) with GOODS multi--band photometry to obtain a field galaxy sample of 1021 galaxies down to magnitude m(4.5$\mu$m)=23mag. They distinguish bimodality in the color--stellar mass plane out to $z=2$. At $z>2$ they find red galaxies ($M > 10^{10} M_{\sun}$), but the bimodality is no longer observed. The fraction of early--type galaxies on the red sequence decreases from 60--70\% at z$<$0.5 to 50\% at $z=2$.

The CMR is usually characterized by a linear relation, defined by a zero point, slope and color scatter. These three parameters depend on stellar population age and metallicity.  The lack of strong evolution in the slope and scatter back to z$\sim$1 suggests that the CMR primarily reflects a metallicity-mass relation (i.e. metallicity--magnitude), while the scatter around the CMR is mainly due to galaxy age variations (e.g., Kodama \& Arimoto 1997; Kauffman \& Charlot 1998; Bernardi et al. 2005). 
From an analysis of SDSS early--type galaxies, Bernardi et al. (2005) and Gallazzi et al. (2006) concluded 
that the relation between galaxy luminosity (magnitude) and stellar population (colors) arises mainly through a dependence 
on galaxy velocity dispersion/stellar mass; both metallicity and luminosity--weighted age increase with stellar mass. 
The intrinsic color scatter  around the CMR, on the other hand, appears driven principally by galaxy age, with a small contribution from metallicity variations. This implies that accurate galaxy color measurements (e.g. when the intrinsic scatter 
can be measured because of small uncertainties on galaxy colors) can be used to constrain galaxy formation ages. 

At both low and high ($z \approx$~1) redshift, wide ground--based surveys have identified some general trends in the CMR.  Using a large sample (55,158 galaxies) of local ($0.08 < z < 0.12$) galaxies selected from the Sloan Digital Sky Survey (SDSS; York et al. 2000), Hogg et al. (2004) find a CMR with remarkably stable parameters for bulge--dominated galaxies in different environments, with changes in the metallicity and age of the red population of less than 20\%, according to Bruzual \& Charlot (2003) stellar population models.  Baldry et al. (2004) studied a different subsample of the SDSS data, at lower redshift ($0.004 < z < 0.08$; 207,654 objects, most with spectroscopic observations), fitting the red and blue peaks of the CMR with a double Gaussian and deriving best fits for both relations over a large range in magnitude, $-23.5<M_r<-15.5$~mag. These authors show that even if a linear fit is a good approximation to the red sequence (and the blue sequence) for bright magnitudes, a linear plus a {\em $\tanh$} function fit is necessary to cover the entire magnitude range of their sample. The mean position of the red sequence does not change significantly with environment in their local sample (Balogh et al. 2004; Baldry et al. 2006). They found a strong dependence of the red fraction with environment and stellar mass (with the galaxy red fraction increasing with both projected neighbor density and galaxy stellar mass), consistent with predictions from semi-analytical models based on the Millennium simulation (Bower et al. 2006, Croton et al. 2006). 

Cucciati at al. (2006) and Franzetti et al. (2007) performed a similar kind of analysis out to $ z \approx$~1.5 with a sample of $\approx$~6000 galaxies from the VIMOS--VLT Deep Survey. By comparing local and high ($z \approx 1$) redshift samples, they show that the CMR distribution is not universal, but rather depends on redshift and environment. While in the local Universe they found (as is commonly found) a dominance of red--sequence galaxies in overdense regions -- with less dense regions mostly populated by blue galaxies -- at higher redshifts they suggest that this trend might possibly be reversed, with a more pronounced presence of blue galaxies in higher density regions. The inversion is mainly observed at $z > 1$, with the red population equally distributed in different environments at $0.9 < z < 1.2 $ (e.g. the fraction of red galaxies does not depend on environment; see also Cooper et al. (2007) for a different interpretation of these results), and an increase of blue galaxies in high density regions at $1.2 < z < 1.5$. A high fraction (35 --40\%) of their red--sequence galaxies turned out to be star--forming galaxies, showing the importance of good morphological or spectroscopical classification to studies of early--type galaxies on the red sequence.

Cooper et al. (2007) performed a similar analysis on a much larger sample of 19,464 field and group galaxies from the DEEP2 Galaxy Redshift Survey ($0.4<z<1.35$). 
They found a highly significant relation between galaxy red fraction and environment at $z \approx 1$, which disappears at $z > 1.3$, contrary to what Cucciati et al. found. Exploring this difference in detail, they pointed out that the two results are consistent if the larger uncertainties inherent in the smaller sample of Cucciati et al. and differences in data analysis techniques are taken into account. With a better sample and a more detailed analysis, they conclude that a significant relation between red fraction and environment still exists  at $z\approx 1 $, demonstrating that a reversal of the color--density relation is not confirmed by the data. While the fraction of galaxies on the red sequence decreases with redshift in overdense environments (mainly groups of galaxies), it remains constant in the field. The two become comparable at $z\approx1.3$ (see also Gerke et al. 2007). Their results support a scenario in which local environment was important in quenching star formation and populating the red sequence in overdense environments.

The present observational situation highlights the importance of detailed studies of galaxies at redshifts $ 0.5 < z < 1.5 $ 
that combine both accurate color measurements and morphological classifications.
Although the wealth of ground--based observations has lead to the identification of significant trends in the color--magnitude relation, only the high resolution, high sensitivity observations afforded by the Advanced Camera for Survey (ACS; Ford et al. 2002) on the Hubble Space Telescope (HST) permit these two essential measurements, otherwise impossible from the ground at high redshift: 1) the morphological classification of galaxies as Hubble types out to $z \approx$~1 (e.g. Postman et al. 2005) and 2) galaxy color measurements to an accuracy of a few percent of a magnitude (Sirianni et al. 2005; Blakeslee et al. 2003a).

This is precisely one of the main goals of the the ACS Intermediate Redshift Cluster Survey 
(Ford et al. 2004; Postman et al. 2005; see also Table~\ref{sample}).  As part of this survey, we observed eight X--ray luminous galaxy clusters with redshifts between 0.8 and $\sim$1.3.  This is now the best sample available in terms of multi--wavelength observations and spectroscopic follow-up of known clusters at $ z \sim$~1. While eight clusters do not constitute a large sample when compared to ground--based galaxy cluster samples in the local universe,  this is the best sample available with ACS morphological classification and high precision color measurements.  And it provides the opportunity to take a closer look at average color trends observed from the ground and find correlations between these trends and galaxy morphology and color--derived ages.

Color--magnitude relations (CMRs) were presented for each cluster in a series of dedicated papers (Blakeslee et al. 2003a, 2006; Homeier et al. 2006; Mei et al. 2006a,b, hereafter the CMR paper series).
The principal aim of the CMR paper series was to constrain galaxy ages and study  variations in CMR parameters as a function of galaxy morphology and structural properties (e.g. effective radii, ellipticities, surface brightness). 
We give the mean luminosity weighted ages derived for the elliptical population in Table~\ref{sample}, using stellar population models from Bruzual \& Charlot (2003; hereafter BC03). We found elliptical population ages ranging from  2.5 to 3.5~Gyr, depending on  cluster redshift, with an average formation redshift $z_f \gtrapprox 2$.  Early--type galaxy masses range from $\approx 10^{11}$ to $\approx 10^{12}  M_{\sun}$  (Holden et al. 2006; Rettura et al. 2006). Galaxy masses were estimated using galaxy color or SEDs, with an error in mass of $\approx 40\%$.
CMR scatter was shown to increase slightly at faint luminosities and with distance from cluster X--ray emission centers. This suggested that fainter (and thus less massive) and more peripheral galaxies have a larger age dispersion than bright central galaxies. This dependence on galaxy luminosity/mass and environment is also observed in local samples (Hogg et al. 2004; Bernardi et al. 2005; McIntosh et al. 2005; Gallazzi et al. 2006).

In this paper, 
we use the full sample to systematically investigate trends in CMR parameters and their dependence on redshift and galaxy cluster properties. Our sample is discussed in Section~2.  In Section~3 we present our CMR measurements, in Section~4 our results, and we conclude with Section~5.

We adopt the WMAP cosmology (Spergel et al. 2007) $\Omega_m h^2 =0.137$,
 $\Omega_{\Lambda} =0.72$, $h=0.70$)  as our standard cosmology.
All ACS filter magnitudes are given in the AB system (Oke \& Gunn 1983; for ACS see Sirianni et al. 2005), while magnitudes in the Johnson system (Johnson \& Morgan 1953; Buser \& Kurucz 1978; Bessel 1990) are given as Vega magnitudes (see also Appendix~II).

\section{The ACS Intermediate Redshift Cluster Survey}

\subsection{The sample}

The ACS Intermediate Redshift Cluster Survey includes eight clusters with redshifts between 0.8 and 1.27.
Five of the clusters were identified from the ROSAT Deep Cluster Survey (Rosati et al. 1998), while MS1054--03 comes from the Einstein Extended Medium Sensitivity Survey (Gioia \& Luppino 1994) and the clusters CL1604+4304 and CL1604+4321 were found in a Palomar deep near--infrared photographic survey (Gunn et al. 1986). Recently, CL1604+4304 and CL1604+4321 were observed in the X-ray by Lubin et al. (2004) and Kocevski et al. (2008), who detected CL1604+4304 and set an upper limit on emission from CL1604+4321. 

Table~\ref{sample} shows the principal properties of this sample (see also Table~1 and the sample description in Ford et al. 2004 and Postman et al. 2005). It spans cluster bolometric X--ray luminosities from $\sim$~1.5 to $\sim$~28~$\times$~$10^{44} h^{-2}$~ergs/sec, velocity dispersions from  $\sim$~600 to $\sim$1200~km/s, and estimated total masses from $\sim$~1.3 to 21~$\times$~$10^{14} M_{\odot}$.  We have measured accurate redshifts with spectroscopic follow--up for a large sample of galaxies in most of these clusters (van Dokkum et al. 2000; Tran et al. 2005, 2007; Demarco et al. 2005, 2007). Where available (all clusters except CL1604+4304 and CL1604+4321), bolometric X--ray luminosities and total cluster mass estimates (dark and visible matter) are taken from Ettori et al. (2004), which gives us as homogeneous a sample as possible.  We use $R_{200}$, defined as the radius at which the cluster mean density is 200 times the critical density, as an approximation for the virial radius; in this paper, it is derived from the cluster velocity dispersion (as per Carlberg et al. 1997).

These eight clusters are still in the process of forming, showing filamentary and clumpy structures (Gal \& Lubin 2004; Gal et al. 2005; Nakata et al. 2005; Tanaka et al. 2005; Tanaka et al. 2007;  Gal et al. 2008). X--ray luminosities and velocity dispersions deviate from the standard $L_X$~vs~$\sigma$ relation for clusters (e.g. Wu et al. 1999; Rosati et al. 2002; Mei et al. 2006a), implying that neither X--ray luminosity nor cluster velocity dispersion can be used as unbiased proxies of cluster mass. In general, X--ray luminosity is very sensitive to processes in cluster cores and can be enhanced by substructure and merging of sub--clumps, while velocity dispersions can be boosted by infalling substructures. We will use both in our analysis, but keeping this {\it caveat} in mind.

Both MS~1054--03 and RX~J0152.7--1357 display complex structure in the X--ray and the optical.  We observe central cluster clumps surrounded by minor satellite groups (Gioia et al. 2004; Demarco et al. 2005; Tran et al. 2005; Jee et al. 2005a; Jee et al. 2005b; Tanaka et al. 2005). In MS~1054--03 the different peaks in the X--ray and optical distributions are not well separated, while in RX~J0152.7--1357 there are two distinct central clumps (a northern and a southern clump; Maughan et al. 2003), contained within well--defined circular regions identified by Demarco et al. (2005). The velocity dispersions in these clusters are higher than expected 
from a simple linear $\sigma - L^X_{bol}$ relation, meaning that
 $R_{200}$ could be overestimated. 
For RX~J0152.7--1357, Jee et al. (2005a) estimated $R_{200} = 1.14 \pm 0.23$~Mpc for the entire cluster, from a weak lensing analysis, while Girardi et al. (2005) derived $R_{200} = 1.3 $~Mpc for the northern clump and $R_{200} = 0.5 $~Mpc for the southern clump (in our adopted cosmology).
For MS~1054--03, a virial radius of  $1.7 \pm 0.2$~Mpc is found from the cluster X--ray emission and an isothermal model, consistent with the virial radius of $1.5 \pm 0.1$~Mpc obtained by the weak lensing analysis  of Jee et al. (2005b).  Both estimates are also consistent with the virial radius estimated from the cluster velocity dispersion in Table~\ref{sample}.

The clusters CL1604+4304 and CL1604+4321 are part of a complex supercluster (Lubin et al. 2004; Gal et al. 2005; Gal et al. 2008) with eight spectroscopically confirmed galaxy clusters and groups.  CL1604+4304 is the more X--ray luminous cluster (Kocevski et al. 2008) and shows a well-established intra-cluster medium, while  CL1604+4321 shows evidence of ongoing collapse and appears to be a less massive structure (Gal et al. 2008). 

RDCS~J0910+5422 has a low velocity dispersion despite its high X--ray luminosity (Mei et al. 2006a). It is also part of a extended supercluster (Tanaka et al., 2008).

The three clusters at $z > 1.2$ show filamentary structures observed in the regions around RDCS~J1252.9-2927 and the two clusters RX~J0849+4452 and RX~J0848+4453 by Tanaka et al (2007) and Nakata et al. (2005), respectively.  
The clusters RX~J0849+4452 (hereafter Lynx E) and RX~J0848+4453 (hereafter Lynx W) define the so--called Lynx Supercluster, the largest superstructure known at these redshifts. Lynx E is likely to be more dynamically evolved than Lynx W.
In fact, Lynx E has a more compact galaxy distribution, while the galaxies in Lynx W are more sparsely distributed in a filamentary structure and it lacks an obvious central bright cD galaxy (e.g. Mei et al. 2006b); X-ray observations by 
Chandra support these dynamical characteristics (Rosati et al. 1999; Stanford et al. 2001; Ettori et al. 2004). Virial radii derived from X--ray profiles and the cluster velocity dispersion in Table~\ref{sample} are consistent with those found by the weak lensing analysis of Jee et al. (2006).

\subsection{Observations and data reduction}

Each cluster was observed in at least two ACS WFC (Wide Field Camera) band-passes chosen to straddle the Balmer break and corresponding approximately to rest--frame Johnson $U$ and $B$ bandpasses (see Table~1 of Postman et al. 2005).  The ACS WFC resolution is 0.05~\arcsec/pixel, and its  field of view is 210\arcsec x 204\arcsec. 

A detailed description of the observations can be found in Ford et al. (2004) and in Table~1 in Postman et al. (2005). We summarize here the main characteristics of the survey. The two most massive clusters at $ z \approx$~0.8 have been observed in more than two band--passes --  MS~1054--03 in the $V_{606} (F606W), i_{775} (F775W), z_{850} (F850LP)$ filters, and RX~J0152.7--1357 in the $r_{625} (F625W), i_{775}, z_{850}$ filters -- both clusters for a total of 24 orbits. These two clusters were observed with a
pattern of 2x2 ACS pointings covering a region of 5'x5'.

The clusters CL1604+4304 and CL1604+4321 were observed for a total of 4 orbits each in the  $V_{606}$ and $I_{814} (F814W)$ bandpasses. The four clusters at $z > 1$ were observed in the $i_{775}$ and $z_{850}$ filters, with mosaics taken over each region for 3 orbits in the $i_{775}$ and 5 orbits in the  $z_{850}$ filter.

 The images were processed with the APSIS pipeline (Blakeslee et al. 2003b), with a {\it Lanczos3}  interpolation kernel.
Our ACS photometry was calibrated to the AB system, with  synthetic photometric zero-points from Sirianni et al. (2005) and reddening from Schlegel et al. (1998).  For most of our clusters, ground--based optical and near--infrared data are available and were used to select galaxies for spectroscopic follow--up.

Spectroscopically confirmed galaxy members were obtained from van Dokkum et al. (2000), Tran et al. (2005, 2007) and Demarco et al. (2005) for MS~1054--03 and  RX~J0152.7--1357 (see Blakeslee et al. 2006);  from Gal \& Lubin (2004) for CL1604+4304 and CL1604+4321 (see Homeier et al. 2006); from Stanford (private communication; see Mei et al. 2006a) for RDCS~J0910+5422; from Demarco et al. (2007) for RDCS~J1252.9-2927; and from Rosati et al. (1999), Stanford et al. (1997) and Holden et al. (in preparation) for the Lynx Supercluster (see Mei et al. 2006b). 
Spectroscopically confirmed interlopers have been excluded from our analysis.


 
\section{Color--Magnitude Relation in the ACS Intermediate Cluster Survey}

\subsection{Galaxy sample selection}

In this analysis we concentrate on the early--type red sequence.
We select early--type CMR galaxy candidates using the visual morphological classification from Postman et al. (2005), ACS galaxy colors and, when available, ground--based infrared photometry and spectroscopy (Blakeslee et al. 2003a, 2006; Homeier et al. 2006; Mei et al. 2006a,b). 

First, we use the catalogs of Postman et al. (2005) to select early-type galaxies. From here on, the terms early--type and late--type galaxy refer, respectively, to galaxies morphologically classified as elliptical and S0, and as spiral galaxies according to Postman et al. (2005). 
Thanks in particular to the high angular resolution and sensitivity of the ACS, the Postman et al. (2005) visual morphological classification distinguishes two different classes of early--type galaxies -- elliptical and S0 -- and different classes of late--type galaxies.  
Of the latter, however, we only considered the S0/a as a separate class in this work.

In MS~1054--03 and RX~J0152.7--1357, we considered only spectroscopically confirmed cluster members.
For all clusters, following the morphological selection, we perform a color cut to further isolate the likely cluster members. 
To this end, we used SExtractor (Bertin \& Arnouts 1996)  photometry in
{\it dual-image mode}, as in Ben\'{\i}tez et al. (2004). This means that object detection employed the two filters simultaneously, and object fluxes were then measured independently in the two filters using the same object coordinates and apertures. 
This color selection was performed to extract galaxies over the color range $0.1 \lessapprox (U-B)_{z=0} \lessapprox 0.8$~mag. 

\subsection{Measurements of galaxy color and magnitude}

In order to accurately determine the early--type CMR, we made precision color measurements on the galaxies selected as 
described above.  Aiming to avoid systematics due to internal galaxy gradients, our final colors
were measured inside a circular aperture scaled by the galaxy average half-light radius $R_e$ (van Dokkum et al. 1998, 2000; Scodeggio 2001).  The primary effect of internal galaxy gradients on this sample would be a steepening of the CMR slope (e.g.,
by $\sim$50\% when isophotal colors from SExtractor are used). Our $R_e$ values were
derived by fitting elliptical Sersic models to each galaxy image  using the program GALFIT (Peng et al. 2002). In the fit we constrained the {\it Sersic} index to $ 1 \leq n \leq $ 4. Our final results do not change (within the uncertainties) if the  effective radii are calculated via a two component 
(Sersic bulge + exponential disk) surface brightness decomposition 
technique using GIM2D (Marleau \&  Simard 1998; Rettura et al. 2006) that better fits 
the galaxy light profile (Mei et al. 2006a).

We removed blurring effects due to different PSFs by deconvolving galaxy images with the CLEAN algorithm (H{\"o}gbom et al. 1974).
Final colors were measured on the deconvolved images within a circular aperture equal to $R_e \sqrt{q}$ (see Blakeslee et al. 2006), where $q=\frac{b}{a}$ is the axial ratio obtained from GALFIT. When $R_e < 3$ pixels, we set it equal to 3 pixels. Our median $R_e$ is $\approx$~5~pixels (0.25$\arcsec$, $\approx$~2~kpc at $ z \approx$~1). 

For most of the clusters, we estimate color errors by adding the uncertainty due to flat fielding, PSF variations, and ACS pixel--to--pixel correlations in quadrature to the flux uncertainties (Sirianni et al. 2005). The images covering MS~1054--03 and RX~J0152.7--1357 were processed both individually and as a large mosaic (see details in Blakeslee et al. 2006) to assess the color measurement uncertainties. With the high sensitivity of ACS, we reach on average color uncertainties of 0.01 and 0.03~mag, an impressive achievement of 
HST for galaxies at these high redshifts. 

We used SExtractor's MAGAUTO as an estimate of galaxy total magnitude. As pointed out by Ben\'{\i}tez  et al. (2004), Giavalisco et al. (2004) and Blakeslee et al. (2006), comparison to other measures suggests that MAGAUTO is an imperfect estimator. Specifically, 
Ben\'{\i}tez  et al. (2004) required a 5th order polynomial to describe the relation between MAGAUTO and the difference between MAGAUTO and the asymptotic isophotal Sextractor magnitude.  Over the magnitude range 
$ 20.5 \lessapprox i_{775} \lessapprox 23.5$~mag, Blakeslee et al. (2006) found a constant shift of 0.2~mag 
between GALFIT and MAGAUTO total magnitudes.
Giavalisco et al. (2004) discovered a similar systematic offset of $\approx$~0.2-0.3~mag between MAGAUTO and simulated spheroid magnitudes for the magnitude range of our sample. Bertin \& Arnouts (1996), Giavalisco et al. (2004), and H{\"a}ussler et al. (2007) found that both GALFIT and SExtractor magnitudes give estimates fainter than real magnitudes by a quantity that depends on galaxy surface brightness and the sky brightness determination.  We will not correct our MAGAUTO ACS magnitudes in this paper but, keeping this in mind, will warn the reader when a comparison to other samples is made.

\subsection{Measurements of galaxy properties and projected density} 

Each galaxy is described by its {\it ellipticity} (defined as 1-$q$),  average half-light radius $R_e$, and {\it Sersic} index $n$.  These parameters are
found by fitting elliptical Sersic models to each galaxy image using GALFIT. As described above, $q=\frac{b}{a}$ is the axial ratio obtained from GALFIT and we constrained $ 1 \leq n \leq $ 4.

Postman et al. (2005) provide the {\it neighbor galaxy projected density} $\Sigma$ for each galaxy.  
These densities were calculated using the distance to
the 7th nearest neighbor (Postman et al. 2005 for details).  Both the nearest--N--neighbor approach and friends--of--friends algorithm gave consistent results (Postman et al. 2005), indicating the robustness of this measurement.  
A good estimate of galaxy densities implies the ability to correct for fore/background galaxy contaminants. The density
estimate is more accurate for some of clusters (MS~1054--0321, RX~J0152--1357, and RDCS~J1252--2927) where spectroscopic or photometric redshift information was available; this enabled us to exclude fore/background objects (see the Appendix in Postman et al. 2005). When redshift information was not available, a statistical background correction was applied with the {\it caveat} that it is only reliable in dense regions ($> 80$~$Gal/Mpc^2$). Statistical uncertainties on the galaxy projected density are estimated to be $\approx$0.2~dex in $Gal/ Mpc^{2}$ ($\approx$~0.2~$log_{10} \Sigma $; Postman et al. 2005).

\begin{figure*}
\centerline{\includegraphics[scale=0.5,angle=180]{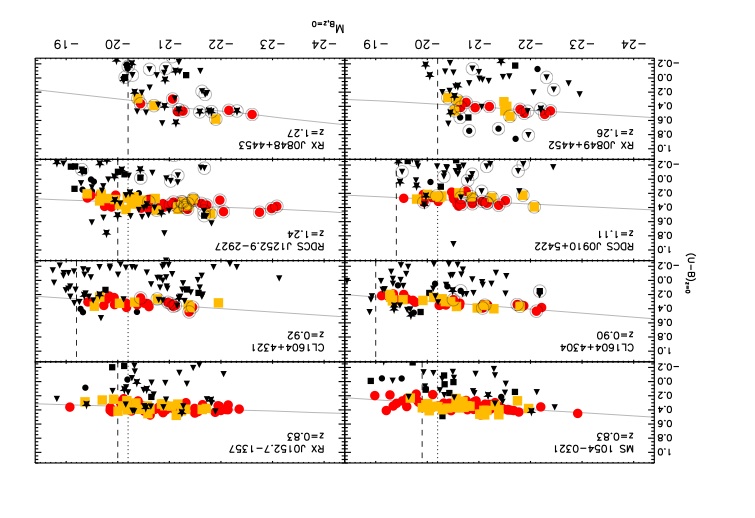}}
\caption {Color--magnitude relations for galaxies in rest--frame $(U-B)_{z=0}$ color vs absolute rest frame $B$ magnitude $M_{B,z=0}$ for the eight clusters of the ACS Intermediate Cluster Survey.  Galaxies within $R_{200}$ are shown. Circles indicate elliptical galaxies; squares, S0s; stars, S0/a; and triangles, spirals. Large colored symbols identify galaxies within three times the scatter of the CMR in each cluster.  The continuous line gives the CMR for the elliptical sample of each cluster calculated within $R_{200}$, at approximatively the same $M_*$ limit (shown by the vertical dashed line). The dotted line shows the same rest--frame $M_{B,z=0}=-20.2$~mag. Galaxies plotted in the two most massive clusters, RX~J0152.7-1357 and MS~1054-0321 are all spectroscopically confirmed. Circles around symbols denote spectroscopically confirmed members in the other clusters. S0 and elliptical galaxies lie on the same CMRs, apart from RDCS~J0910+5422, in which the S0 CMR has a bluer zero point. {\label{cmr}}}
\end{figure*}

\subsection{CMR parameter estimation}

We employ three parameters in our CMR fits:  the zero point, slope and scatter around the mean CMR:
\begin{equation}
ACS \ Color = c_0 + Slope \times (ACS \ mag - 22.5)
\end{equation}
Table~\ref{fitcmr_mlm} lists the ACS colors (second column) and magnitudes (third column) that were used to derive our CMR parameters. For MS~1054--03 and RX~J0152.7--1357,  we used $(V_{606}-z_{850})$ and the $(r_{625}-z_{850})$ colors respectively, because they are more sensitive to stellar population changes and less sensitive to photometric errors and small changes with redshift, as pointed out in Blakeslee et al. (2006).

The color-magnitude relation was fitted using a robust linear fit based on Bisquare weights (Tukey's biweight; Press et al. 1992), and the uncertainties on the fit coefficients were obtained by bootstrapping on 1,000 simulations. The scatter around the fit was estimated from a biweight scale estimator (Beers, Flynn \& Gebhardt 1990) that is insensitive to outliers in the same set of bootstrap simulations.
A linear least--squares fit with three-sigma clipping and standard rms scatter gives similar results to the biweight scale estimator within $\approx$~0.001-0.002~mag for the slope and the scatter. 

\begin{figure*}
\centerline{\includegraphics[scale=0.5,angle=180]{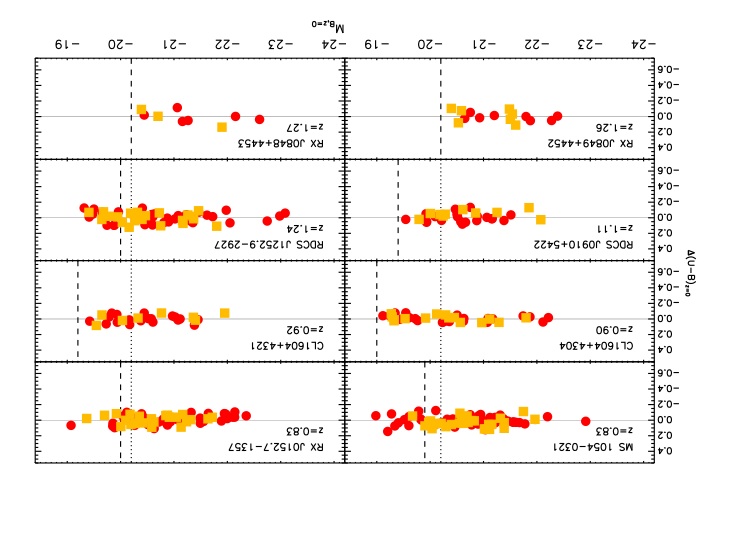}}
\caption {Color residuals (the individual cluster CMR was subtracted from the rest frame $(U-B)_{z=0}$ color)  vs $M_{B,z=0}$ for the early--type galaxies analyzed in this paper.  Red circles and yellow squares show elliptical and S0 galaxies contained within three times the scatter of each cluster's CMR. The continuous line indicates the zero level.  {\label{res_cmr}}}
\end{figure*}

To estimate the {\it intrinsic} galaxy scatter (i.e., not due to galaxy color measurement uncertainties), we estimated the additional scatter
needed beyond the measurement error to make the observed  $\chi^2$ per degree of freedom of the fit equal to one.
Again, the uncertainty on the internal scatter was calculated by bootstrapping on 1,000 simulations. 

CMR zero points, scatter and slopes were calculated for the elliptical and the early--type (elliptical plus S0) galaxy population in all of our clusters for a variety of sub--samples taken from different spatial regions, local densities, and absolute magnitude intervals. We considered:

\begin{enumerate}

\item {\it Two projected radial regions}, one within $0.5R_{200}$  and the other within $R_{200}$ (this differs from the CMR paper series, where CMRs were fitted  within regions scaled by a radius of two arcminutes from the center of the cluster [taken as the center of the X--ray emission], a scale that corresponds to $\approx 1 Mpc$ at these redshifts in the WMAP cosmology); 

\item  {\it Galaxies from dense and less dense regions}.   We define dense regions as those with $\Sigma > 100$~$Gal/Mpc^2$ and compare them to the full sample, corresponding to $\Sigma > 10$~$Gal/Mpc^2$; 

\item  {\it Two different magnitude ranges.} We define two different magnitude ranges.
The first one corresponds to about one magnitude fainter than the characteristic magnitude $M_*$ at the cluster redshift (that corresponds to $\approx 0.5 L_*$, e.g., we are probing the same range in $M_*$) and it is the standard magnitude range used in our CMR paper series. For clusters at $z \approx 0.8$ the magnitude limit was $i_{775} = 23$~mag  ($m_*$ is equal to $i_{775} = 22.3$~mag at these redshifts from Goto et al. 2005), at $z \approx 0.9$ it was $I_{814}=24$~mag and  for clusters at $z > 1$ it was $z_{850} = 24$~mag  ($m_*$ equal to $z_{850} = 22.7$~mag in RDCS~J1252.9-2927 from Blakeslee et al. 2003a,  and $m_*$ equal to $z_{850} = 22.6^{+0.6}_{-0.7}$~mag for RDCS~J0910+5422 from Mei et al. 2006a). These values correspond to an evolution of the galaxy luminosity function as published in the recent literature (e.g. Norberg et al. 2002; Bell et al. 2004; Faber et al. 2007, and references therein).  We also fitted the CMR at the same rest--frame limiting magnitude $M_B = -20.2$~mag, which corresponds to $z_{850} = 24$~mag at $z = 1.26$ (see Appendix~II), the limiting magnitude for the Lynx clusters. Using two different magnitude ranges permits us to understand how our results depend on our assumptions concerning the uncertain evolution of the Schechter function (e.g. Faber et al. 2007).  For example, for RDCS~J0910+5422 Mei et al. 2006b derived a $m_*$  equal to $z_{850} = 22.6^{+0.6}_{-0.7}$~mag, corresponding to an absolute rest--frame B magnitude equal to $M_* \approx -21$~mag.  From Faber et al. (2007), we would expect  $M_* \approx -21.4$~mag at this redshift ($z=1.106$). 

\end{enumerate}

\section{Results}


Tables ~\ref{tcmr1} to~\ref{tcmr8} in Appendix~I summarize the results for our different samples (described in Sect. 3.4).  
They list fits to the original ACS color CMR and the conversion of those fit parameters to the Johnson Vega rest--frame $(U-B)_{z=0}$ color and absolute B magnitude $M_{B,z=0}$. Details of this conversion, using BC03 stellar population models, are given in Appendix~II. 
In most of our analysis, and when not stated otherwise, we use  the CMR parameters fitted over regions within the virial radius $R_{200}$ and over the same range in terms of $M_*$.
These results  are collected in Table~\ref{fitcmr_mlm}.

The $(U-B)_{z=0}$ rest--frame CMR is defined as:
\begin{equation}
(U-B)_{z=0} = c_{0,(U-B)} + Slope_{(U-B)} \times (M_{B,z=0} + 21.4)
\end{equation}
The zero point is very stable to changes in limiting magnitude and region (differing local densities and radii) used for the fit, 
while the slope and scatter show greater differences.  For example, 
the average difference in the CMR slope and scatter for most clusters when changing limiting magnitude is of  order 0.01 to 0.02~mag, the same as the uncertainties on the parameters estimated from our bootstrap procedure. The largest average difference ($\approx 0.03$~mag) in the slope is observed in CL1604+4304,  CL1604+4304 and RDCS~J0910+5422.
These results give us confidence in the stability of our analysis.

\begin{figure}
\centerline{\includegraphics[scale=0.37,angle=180]{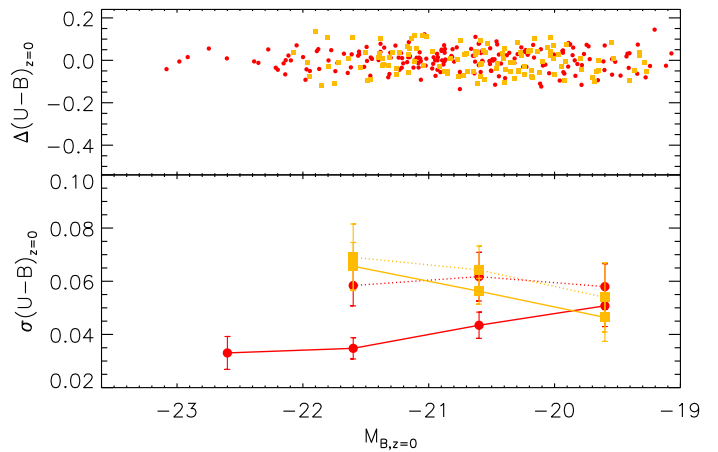}}
\caption {Top panel: Early--type color residuals in rest--frame $(U-B)_{z=0}$  vs $M_{B,z=0}$.   Red circles indicate ellipticals and yellow squares S0 galaxies within 3 times the CMR scatter of each cluster sample. Bottom panel: Intrinsic scatter averaged in bins of one magnitude. Red circles connected by lines show the elliptical galaxy intrinsic scatter, and yellow squares connected by lines, the S0 scatter. Intrinsic scatter within $0.5 R_{200}$ and over $ 0.5 R_{200} < R < R_{200}  $ are shown by continuous and dotted lines, respectively. The error bars give the uncertainties in each magnitude bin (see text for details).  Bright S0s, faint galaxies and elliptical galaxies in more peripheral regions all exhibit larger scatter than bright ellipticals in the core, suggesting younger ages (by $\approx 0.5$~Gyr, according to a simple single burst solar metallicity BC03 stellar population model).    {\label{averes_cmr}}}
\end{figure}

In Fig.~\ref{cmr} we show the early--type CMRs for the  sample of eight clusters in rest--frame $(U-B)_{z=0}$ color versus rest--frame $M_{B,z=0}$ magnitude. The continuous line traces the fit to the elliptical galaxy sample of each cluster taken from Table~\ref{fitcmr_mlm}.  Spectroscopically confirmed members are indicated by large circles around the galaxy symbols. For the two most massive clusters at $z$=0.8 we show only spectroscopically confirmed members. 

The primary characteristics of our sample are already visible from this overview of the CMR in rest--frame magnitudes:
The early--type red sequence is well defined and tight out to redshifts z~$\approx$~1.3. Elliptical and lenticular galaxies lie on similar CMRs. We observe the emergence of bright, blue late--type galaxies at higher redshifts and in less massive clusters, and which are not observed in local samples.

The elliptical and S0 CMR zero points in RDCS~J0910+5422 present an interesting case.  As already pointed out by Mei et al. (2006a), 
the S0 CMR zero point in  $(i_{775}{-}z_{850})$ for this cluster is bluer by $0.07 \pm 0.02$~mag with respect to the ellipticals. This corresponds to an age difference of $\approx$~1~Gyr, for a BC03 single-burst solar metallicity model, and suggests a transitional S0 population still evolving towards the bulk of the red sequence already defined by the elliptical galaxies. Alternatively, this offset could be the result of
a different star formation history. For example, when considering a solar metallicity model 
with an exponentially decaying star formation, we find an age of 3.5 Gyr, i.e., the S0s have evolved gradually from star forming progenitors  (Mei et al. 2006a).

\begin{figure}
\centerline{\includegraphics[scale=0.35,angle=180]{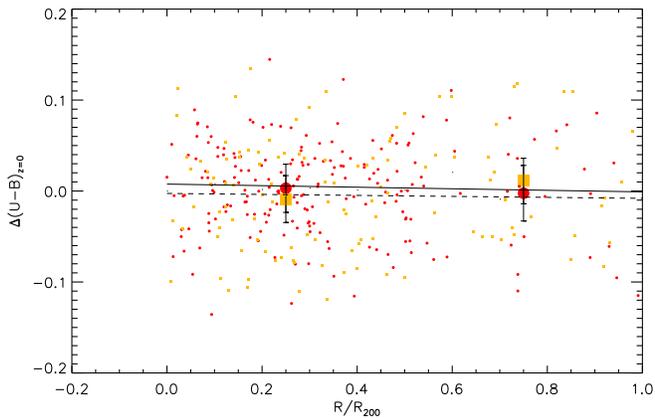}}
\caption {Color residuals in rest--frame $(U-B)_{z=0}$ vs $R/R_{200}$. Small red circles and yellow squares show individual elliptical and S0 galaxy residuals (within 3 times the CMR scatter of each cluster sample), respectively. The larger symbols give the average residuals for two radii bins:  $R < 0.5 R_{200}$ and $ 0.5 R_{200} < R < R_{200}  $. The continuous and dashed lines show the fit of this dependence for elliptical and S0 galaxies, respectively. The error bars represent the uncertainty in the average residual for each radius bin. The average CMR residuals do not change significantly with distance from the cluster center, or with early--type galaxy morphology.
   {\label{averes_cmr_bis}}}
\end{figure}

\subsection{CMR scatter}

The first parameter we will study in detail is the CMR scatter. As discussed in the Introduction, the scatter in the CMR gives us information on the average age of the cluster early--type galaxies  (e.g., Kodama \& Arimoto 1997; Kauffman \& Charlot 1998; Bernardi et al. 2005; Gallazzi et al. 2006; Tran et al. 2007; see the Introduction).

\begin{figure}
\centerline{\includegraphics[scale=0.35, angle=180]{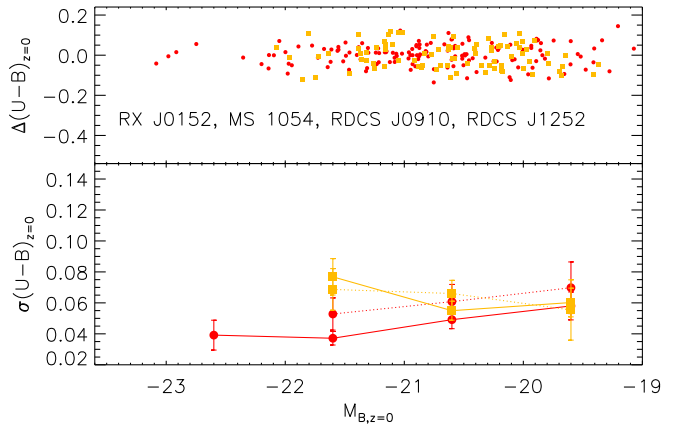}} 
\caption {Similar to Fig.~3: Color residuals and intrinsic scatter for a sample limited to clusters with spectroscopically confirmed members complete to magnitudes brighter than $M_{B,z=0} \approx -20.5$~mag, corresponding to the brighter three ranges of magnitude in the bottom panel. The difference between elliptical and S0 scatters is still significant.
{\label{fig3_sp}}}
\end{figure}

\subsubsection{CMR scatter as a function of galaxy magnitude and distance from the cluster center}

We first examine CMR scatter as a function of galaxy magnitude.
By subtracting the CMR--predicted colors from our individually measured galaxy colors in the $(U-B)_{z=0}$ rest--frame, 
we essentially eliminate a metallicity--mass dependence (e.g., Kodama \& Arimoto 1997). 
We can then measure the scatter of the $(U-B)_{z=0}$ rest--frame residuals, which is mainly driven by galaxy age (e.g., Kodama \& Arimoto 1997; Kauffman \& Charlot 1998; Bernardi et al. 2005; Gallazzi et al. 2006; Tran et al. 2007).
In Fig.~\ref{res_cmr}  the residuals of the early--type CMR (rest--frame colors from which we subtract the CMR fit for each cluster) are shown as a function of galaxy magnitude. 
\begin{figure}
\centerline{\includegraphics[scale=0.35, angle=180]{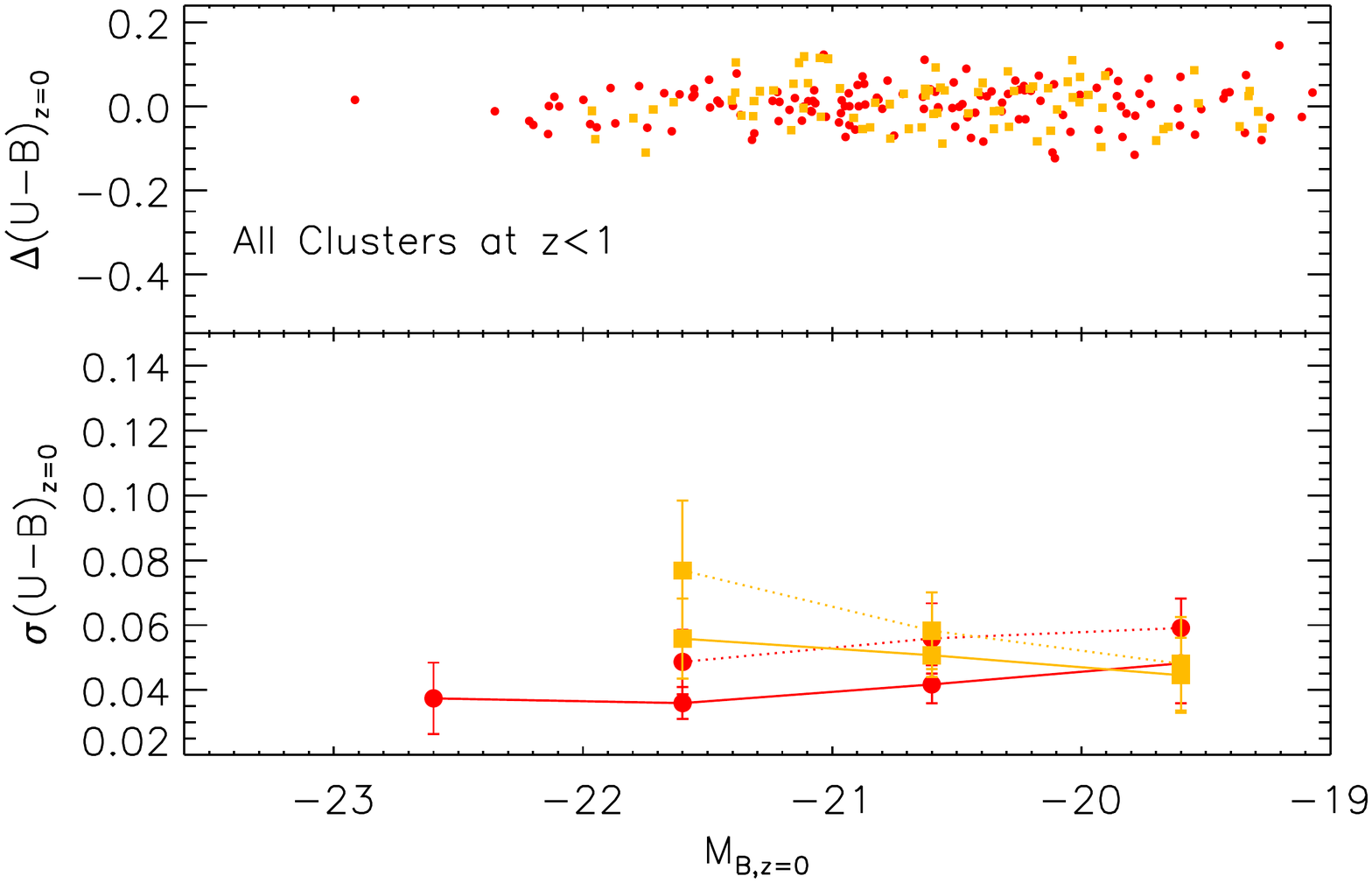}} 
\centerline{\includegraphics[scale=0.35, angle=180]{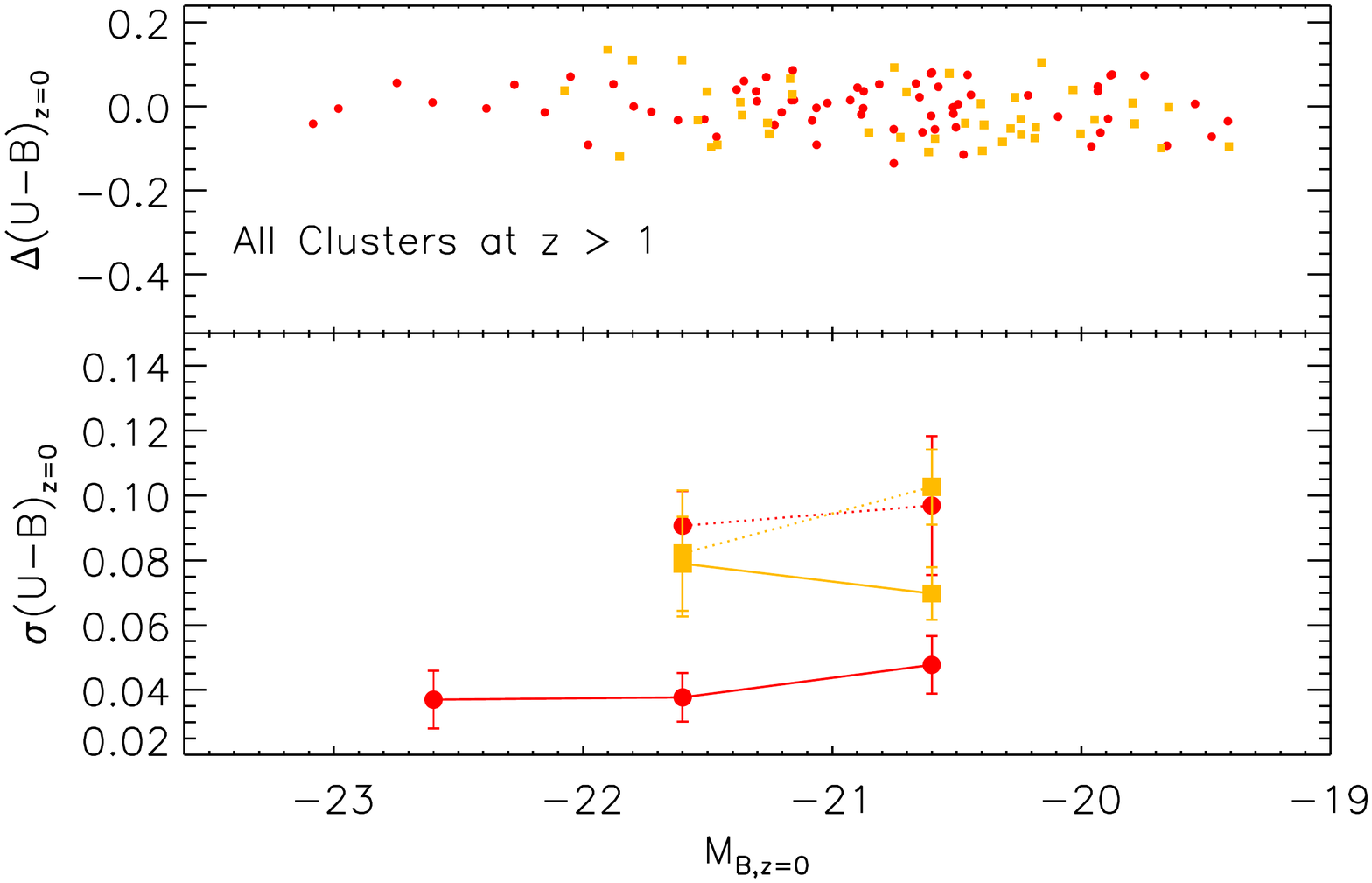}}
\caption {Similar to Fig.~3: All clusters at $z < 1$ (top) and all clusters at $z > 1$  (bottom). The S0 scatters are always significantly different. The elliptical scatter increases at higher redshifts in regions further from the cluster center. {\label{f3_z}}}
\end{figure}

The trend of average intrinsic scatter for different early--type populations (ellipticals and lenticulars) in the entire sample is shown in Fig~\ref{averes_cmr}. 
The top panel of this figure shows the early--type galaxy color residuals (within 3 times the CMR scatter) in rest--frame $(U-B)_{z=0}$ color vs absolute  rest frame $M_{B,z=0}$ magnitude. The bottom panel shows the intrinsic scatter (calculated as described in Sect.~3.4) averaged in bins of one magnitude as a function of $M_{B,z=0}$ and of distance from the cluster center (for $R < 0.5 R_{200}$ and over $ 0.5 R_{200} < R < R_{200} $).
The error bars give the uncertainty on the average intrinsic scatter, calculated as a standard deviation by bootstrapping on 1,000 simulations (Sect.~3.4).

\begin{figure}
\centerline{\includegraphics[scale=0.35, angle=180]{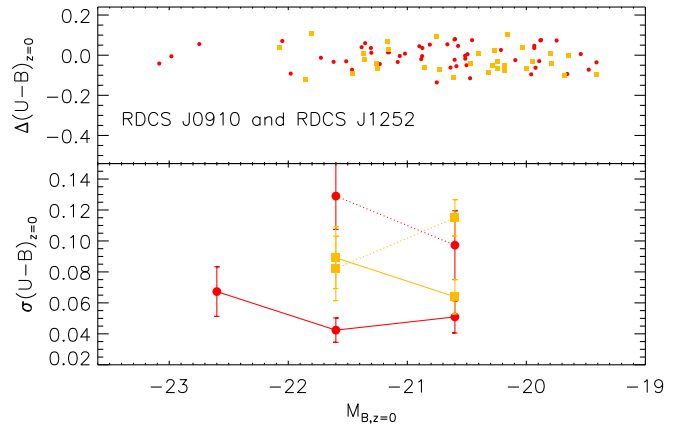}} 
\caption {Similar to Fig.~3: Color residuals and intrinsic scatter for a sample at $z > 1$ limited to clusters with spectroscopically confirmed members complete to magnitudes brighter than $M_{B,z=0} \approx -20.5$~mag, corresponding to the brighter three ranges of magnitude in the bottom panel. This shows that the larger scatter observed in S0 and peripheral elliptical populations is not due to interlopers, since it is observed over the magnitude range where our CMR sample is only composed of spectroscopically confirmed members.  {\label{f3_z_2}}}
\end{figure}

Fig.~\ref{averes_cmr} displays the main results from this analysis. 
First, for bright galaxies ($M_{B,z=0} < -21$~mag) in the central cluster regions ($R \lessapprox$~$0.5 R_{200}$), 
the elliptical CMR scatter is smaller (at $\approx 2 \sigma$) than that of 
the S0 population. Hereafter, we compare our estimated scatters to this elliptical measurement (e.g. larger scatters are those that are larger than the elliptical scatter in the central cluster regions).
 Second, the elliptical scatter increases with distance from the cluster center,
approaching that of the S0 population in the outer regions.    
In other words, the S0 population displays similar scatter at all distances within the virial radius 
for all magnitudes, while bright elliptical galaxies exhibit smaller scatter in the cluster core (at $\approx 2\sigma$).
Third, at faint magnitudes ellipticals and S0s show larger scatters  (at $\approx 2 \sigma$).

\begin{figure*}
\begin {tabular} {r@{\hspace*{-3mm}}r}
\includegraphics[scale=0.25, angle=180]{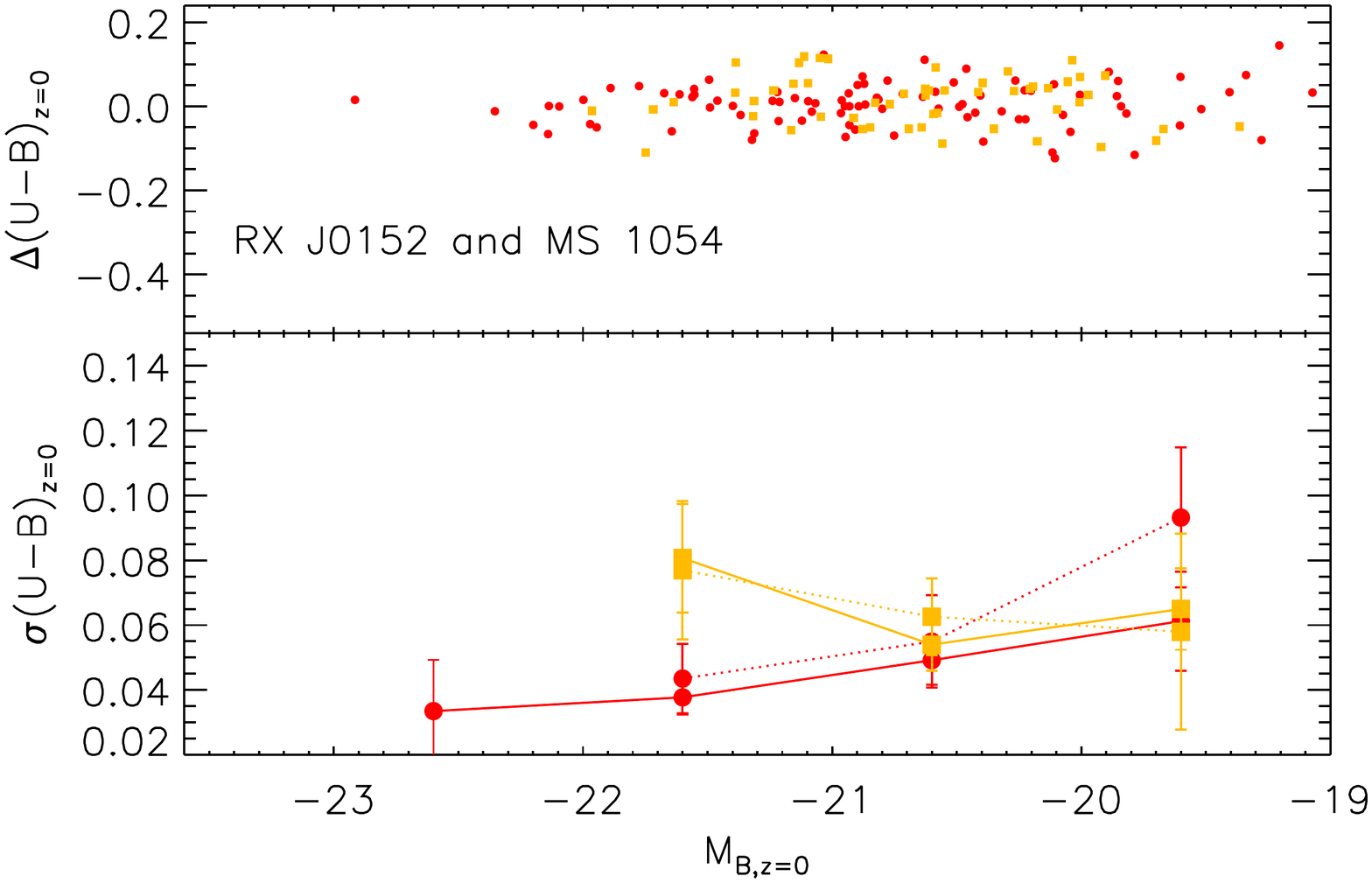} &\includegraphics[scale=0.25, angle=180]{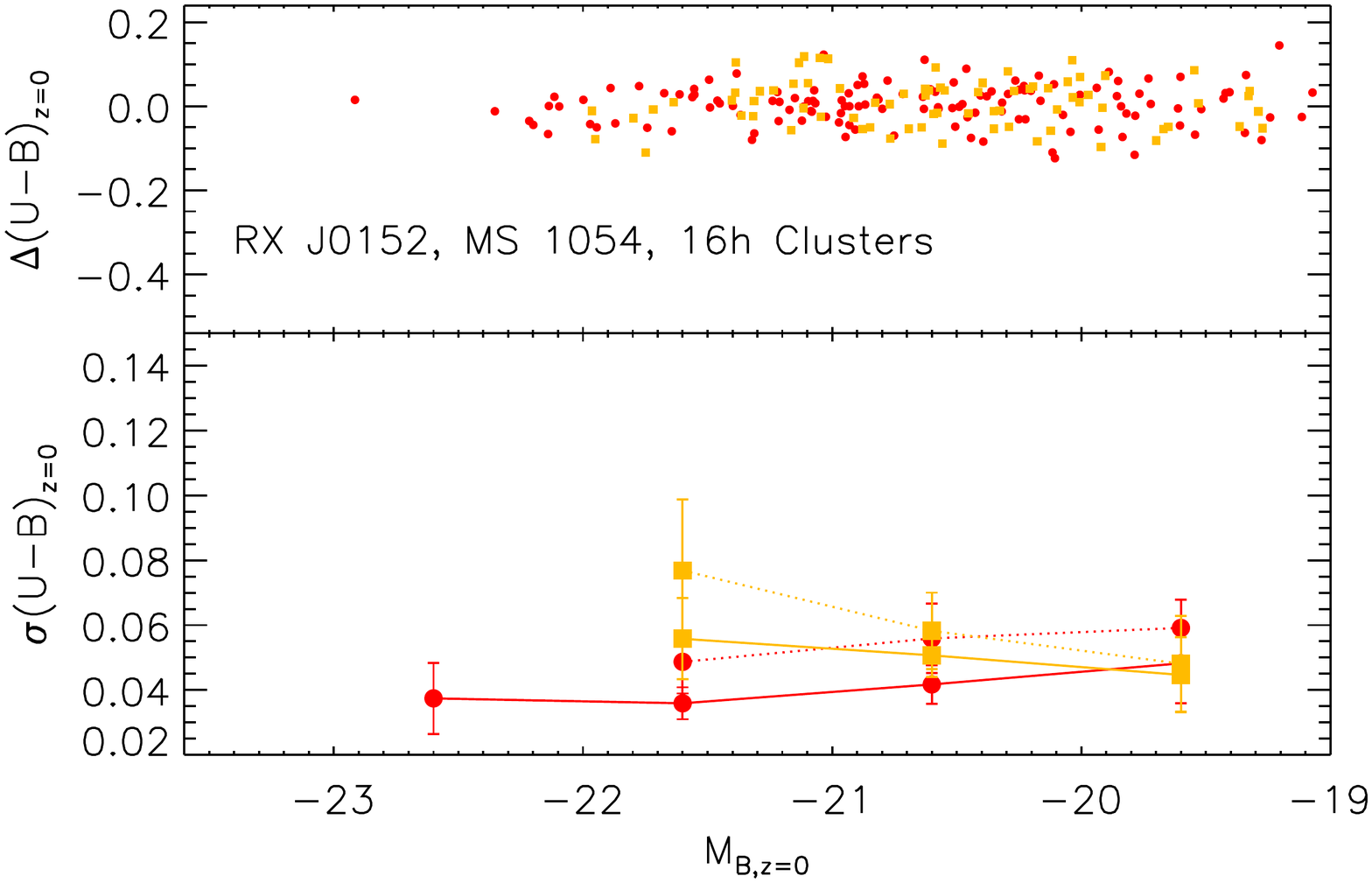}\\
\includegraphics[scale=0.25, angle=180]{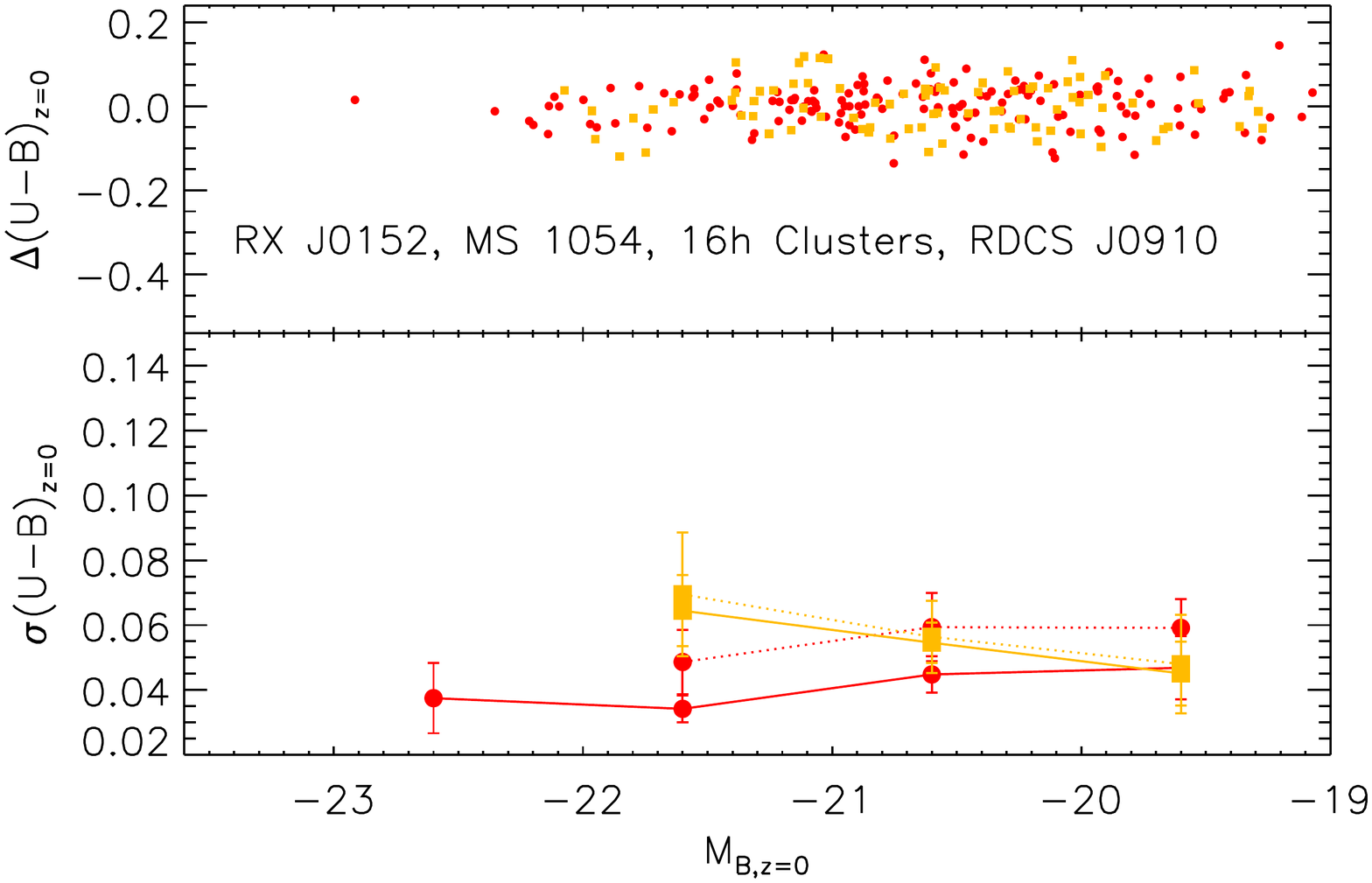} &\includegraphics[scale=0.25, angle=180]{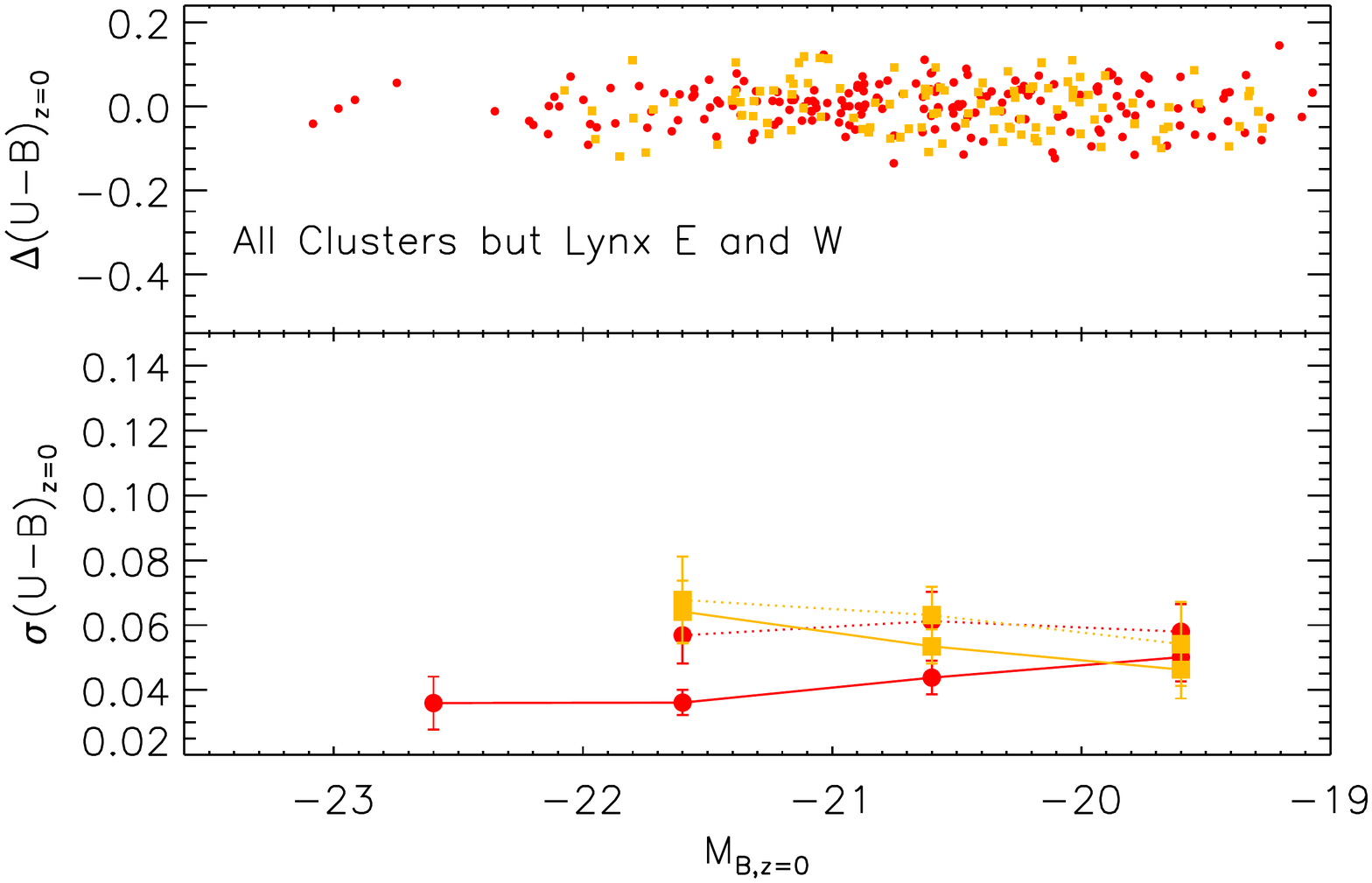}\\
\end{tabular}
\caption {Similar to Fig.~3, but including clusters at progressively higher redshift.  The scatter of elliptical galaxies in the peripheral regions increases when adding clusters at higher redshift. The S0 scatter is always larger than that of the ellipticals {\label{f3_3}}}
\end{figure*}

Fig.~\ref{averes_cmr_bis} shows early--type galaxy residuals in rest--frame $(U-B)_{z=0}$ color as a function of distance from the cluster center, $R/R_{200}$.  Small symbols indicate individual elliptical and S0 galaxy residuals, while large symbols represent the average residuals for two distance ranges:  $R < 0.5 R_{200}$ and $ 0.5 R_{200} < R < R_{200}  $.  The average CMR residuals do not change significantly with distance from the cluster core or with early--type galaxy morphology.
The average residual for the elliptical and S0 populations are $0.002 \pm 0.027$~mag and $-0.002 \pm 0.025$~mag, respectively -- the same within the uncertainties. The continuous and dashed lines show fits to the elliptical and S0 residuals as a function of $R/R_{200}$. We calculate Pearson Coefficients $PC$ and the probability of correlation between two variables as $PC^2$. 
 The Pearson coefficient for these relations is $<0.10$: residuals do not show a trend as a function of $R/R_{200}$.

Qualitatively, we see from the Figure that galaxies within $\approx 0.5 R_{200}$ display less scatter around the average, when compared to galaxies at larger distance. Furthermore, most of the early--type galaxies lie within $R < 0.5 R_{200}$ (235 galaxies; recall that our regions are projected radial regions), the more distant sample being more sparse (80 galaxies). While the central ($R<0.5 R_{200}$) early--type population falls tightly around the CMR fit, the more sparse population at $R > 0.5 R_{200}$ is more dispersed (e.g., shows larger intrinsic scatter around the CMR, as quantified in Fig.~\ref{averes_cmr}). 

Interlopers could artificially increase the measured intrinsic scatter.
It is interesting to note that almost all galaxies in our CMR sample are spectroscopically confirmed members for magnitudes brighter than $M_{B,z=0} \approx -20.5$~mag in MS~1054--0321, RX~J0152--1357, RDCS~J0910+5422 and RDCS~J1252--2927. We plot the intrinsic scatter from these clusters in Fig.~\ref{fig3_sp} and see that the difference between the elliptical and the S0 galaxy scatter remains significant.

\begin{figure}
\centerline{\includegraphics[scale=0.35, angle=180]{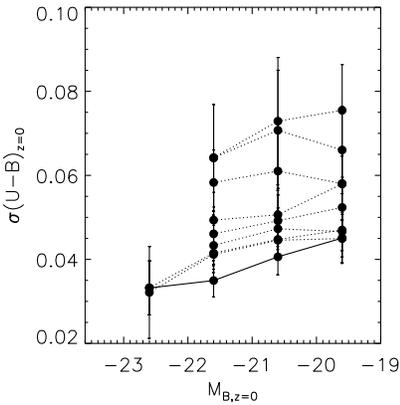}}
\caption {The elliptical CMR scatter at radii $ R < {R_{200}}$, with $\frac{R}{R_{200}}$ 
varying from 0.2 to 0.8 from the bottom to the top, in intervals of 0.1. The continuous line shows $R=0.2 R_{200}$. The scatter increases at larger radii. {\label{f3_0}}}
\end{figure}

\begin{figure}
\centerline{\includegraphics[scale=0.36,angle=180]{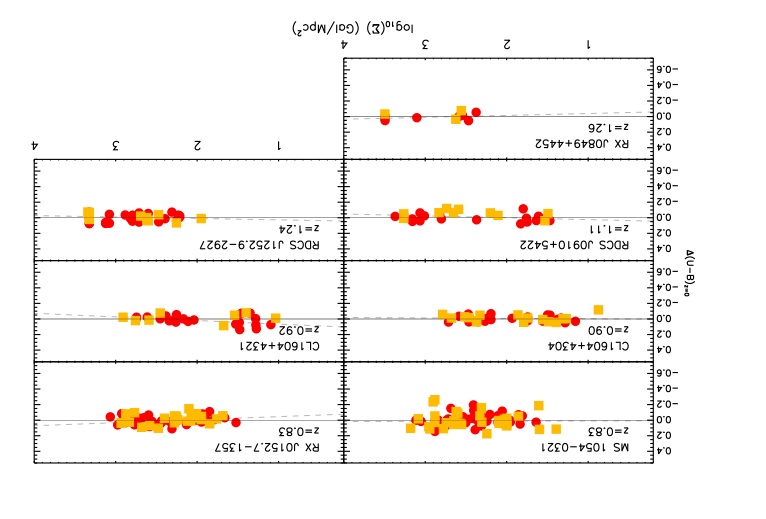}}
\caption {Color residuals (as in Fig.~\ref{res_cmr}) vs local galaxy density. The continuous line shows the zero level and the dashed line gives a linear fit. We do not observe significant correlations between colors and local galaxy density. {\label{rescmr_den}}}
\end{figure}

\begin{figure}
\centerline{\includegraphics[scale=0.35,angle=180]{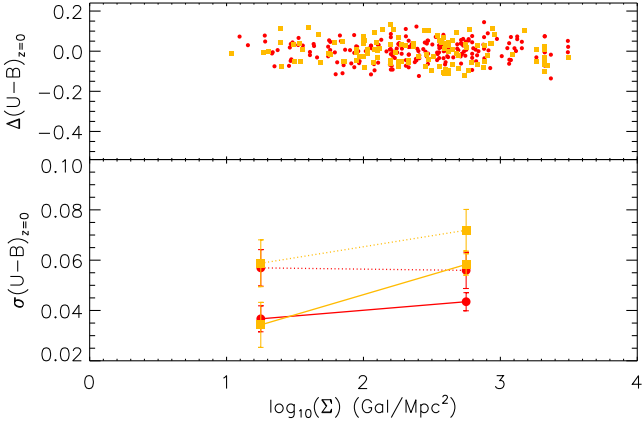}}
\caption {Color residuals vs local galaxy density. Symbols are as in Fig.~\ref{averes_cmr}. We do not observe any trends in scatter correlated with galaxy density. {\label{averescmr_den}}}
\end{figure}

\begin{figure}
\centerline{\includegraphics[scale=0.36,angle=180]{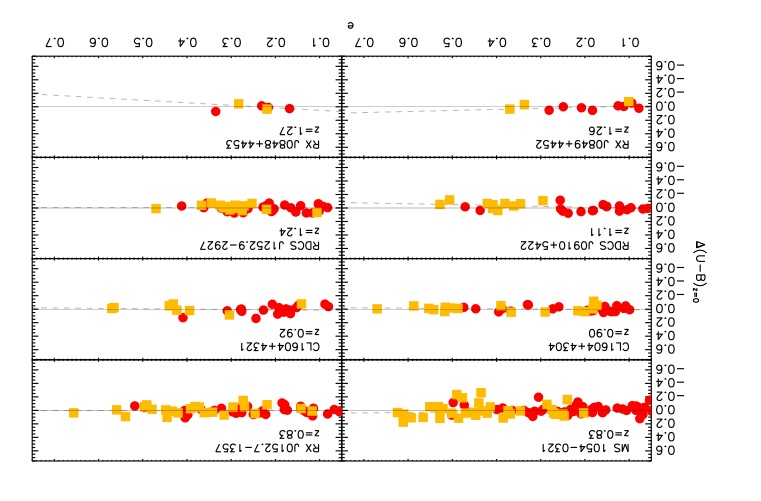}}
\caption {Color residuals (as in Fig.~\ref{res_cmr}) vs galaxy ellipticity. We do not observe any significant correlations between color and ellipticity. This implies that the
S0 galaxies (that have on average higher ellipticity) have the same color as elliptical galaxies, and that the
increased scatter we find in the S0s can be ascribed to a younger
population and not to contamination by bluer, later-type galaxies (see text for discussion). We also notice that the elliptical galaxies have on average lower ellipticities than the S0 galaxies.  {\label{rescmr_el}}}
\end{figure}

\begin{figure}
\centerline{\includegraphics[scale=0.35,angle=180]{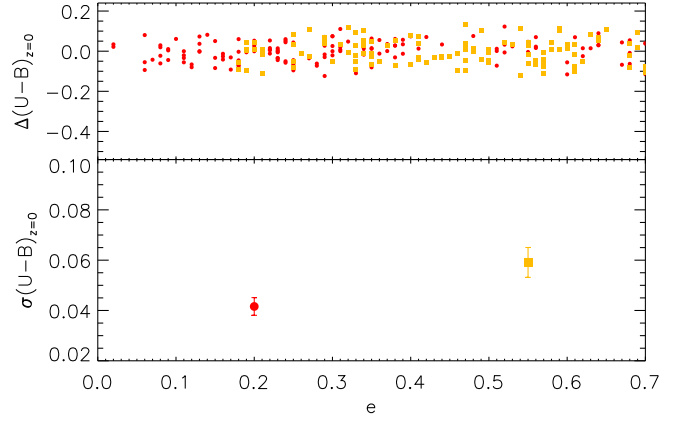}}
\caption {Color residuals vs galaxy ellipticity. Symbols are as in Fig.~\ref{averes_cmr}. S0 galaxies have on average larger ellipticities than the ellipticals. The elliptical and S0 overall average scatters shown in the figure are $0.042 \pm 0.003$ (in the range $0<e<0.4$) and $0.059 \pm 0.006$ (in the range $0.4<e<0.7$), respectively. In these ellipticity ranges, the S0 have a larger scatter than the ellipticals at $\approx 2.5\sigma$. {\label{averescmr_el}}}
\end{figure}

\begin{figure}
\centerline{\includegraphics[scale=0.35,angle=180]{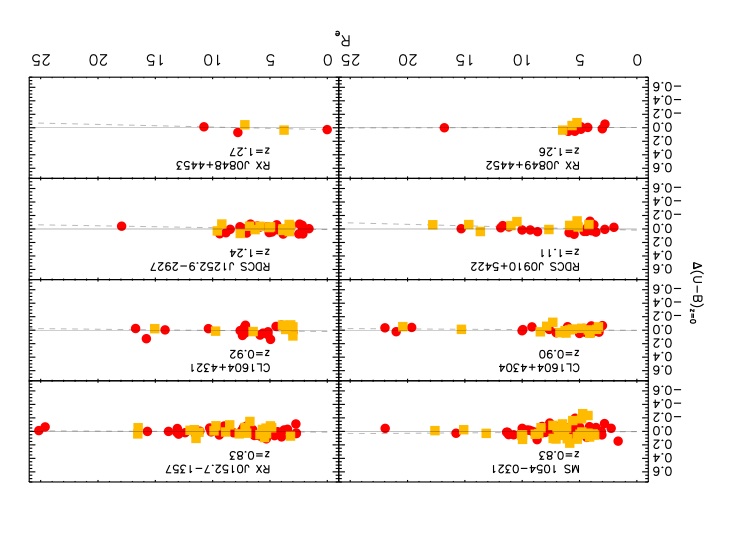}}
\centerline{\includegraphics[scale=0.35,angle=180]{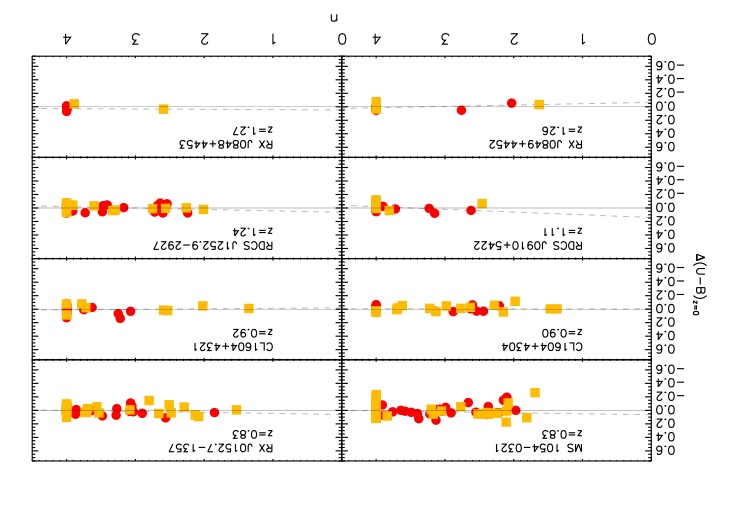}}
\caption {Color residuals (as in Fig.~\ref{res_cmr}) vs galaxy effective radius $R_e$ and Sersic index $n$. We do not observe any significant correlation between color residuals and these two parameters.   {\label{rescmr_re_nser}}}
\end{figure}

It is also interesting to examine the behavior of the scatter in the lower ($0.8<z<1$) and higher ($z>1$) redshift samples to see if there is any evolution.  In Fig.~\ref{f3_z}, we show the low and high redshift samples separately. The observed difference in scatter is larger in the high redshift sample. To verify that this is not due to a higher presence of interlopers in the high redshift sample, we show in Fig~\ref{f3_z_2} the same for RDCS~J0910+5422 and RDCS~J1252--2927, whose early-type CMR galaxies are all spectroscopically confirmed members up to magnitudes $M_{B,z=0} \approx -21.5$~mag. 

 In Fig.\ref{f3_3} we gradually add higher redshift clusters to the sample. In MS1054-0321 and RX J0152-1357, the lenticulars show larger scatter than that of the full elliptical sample, and there is no difference between the scatter of the ellipticals in the central and external regions.  As higher redshift clusters are added, the lenticulars still show higher scatter, and difference between the scatter of ellipticals in central and external regions increases.  

Fig.~\ref{f3_0} shows how the elliptical scatter increases when adding galaxies progressively farther from the cluster center.  This suggests that ellipticals in the outer regions define a tighter red sequence as they evolve to lower redshifts, while those in the center of the cluster already lie on the red sequence at z$\approx$1.3. A larger sample of clusters at $z > 1$ is needed to establish the statistical significance of these suggestive trends.

Since most of our clusters are part of superclusters, filamentary structure around the cluster and infalling groups might also be a source of contamination in this analysis. However, while we do observe an infalling group in  RX J0152-1357 and filamentary structures around MS1054-0321 and CL1604+4321, the higher redshift clusters show at present negligible contamination within one virial radius (from photometric and spectroscopic observations: Nakata et al. 2005; Tanaka et al. 2007; Gal et al. 2008; Tanaka et al. 2008). 

\subsubsection{CMR scatter dependence on galaxy projected density and galaxy properties}

In this section we study CMR color residuals and scatter as a function of  galaxy projected density, ellipticity, effective radius $R_e$ and Sersic index $n$, in order to see if our results depend on galaxy environment or intrinsic properties.
Fig.~\ref{rescmr_den} and Fig.~\ref{averescmr_den} provide insight into the dependence of color and CMR intrinsic scatter on galaxy environment, quantified here as neighbor galaxy density (from Postman et al. 2005; see Sect. 3).
Dashed lines show linear fits. The absolute Pearson coefficients are always less than 0.3 and the probability of correlation always less than $8\%$, showing no significant correlation of early--type galaxy color with environment within the clusters.
Intrinsic scatters in low and high density regions do not differ as much as the intrinsic scatters of the S0 and elliptical populations, or as much as scatters from regions at different radii.

We next consider galaxy ellipticity.  Potentially, the larger CMR scatter that we observe in lenticulars and peripheral ellipticals could be caused by either a misclassification of face--on S0 galaxies as ellipticals or of flattened late--type galaxies as S0s, both of which would increase our sample average age. Both effects, if present, should be larger in the cluster external regions ($R > 0.5 R_{200}$), where elliptical galaxy fraction decreases and late--type fraction increases (Postman et al. 2005).
Postman et al. (2005), Blakeslee et al. (2006) and Mei et al. (2006a) have shown that a misclassification of face--on S0 as ellipticals
would be detected as a predominance of flattened S0s  in our sample, while a misclassification of  flattened late--type galaxies as S0s would result in bluer galaxy colors at higher axial ratios.
Fig.~\ref{rescmr_el} and Fig.~\ref{averescmr_el} show the dependence of color residuals and CMR intrinsic scatter on galaxy ellipticity. 

These figures are revealing:  We observe that CMR ellipticals have on average lower ellipticity than the S0 population. This different distribution in axial ratios would suggest that some of the face--on S0s have been classified as ellipticals (a complete analysis of our sample ellipticity is performed in Holden et al. 2008).
Blakeslee et al. (2006) and Mei et al. (2006a) already noted this trend when studying the ellipticity distribution of S0 and elliptical galaxies in the two massive clusters at $z$=0.8 and in RDCS~J0910+5422, respectively.  In these analyses, we concluded that there is a lack of round S0s on the CMR with respect to simple axial distribution models.   
This observed lack of round S0s could indicate that face--on S0s have been
misclassified as ellipticals and/or that edge--on spirals have been misclassified as S0s.  In the latter case, we should observe bluer colors in objects with higher axial ratios.

The dashed lines in Fig.~\ref{rescmr_el}  show linear fits, and we do not observe any correlation of color residuals with ellipticity.
The most pronounced trends are found in RDCS~J0910+5422 (as already discussed in Mei et al. 2006a) and in Lynx~W. 
With respective Pearson coefficients of -0.4 and 0.4 (both corresponding to a correlation probability of $\approx 15\%$), the color--ellipticity correlation is not significant even in  these two objects. The trend in Lynx~W is very probably due to the paucity of early--type CMR galaxies in this cluster.  RX~J0152.7--1357 has $PC = -0.3$, corresponding to a probability of correlation of only $7\%$. 

We can conclude that, most likely, face-on S0s are misclassified as ellipticals in our sample.  Holden et al. (2008) reached the same conclusion using our same high redshift sample and ellipticity measurements. Comparing our sample to low redshift samples in detail, they found an apparent deficit of low ellipticity S0s at high redshift.
Since S0 scatters are larger or similar to ellipticals at all luminosities (e.g., from Fig.~3), S0s misclassified as ellipticals would tend to  increase the measured elliptical scatter. Our elliptical scatter is, however, smaller, so it is improbable that morphological misclassification is significant in driving our results in the cluster core.  On the other hand, we cannot exclude the hypothesis that some S0 misclassified as ellipticals might be the cause of the larger elliptical scatter in cluster peripheral regions and at fainter magnitudes.

Fig.~\ref{rescmr_re_nser} shows the dependence of color residuals and CMR intrinsic scatter on galaxy effective radius $R_e$ and  Sersic index $n$, derived from our GALFIT fit (see Sect. 3). The probability of correlation with each one of the two parameters is always less than 6$\%$.  In  Lynx~W the probability of correlation between color and $R_e$ and $n$ is 36$\%$ and 20$\%$, respectively -- not significant.
Six galaxies have $R_e > 25$~pixels and were not considered when calculating correlations.

\subsubsection{E and S0 galaxy age}

As discussed in the Introduction, CMR intrinsic scatter is driven principally by galaxy age (e.g., Kodama \& Arimoto 1997; Kauffman \& Charlot 1998; Bernardi et al. 2005; Gallazzi et al. 2006).  Accordingly, if we assume that higher scatter corresponds to younger ages (Kodama and Arimoto 1997; Kauffman \& Charlot 1998; Bernardi et al. 2005; Gallazzi et al. 2006), then bright elliptical galaxies contain stellar populations older on average than S0 galaxies; and elliptical galaxies in cluster cores have stellar populations that are on average older than galaxies at the virial radius. At faint magnitudes ($M_{B,z=0} > -21$~mag) all populations present similarly larger scatters and, under these assumptions, younger stellar populations. 

We use simple BC03 stellar population models to quantify this.  As in the CMR paper series, we consider three models: 1) a simple, {\it single burst} solar metallicity BC03 stellar population model; 2) a model with solar metallicity and {\it constant star formation rate} over a time interval $t_1$ to $t_2$, randomly chosen to lie between the age of the cluster and
the recombination epoch; 3)  a model with solar metallicity
and with an {\it exponentially decaying star formation rate}.

With both the simple, single burst solar metallicity BC03 stellar population and the constant star formation rate model, we find the average luminosity-weighted age of bright ellipticals in the core to be $\sim 0.5$~Gyr older than the S0 galaxies, and the faint early--type and peripheral ellipticals.   A model with solar metallicity
and an exponentially decaying star formation rate predicts that if galaxies with larger scatters had a different star formation history (exponential decay versus single burst), they would have an average luminosity-weighted age similar to the bright ellipticals. This scenario, however, also predicts that those galaxies would exhibit bluer color residuals, which we do not observe in Fig.~4. 

We therefore believe that a single burst model is a reasonable approximation for our data.  We will use it in different sections of this paper: it will not give us a precise estimation of the galaxy star formation history; however, we expect it to provide good estimates of the average luminosity--weighted galaxy age. In recent work, for instance, Thomas et al. (2005) (from observations of local galaxies) and De Lucia et al. (2006) (from numerical simulations) have shown that both current observations and predictions from $\Lambda$CDM suggest that the duration of galaxy star formation history depends on galaxy mass. Massive ellipticals, like those in our sample, are predicted to form most of their stars in a short episode of star formation,  which we are approximating here as a single burst. 

Concerning metallicity, the CMR scatter at the same average luminosity--weighted galaxy age is predicted to be smaller for lower metallicities, which correspond to fainter magnitudes (e.g. Kodama \& Arimoto 1997 and Fig. 8 in Mei et al. 2006a).  Since we observe larger scatter for S0 galaxies and at fainter magnitudes, taking this into account would only increase the deduced difference in age.

\begin{figure}
\centerline{\includegraphics[scale=0.36,angle=180]{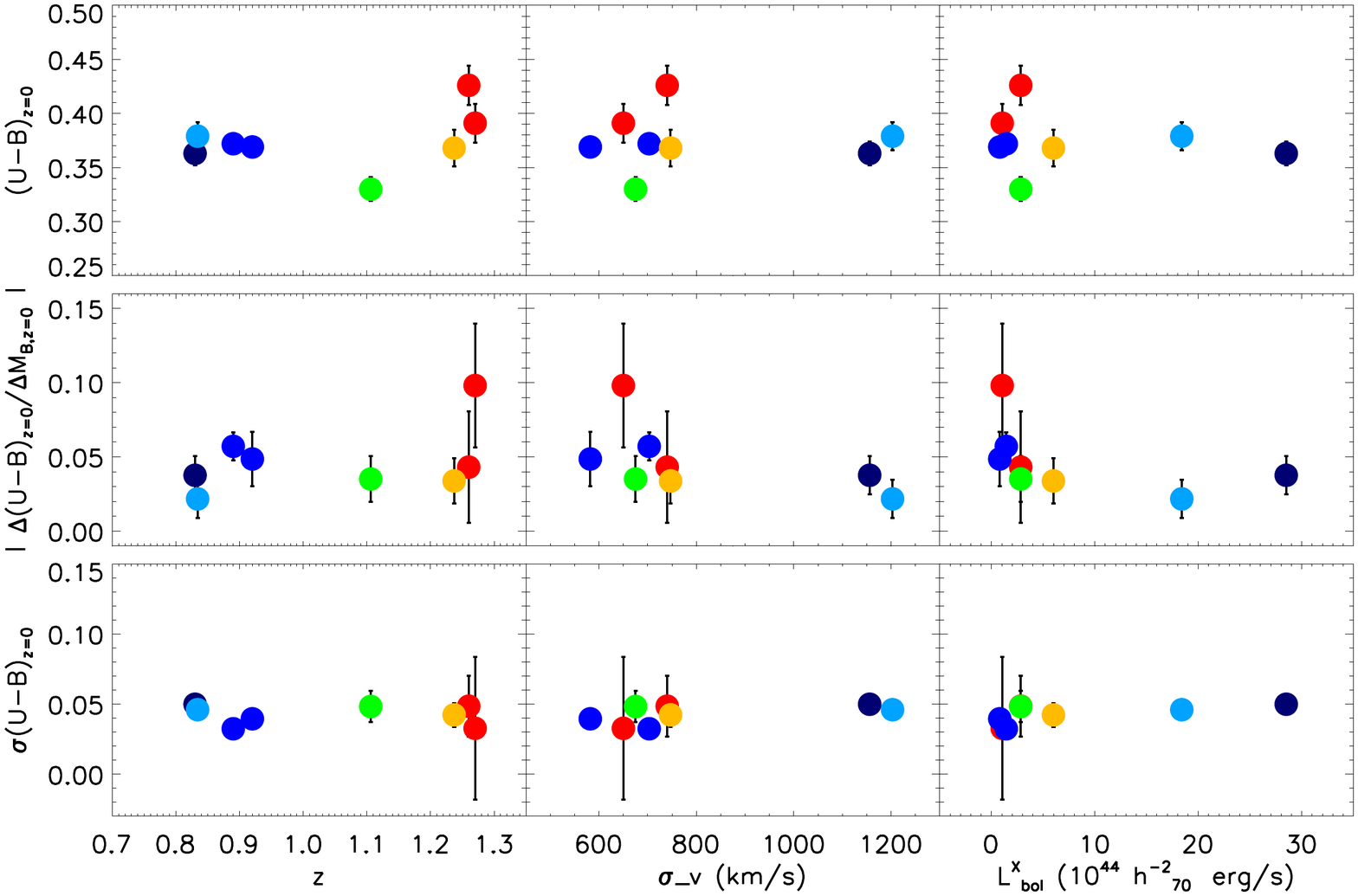}}
\centerline{\includegraphics[scale=0.36,angle=180]{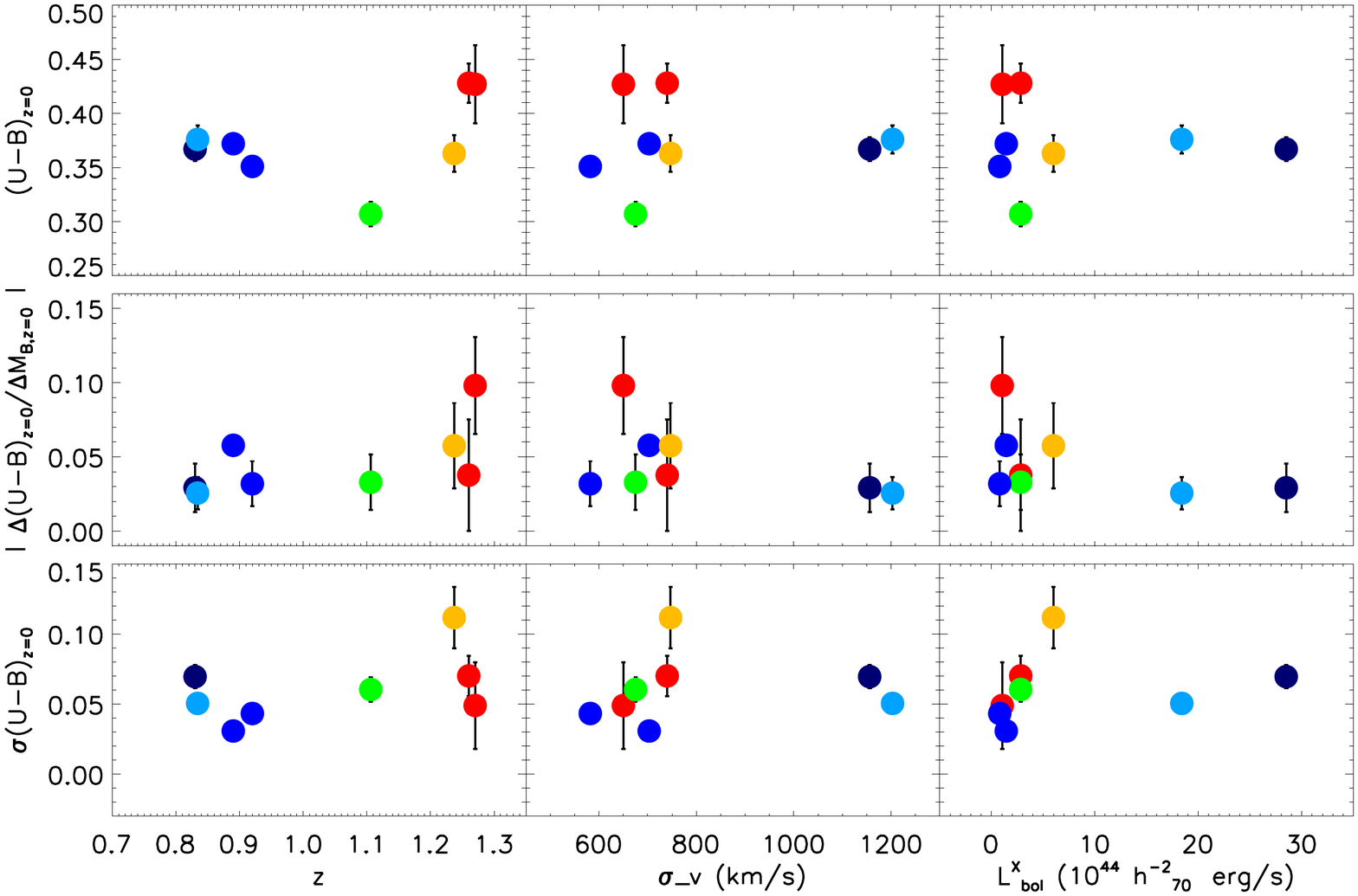}}
\caption {CMR parameters for the elliptical (top set of figures) and the early--type (bottom set of figures) galaxy sample in the ACS Intermediate Cluster Survey clusters. From top to bottom in each set, we plot the CMR zero point $(U-B)_{z=0}$, slope $| \delta (U-B)_{z=0} / \delta M_{B,z=0} |$, and scatter $\sigma (U-B)_{z=0}$  as a function (left to right) of redshift $z$, cluster velocity dispersion $\sigma_v$, and X--ray luminosity $L^X_{bol}$.  Our clusters are shown in different colors: MS 1054--0321 in dark blue, RX~J0152.7--1357 in light blue, CL1604+4304 and CL1604+4321 in navy blue, RDCS~J0910+5422 in green, RDCS~J1252.9-2927 in yellow, the two Lynx clusters in red. We do not find any significant correlation of CMR parameters with cluster mass or any significant evolution with redshift.   {\label{cmrpar_mlm}}}
\end{figure}

\begin{figure}
\centerline{\includegraphics[scale=0.40]{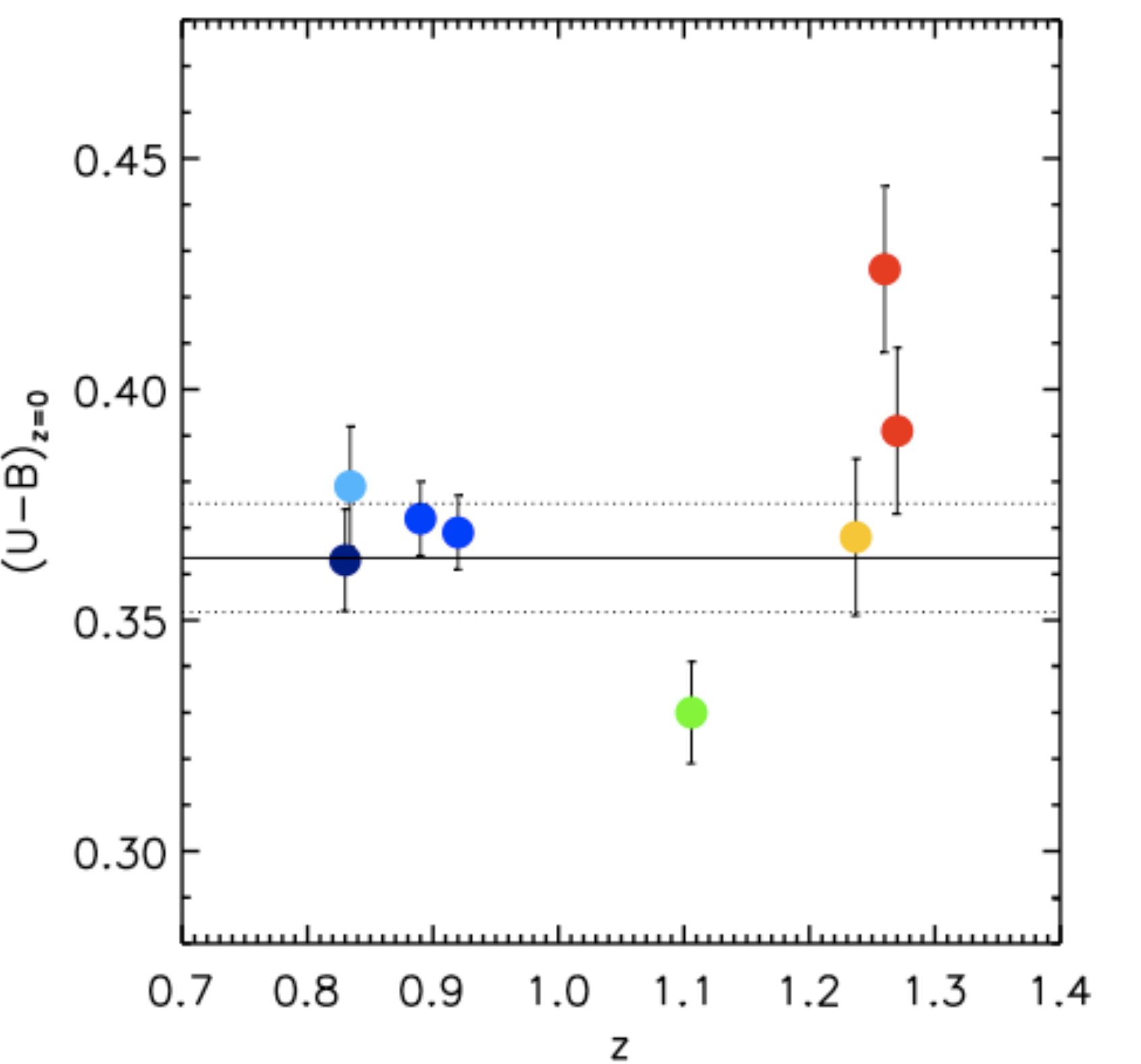}}
\caption{The elliptical CMR zero point $(U-B)_{z=0}$ as a function of redshift.  Our clusters are shown in different colors: MS 1054--0321 in dark blue, RX~J0152.7--1357 in light blue, CL1604+4304 and CL1604+4321 in navy blue, RDCS~J0910+5422 in green, RDCS~J1252.9-2927 in yellow, the two Lynx clusters in red.  The continuous line is the average zero point and the dotted line shows the 1~$\sigma$ range. The two Lynx clusters show a zero point that is redder than the average (see discussion in the text).  {\label{zerop}}}
\end{figure}

\subsection{CMR parameters as a function of redshift and total cluster mass }

In this section, we analyze CMR parameters as a function of redshift.  Because our higher redshift clusters are also the least massive of our sample,  we examine CMR parameters as a function of cluster velocity dispersion and X--ray luminosity to understand the influence of cluster properties on our interpretation of the evolution of the CMR with redshift.  These relations also give us additional information about galaxy evolution:  different galaxy histories in clusters with different physical properties, in particular cluster total mass,  can be brought out by studying the dependence of CMR parameters on those physical properties (e.g. Wake et al. 2005; Poggianti et al. 2006).  In current cosmological models, for example, we expect that galaxies form later (and as a consequence might have younger ages/larger CMR scatter at z$\approx$1) in low--mass clusters.
When comparing our CMR parameters to cluster mass, we will consider both X--ray luminosity and velocity dispersion as proxies for cluster total mass, keeping in mind the potential systematics associated with each of these quantities (see section~2.1).

In Fig.~\ref{cmrpar_mlm}, from top to bottom,  we plot the elliptical (top) and early--type (bottom) CMR zero point  $(U-B)_{z=0}$, slope $| \delta (U-B)_{z=0} / \delta M_{B,z=0} |$, and scatter $\sigma (U-B)_{z}$  as a function of redshift $z$, cluster velocity dispersion $\sigma_v$, and X--ray luminosity $L^X_{bol}$ (from left to right).  
We calculate the Pearson coefficient and correlation probabilities for the CMR zero point, slope and scatter as a function of redshift, cluster velocity dispersion and X--ray luminosity.  
All relations have $PC \leq 0.5$, $PC^2\leq 30\%$, and we do not find any significant trend with the cluster mass proxies. The lower mass, higher redshift clusters show more dispersion in their CMR slopes and zero points, as predicted by some semi--analytical models (Menci et al. 2008). The size of our sample does not, however, permit us to establish a general trend at high redshift.

The average CMR zero point, slope and scatter in our sample in rest--frame  $(U-B)_{z=0}$ color are $0.36 \pm 0.01$~mag (not including the two Lynx clusters), $-0.047 \pm 0.023$, and $0.042 \pm 0.021$~mag, respectively, which are the values plotted in Fig.~\ref{zerop} and  Fig.~\ref{cmrpar_mlm}. We applied a 3~$\sigma$ clip to derive the average, and the error is the uncertainty on the average. When we consider the total early--type (ellipticals plus S0s) sample, we obtain $0.36 \pm 0.01$~mag, $-0.046 \pm 0.023$, and $0.061 \pm 0.015$~mag, respectively. 

\begin{figure}
\centerline{\includegraphics[scale=0.40]{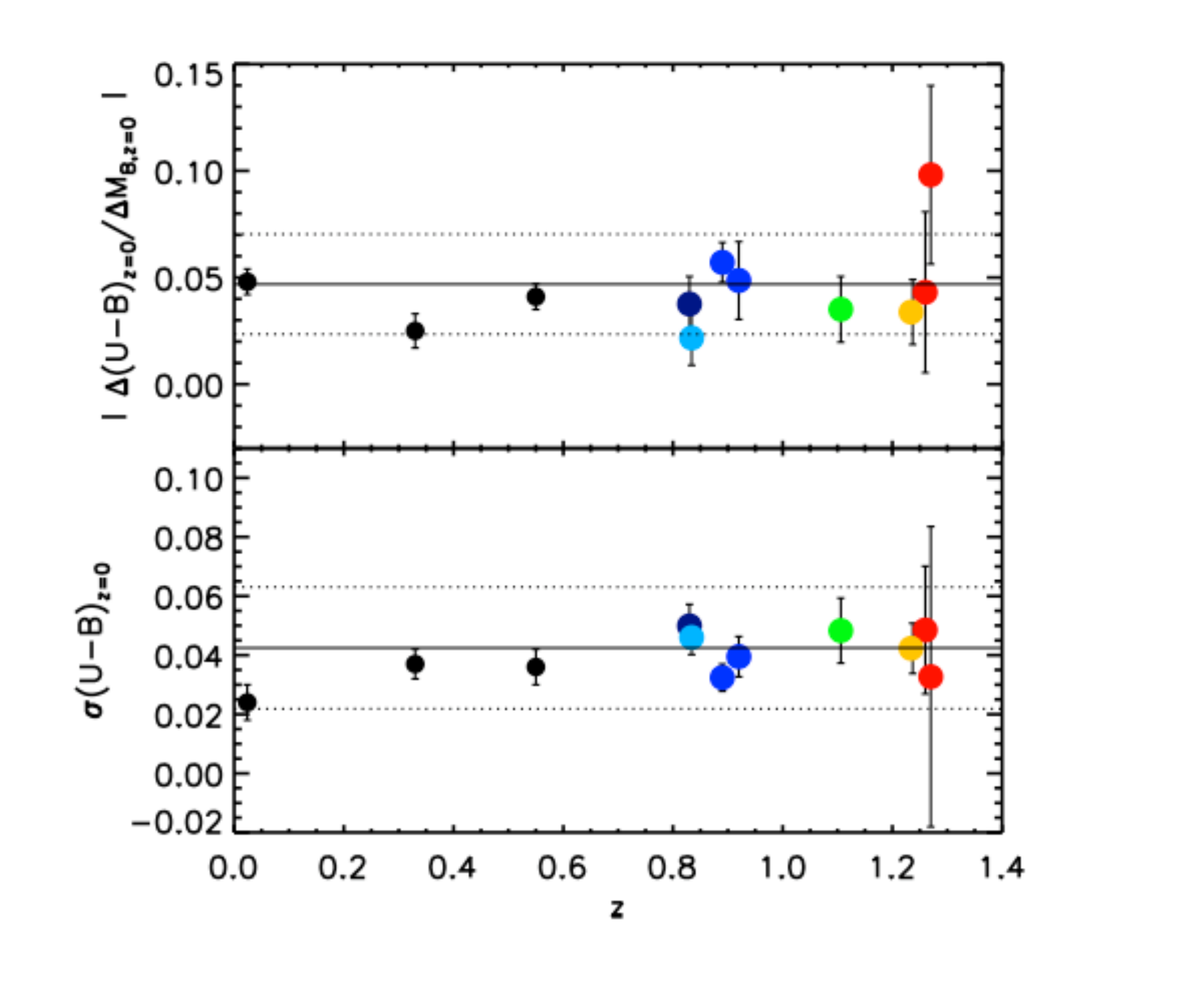}}
\caption {CMR evolution for the elliptical galaxy sample as compared to local cluster CMR parameters.
We show CMR absolute slope  $|\frac{\delta (U-B)_{z=0}}{\delta M_{B,z=0}}|$ and scatter $\sigma  (U-B)_{z=0}$ for ellipticals as a function of redshift. Data points are from Bower, Lucey, Ellis (1992) for the Coma and Virgo clusters; van Dokkum et al. (1998) for MS 1054-03; Ellis et al. (1997) for a sample of nearby clusters of galaxies  (from left to right, in order of increasing redshift).  Our clusters are shown in different colors: MS 1054--0321 in dark blue, RX~J0152.7--1357 in light blue, CL1604+4304 and CL1604+4321 in navy blue, RDCS~J0910+5422 in green, RDCS~J1252.9-2927 in yellow, the two Lynx clusters in red. These results do not indicate a significant dependence of absolute slopes and scatters with redshift.  {\label{cmrpar_ev}}}
\end{figure}

Fig.~\ref{zerop} shows CMR zero points as a function of redshift.
The Lynx cluster zero points show redder  $(U-B)_{z=0}$ colors even when compared to lower redshift clusters of similar X--ray luminosity and velocity dispersion (central and right panel). The difference between their average zero point and the average zero point from our other clusters is $0.09 \pm 0.04$~mag (a $\approx 2 \sigma$ difference). This might suggest a different stellar formation history in these two clusters, and perhaps a higher spread in the CMR zero point at higher redshifts, which is predicted by recent semi--analytical models of galaxy formation (Menci et al. 2008). This observation has been widely discussed in Mei et al. (2006b), who point out that observed  $(i_{775} - z_{850})$ colors in these clusters are redder  (by $0.07 \pm 0.04$~mag) than theoretical predictions from a simple single burst, solar metallicity stellar population model from BC03. While there might be an indication of a different star formation history in these two clusters, this conclusion is highly uncertain due to the uncertainty of a few hundreds of a magnitude in the calibration of the ACS $z_{850}$ filter (Sirianni et al. 2005). The ACS bandpass responses have been calibrated to 9000\AA, with an uncertainty on ACS bandpass zero points of 0.01~mag.
Since the local 4000\AA~break observed in  galaxy templates with age 4~Gyr and solar metallicity is redshifted to around 9000\AA~at $z$=1.26, most of the galaxy light at the Lynx cluster redshift lies at wavelengths larger than 9000\AA, where the ACS $z_{850}$ bandpass calibration is more uncertain.

The variation of CMR scatter and slope with redshift is compared to local clusters in Fig.~\ref{cmrpar_ev}  for both the ellipticals and  the full early--type galaxy sample.  Published colors were transformed to rest--frame slopes, $| \delta (U-B)_{z=0} / \delta M_{B,z=0} |$, and scatters, $\sigma (U-B)_{z=0}$, using single burst solar metallicity stellar population models from BC03, as described in Appendix~II.
For the elliptical galaxy CMR, we considered the Bower, Lucey, Ellis (1992) results for the Coma and Virgo clusters; Ellis et al. (1997) results for a sample of clusters of galaxies at z~$\approx$~0.5, and van Dokkum et al. (1998) results for MS~1054--0321. 
The continuous line shows the average parameter value in our sample and the dotted line the 1~$\sigma$ range.

CMR parameters do not exhibit significant evolution up to redshift z~$\approx$~1.3.
 This remarkable constancy of the CMR with redshift might be due to the fact that, when selecting galaxies within three times the scatter around the CMR, we are not comparing the same galaxy populations at low and high redshift. As pointed out by van Dokkum \& Franx (2001), we might be affected by a {\it progenitor bias}: the high redshift sample would not include the bluer progenitors of the low redshift CMR sample, but only their oldest progenitors.

\begin{figure}
\centerline{\includegraphics[scale=0.45]{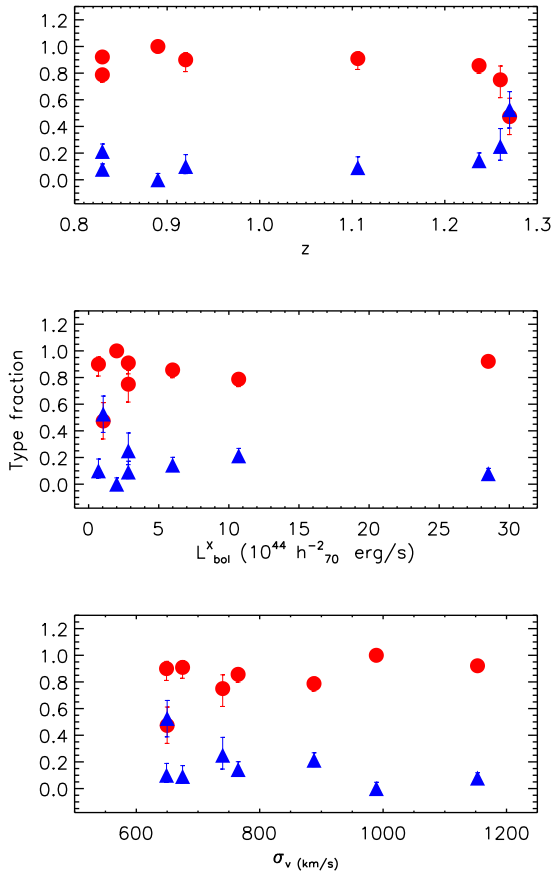}}
\caption {Galaxy--type fraction on the red sequence as a function of redshift $z$,  X--ray luminosity $L^X_{bol}$ and cluster velocity dispersion $\sigma_v$. The early--type galaxy fraction is shown by circles, and the spiral fraction by triangles. We note that the fractions of early--type and late--type galaxies are much closer in the Lynx clusters than for the rest of the sample. This might be due to an evolution of the early--type fraction at higher redshift (in fact, it is not correlated with their X--ray luminosity or dispersion velocity) or to the presence of a larger number of late--type CMR interlopers. {\label{f13}}}
\end{figure}

\subsection{CMR galaxy type fraction evolution}

In this section we focus on the morphological make--up of the red sequence.
Even if our small sample size and lack of a complete spectroscopic sample (especially for blue galaxies) do not permit us to quantify in detail the evolution of blue and red galaxies, we can study evolution of the morphological distribution of the galaxies on our CMR. 

The CMR galaxy early--type and late--type fractions are shown in Fig.~\ref{f13} as a function of redshift, X--ray luminosity and velocity dispersion. We have considered all galaxies within three times the CMR scatter. The uncertainties on morphological fractions are calculated following Gehrels (1986; see Section III for binomial statistics). These approximations apply even when ratios of different events are calculated from small numbers, and yield the lower and upper limit of a binomial distribution within the 84$\%$ confidence limit, which corresponds to 1~$\sigma$.

 The majority of the clusters in our sample show little evidence of
evolution, suggesting that the increase in the late-type fraction
observed at this redshift (Postman et al. 2005; Desai et
al. 2007) might come from an increase in bluer, star forming
galaxies (see also van der Wel et al. 2007).
In Lynx~W the late--type/early--type fractions are similar (around 50\%). Even if the fractions are not significantly different (the difference between 80\% and 50\% is only 1~$\sigma$ at $z > 1$) because of the large Poissonian errors on such a small galaxy sample, the difference becomes more significant (2~$\sigma$) when the Lynx W early--type fraction ($0.47\pm0.19$) is compared to the average early--type fraction of the other clusters ($0.87\pm0.07$).  This trend is not correlated with the cluster X--ray luminosity, nor with its velocity dispersion, and is shown to be a real evolution with redshift in less luminous clusters in our sample. We cannot exclude, however, the presence of a larger number of late--type CMR interlopers in this cluster.

\begin{figure}
\centerline{\includegraphics[scale=0.45]{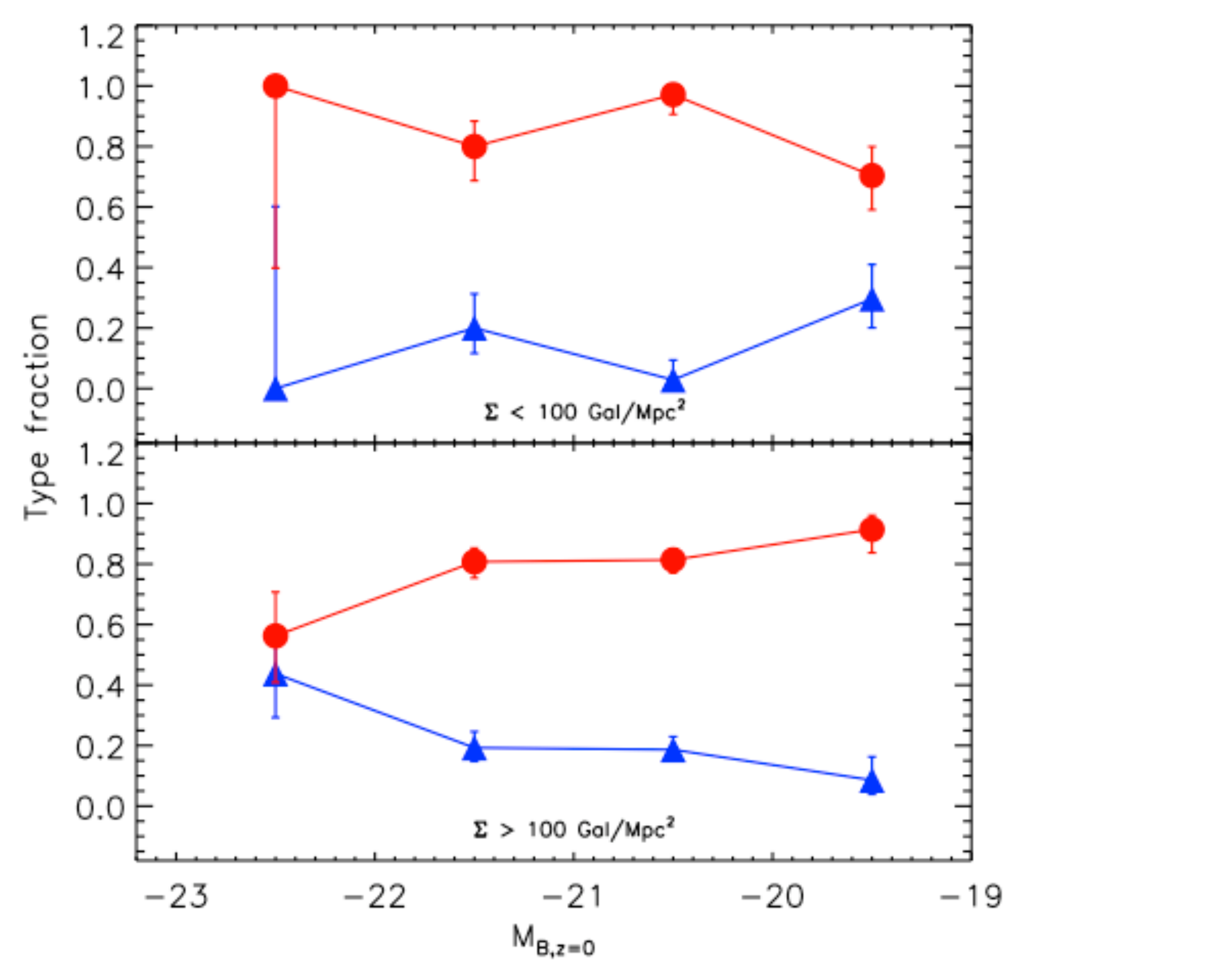}}
\caption { Galaxy--type fractions on the red sequence as a function of rest frame absolute magnitude $M_{B,z=0}$.  On the top panel, we show galaxy fractions in lower density regions (galaxy density between 10 and 100 Galaxies/Mpc$^2$), and on the bottom panel the sample in denser regions (galaxy density between 100 and 1000 Galaxies/Mpc$^2$). In the
lower density regions, bright galaxy fractions are stable at intermediate luminosity for magnitudes  $-22.5 < M_{B,z=0} <  -21.5$~mag. At brighter magnitudes there is a lack of late--type galaxies, while at fainter magnitudes early and late--type fractions are similar. In denser regions, the opposite is observed: the fraction of bright late--type is higher than the late--type fraction at faint magnitudes. {\label{f14}}}
\end{figure}

\begin{figure}
\centerline{\includegraphics[scale=0.50]{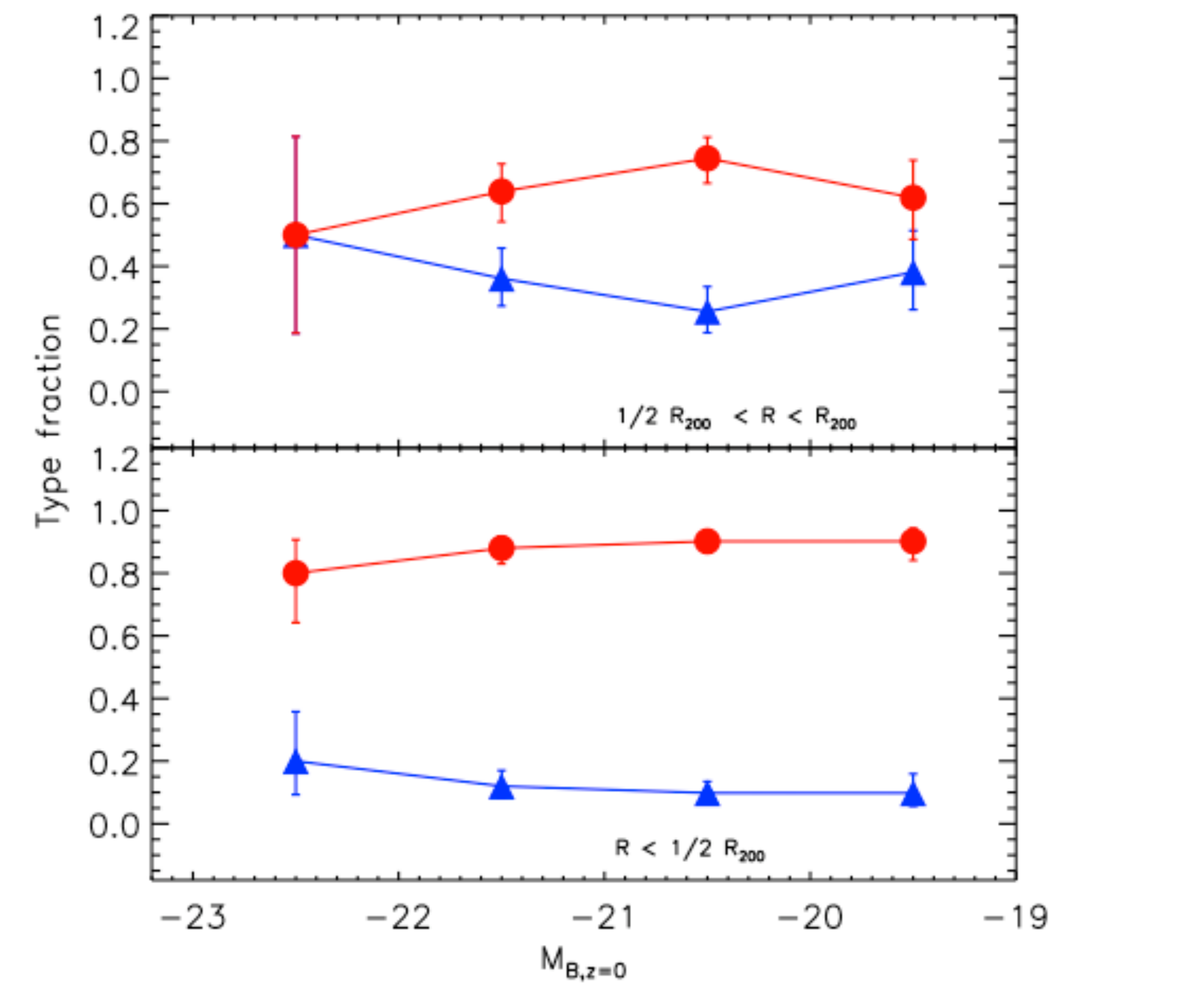}}
\caption { Similar to Fig.~\ref{f14}, but for two different radii bins: within $0.5 R_{200}$ and over the range $ 0.5 R_{200}<R< R_{200}$. The early--type and late--type fractions change in these two different regions, with the fraction of early-type galaxies in the CMR increasing within $0.5 R_{200}$. The fractions are constant with galaxy magnitude. {\label{f14_vir}}}
\end{figure}
To find out which late--type galaxies are populating the red sequence in the Lynx clusters, we return
to Fig.~\ref{cmr}. We observe an important population of  S0/a galaxies (shown as stars in the figure) on the Lynx~W red sequence. Such a large fraction of S0/a at bright/intermediate magnitudes is not observed in any other cluster in our sample. We interpret this population in Lynx~W as S0 galaxies with more extended disks that, after acquiring red colors, are still going through a morphological transformation. This would be consistent with other work interpreting the presence of passive (red) late--type galaxies on the red sequence as evidence that galaxy color evolution precedes morphological transformation (e.g. Couch et al. 1998; Poggianti et al. 1999; Dressler et al. 1999; see however also McIntosh et al. 2004). 

In Fig.\ref{f14} we study morphological fractions as a function of magnitude in two different density regions, a less dense region that includes galaxy densities $ < $~100~Galaxies/Mpc$^2$ (top) and a denser region with galaxy density $>$~100~Galaxies/Mpc$^2$ (bottom).
The total fractions in the low/high density regions are $0.8 \pm 0.1$/$0.81 \pm 0.06$ and $0.17 \pm 0.05$/$0.19 \pm 0.03$ for early--type and late--type, respectively. While the total type fractions are  indistinguishable in the total CMR population, the galaxies show apparently different behavior at different luminosities.
In lower density regions, galaxy fractions are stable at intermediate luminosity for magnitudes  $-22.5 < M_{B,z=0} <  -21.5$~mag. At brighter magnitudes there is a lack of late--type galaxies, while at fainter magnitudes early and late--type fractions are similar. In denser regions, we observe the opposite:  a lack of red late--type galaxies at fainter magnitudes and a higher fraction of red late--type galaxies at bright magnitudes. However, note the large uncertainty due to the small statistics.

In Fig~\ref{f14_vir}, we show the same fraction for different radial regions. In regions closer to the cluster center ($r < R_{200}$), the fraction of CMR early--type galaxies is constant with luminosity and always higher than the late--type fraction. The early--type fraction is smaller at distances $ 0.5 R_{200}< r < R_{200}$, though this
    could in part be a projection effect.

We remind the reader that we are studying the fraction of morphological types on the red sequence (galaxies within three times the CMR scatter) and not the general morphology--density relation in our sample. In particular,  Fig.\ref{f14} and Fig~\ref{f14_vir} do not imply the absence of a morphology--density relation. We do observe such a relation and studied it in detail for our sample in Postman et al. (2005). What our current results tell us is that: 1) from Fig~\ref{f14}, most of the bright galaxies on the red sequence are early--type galaxies. We find a population of bright red late--type galaxies in the denser regions of our clusters. As specified above, these are S0/a galaxies in the Lynx-W cluster. We also do not find many faint red late--type galaxies in dense regions. 2) from Fig~\ref{f14_vir}, we observe that even on the red sequence the fraction of early--type galaxies is larger when closer to the cluster center. 
 Our sample is the best presently available at these redshifts in terms of morphological classification and high precision colors; however, as specified above, a larger sample of  of clusters at $z > 1$ is needed to establish these trends.

\section{Discussion and Conclusions}

We have performed a comparative analysis of CMR evolution on the ACS Intermediate Cluster Sample (Ford et al. 2004; Postman et al. 2005), which comprises eight galaxy clusters spanning redshifts between 0.8 and 1.27 and total cluster masses between $\approx$~1 and $\approx$~20~$\times$~$10^{14}$~$M_{\odot}$. In terms of multiwavelength and spectroscopic follow-up, this is the best cluster sample available today at these redshifts. The single cluster CMRs were studied in a series of previous papers (Blakeslee et al. 2003a, 2006; Homeier et al. 2006; Mei et al. 2006a,b) to derive mean luminosity--weighted early--type galaxy ages and to study CMR parameters as a function of galaxy morphology and structural properties (e.g. effective radii, ellipticities, surface brightness). 

In this paper, the cluster CMR measurements, performed in ACS bandpasses, were converted to the same rest--frame $(U-B)_{z=0}$ color and $M_{B,z=0}$ magnitude, and calculated within similar cluster regions (using $R_{200}$ as the physical scale) down to $\approx 0.5L_*$.
The high angular resolution and sensitivity of the ACS permitted us to obtain visual morphologies for two classes of early--type galaxies, elliptical and S0, and different classes of late--type galaxies, which we regrouped into general late--type and S0/a (from Postman et al. 2005). We systematically examined general trends of the three CMR parameters -- zero point, scatter and slope -- for elliptical and lenticular galaxies as a function of galaxy magnitude and structural characteristics, and of cluster mass.

We find that for bright galaxies ($M_{B,z=0} < -21$~mag) in the cluster cores ($R < 0.5 R_{200}$), the elliptical population shows smaller CMR scatter than the S0 population. The elliptical scatter increases with distance from the cluster center, an effect that becomes larger at high redshift. At faint magnitudes, on the other hand, the early--type populations all display similarly larger scatter.
If CMR scatter is primarily an age effect (e.g., Kodama \& Arimoto 1997; Kauffman \& Charlot 1998; Bernardi et al. 2005; Gallazzi et al. 2006), bright ellipticals in cluster cores have on average older stellar populations than S0 galaxies, peripheral elliptical galaxies, and in general faint early--type galaxies. From the difference in CMR scatter, we deduce an average luminosity--weighted galaxy age difference of $\approx 0.5$~Gyr, using a simple, single burst solar metallicity Bruzual \& Charlot (2003) stellar population model. Similar difference in ages for the elliptical and S0
populations have also been found by measuring the rate of surface
brightness evolution from the size-magnitude relation (Holden et
al. 2005a; Blakeslee et al. 2006) and from evolution of the mass-to-light ratio
of the fundamental plane (Holden et al. 2005b). Note, however, that by redshift $z \approx 0.3$, Kelson et al. (2006) find that ellipticals and S0s formed their stars at about the same epoch.

Our results are also consistent with recent S0 age estimations by Tran et al. (2007) in MS~1054--0321 at $z$=0.83.  These authors analyzed galaxy spectra taken with 
Keck/LRIS (see their paper for details) for a magnitude--selected sample of cluster members on the red sequence.
Using the morphological classification of Postman et al. (2005),  they created composite galaxy spectra for three classes of objects:  elliptical, S0 and late--type cluster members. S0 galaxies exhibit a stronger $H\delta_A$ absorption with respect to the elliptical sample and weak [$O II$] emission, a sign of on--going star formation. S0 galaxies also show larger CMR scatter than the elliptical sample (see also Blakeslee et al. 2006). Analyzing the trend of $H\delta_A$ with galaxy color offset with respect to the CMR, the authors argue that it can be explained by age variations alone (e.g. metallicity trends are negligible; see also Kelson et al. 2001), they concluded that the scatter around the red sequence in MS~1054--0321 is indeed mainly due to differences in mean stellar age.  Their results do not depend on luminosity, and favor S0 galaxy ages that are 0.5-1~Gyr younger than the ages of the ellipticals, according to single burst solar metallicity Bruzual \& Charlot (2003) stellar population models.
Their results confirm the correlation between large CMR scatter and young galaxy age.
We note here that, using a larger statistical sample, we obtained a similar result without using galaxy spectra.

For our sample, average color residuals do not show any significant correlation with distance from the cluster center, galaxy neighbor density, ellipticity, $R_e$ or Sersic index $n$. We find that elliptical galaxies have lower average ellipticity than S0s, and concluded that some S0s might have been misclassified as elliptical galaxies (for a detailed analysis, refer to Holden et al. 2008). Although these misclassified galaxies might contribute to the larger scatter observed in the peripheral elliptical population, the bright ellipticals in the core exhibit smaller scatter despite this potential misclassification.

We do not find any significant evolution in the CMR zero point, slope or scatter up to redshift z~$\approx$~1.3. 
The average CMR zero point, slope and scatter in our sample in rest--frame  $(U-B)_{z=0}$ color are $0.36 \pm 0.01$~mag (not including the two Lynx clusters), $-0.047 \pm 0.023$, and $0.042 \pm 0.021$~mag, respectively. When we consider the total early--type (ellipticals plus S0s) sample, we obtain $0.36 \pm 0.01$~mag, $-0.046 \pm 0.023$, and $0.061 \pm 0.015$~mag, respectively. 
The two highest redshift clusters -- the Lynx clusters -- show a CMR zero point in rest--frame $(U-B)_{z=0}$ that is redder than the average CMR zero point (at $\approx 2\sigma$; see also Mei et al. 2006b). When compared to theoretical predictions from a simple, single burst, solar metallicity stellar population model from BC03, the observed  $(i_{775} - z_{850})$ colors in these clusters are also redder by $0.07 \pm 0.04$~mag. 
While this might be an indication of a different star formation history in these two clusters and possible larger spread in the CMR zero point at higher redshift (Menci et al. 2008), this conclusion remains debatable due to the uncertainty of a few hundreds of a magnitude in the calibration of the ACS $z_{850}$ filter (Sirianni et al. 2005) and the small size of our cluster sample. 

Another special case of the elliptical and S0 CMR zero points is presented by RDCS~J0910+5422. In this cluster, the S0 CMR zero point in  $(i_{775}{-}z_{850})$ is bluer by $0.07 \pm 0.02$~mag with respect to the ellipticals. This could indicate a transitional S0 population still evolving towards the bulk of the red sequence already defined by the elliptical galaxies (Mei et al. 2006a).

Wake et al. (2005) analyzed a sample of 12 X--ray selected clusters spanning a large range in X--ray luminosities (and hence masses) from $L_X \approx  10^{43}$~erg~$s^{-1}$ to  $L_X \approx  10^{45}$~erg~$s^{-1}$ at z$\approx$~0.3. They found that CMR slope and zero point depend strongly on cluster X--ray luminosity and that the CMR zero point becomes bluer at large radii. 
We do not observe any of these trends, perhaps also because of the small size of our sample. The bluer CMR population seen at large cluster radii at z$\approx$~0.3 might not yet be present in our higher redshift sample.

Kodama et al. (2007) studied the CMR in proto-clusters at $2<z<3$ (see also Zirm et al. 2008), and suggested that the CMR is assembled between $z\approx3$ and $z\approx2$ for these structures, based on photometrically selected red galaxies (see the Introduction). Candidate proto-cluster galaxies were observed in near-infrared bandpasses $J$ and $K_s$ with MOIRCS on the SUBARU Telescope and SOFI on the NTT. If we use our BC03 simple stellar population model and the MOIRCS $J$ and $K_s$ bandpass responses to passively de-evolve our average CMR parameters measured at $z\sim1$ up to these redshifts (see Appendix II), we find a CMR zero point, slope and scatter in $(J-K_s)_{z=2}$ color of 2.60 mag, 0.053 and 0.053 mag, respectively, all consistent with the results of Kodama et al. (2007) and Zirm et al. (2008).  Although this is consistent with the hypothesis that bright galaxies reached the red sequence between $z=3$ and $z=2$, we require much better knowledge of the exact forms of the star formation and assembly histories of CMR galaxies in order to relate galaxies observed at these high redshifts to lower redshift CMR galaxies (De Lucia et al. 2006; Overzier et al. 2008).

We also considered the morphological composition of the CMR in terms of class fractions for galaxies within three times of scatter of each cluster CMR.
While in the majority of the clusters the CMR consists mainly of early--type galaxies (with fractions varying around 80\% -- 90\%), in the higher redshift, lower mass cluster of our sample -- Lynx~W -- the late--type/early--type fractions are similar ($\approx 50\%$), with most of the late--type population being composed of galaxies classified as S0/a (Fig.~\ref{cmr}). This trend is found to be a real evolution with redshift in our sample in the sense that it is not correlated with the cluster X--ray luminosity, nor with its velocity dispersion.  This S0/a population might be a population of S0 galaxies with more extended disks that after becoming red are still undergoing morphological transformation. We cannot exclude, however, a larger presence of interlopers in this cluster.

We also studied CMR morphological class fractions as a function of galaxy magnitude and environment. In less dense regions (galaxy density less than 100~Gal/$Mpc^2$), we find a lack of late--type galaxies at bright magnitudes, while at faint magnitudes early and late--type fractions are similar. 
In denser regions, the contrary is observed: there is a lack of late--type galaxies at fainter magnitudes and a higher fraction of late--type galaxies at bright magnitudes.  Cooper et al. (2006) identify a similar trend when comparing red and blue fractions over regions of different density in the DEEP2 galaxy group sample. They found that bright blue galaxies prefer regions of high density. They discussed a scenario in which quenched galaxies migrate from the blue cloud to the red sequence (Bell et al. 2004; Faber et al. 2007), arguing that a process arresting star formation in these bright blue galaxies within groups would move them onto the present--day red sequence. We do observe bright blue late--type galaxies in our clusters, mainly in the less massive objects, in agreement with Cooper et al.'s work (see Fig.~1), although in this analysis ACS morphologies permit us to distinguish early and late--type galaxies on the red sequence. 
In the above scenario, the bright, red late--type galaxies observed in our sample might have originally been bright blue late--type objects in which star formation was quenched and that have already migrated onto the red sequence. Or they might be obscured by dust. 
Our observations show a faint, red late--type galaxy population that is present only in less dense regions, and bright red late--type galaxies that are observed only in dense regions. 
Bright blue (outside three times the CMR scatter) late--type galaxies also start appearing at higher redshifts and in less massive clusters, as also observed by Cooper et al. (2006) in both group and field environments.

How do our results compare with the current view of galaxy formation and evolution?
In the favored cosmological model, the $\Lambda CDM$ model (e.g., Spergel et al. 2007), galaxies form via the hierarchical gravitational collapse of dark matter fluctuations. Galaxy clusters formed at the peaks of dark matter fluctuations, and evolve by accreting galaxy groups and galaxies in filaments around them. Their pristine population, e.g.,  galaxies forming in the denser dark matter concentrations, is thought to have collapsed earlier and  have formed most of its stellar mass in a early short event of star formation (Springel et al. 2005; Thomas et al. 2005; De Lucia et al. 2006).  Galaxies that are accreted from the surrounding field and groups are thought to interact with the cluster environment, quenching their star formation and driving a morphological transformation (Diaferio et al. 2001; De Lucia et al. 2006; Poggianti et al. 2006; Faber et al. 2007; Moran et al. 2007). Different environmental processes can be responsible for these two events (merging, galaxy harassment, gas stripping, etc), and we do not yet understand their respective roles in the assembly of galaxy clusters and galaxy evolution. Most of the morphological transformation and star formation quenching is thought to occur at redshifts $z < 1$ (e.g. Poggianti et al. 2006). Galaxies with old stellar populations lie around a well defined red sequence, the CMR in this paper.  A small dispersion in galaxy age leads to a tighter sequence, i.e., the CMR scatter is small (e.g., Kodama \& Arimoto 1997; Kauffman \& Charlot 1998; Bernardi et al. 2005). Galaxies lying on the red sequence originate in part from the pristine galaxy population that formed in the cluster core, and in part from galaxies in which star formation was quenched when they became cluster members (e.g. Poggianti et al. 2006; Faber et al. 2007).  Some of them might have gone through merging with fainter red objects (Bell et al. 2004; van Dokkum 2005; Faber et al. 2007).

In such a scenario where cluster early--type galaxies are made up of both  pristine and accreted/quenched galaxy populations (e.g. Poggianti et al. 2006), we can interpret our results in the following way:

\begin{itemize}

\item Elliptical galaxies at a distance $R \lessapprox 0.5 R_{200}$ from the cluster center are mainly galaxies  with old stellar populations. They fall tightly on the CMR (smaller scatter) and 
could be identified with a population that is mainly composed of galaxies that formed early in the cluster formation process. 

\item Larger CMR scatters are observed in the elliptical population at $R \gtrapprox 0.5 R_{200}$. The difference between peripheral elliptical scatter and that of the central elliptical population is larger in clusters at $z > 1$. These peripheral galaxies might be identified as elliptical galaxies that had a more extended star formation history, and have on average younger stellar populations (by $\approx 0.5$~Gyr, using a simple solar metallicity single burst BC03 stellar population model). If they are being accreted from groups and filaments around the cluster, it would imply that  accreted ellipticals at $z > 1$ have on average younger ages than the pristine cluster elliptical population. Since we observe this difference to be significant only at $z > 1$,  this accretion might be less important in our sample at $z < 1$, or accreted galaxies are already quenched in filaments and groups around the clusters at these lower redshifts.

\item The S0 population shows larger scatter than the central elliptical population even
        within $R \lessapprox 0.5 R_{200}$. The difference in scatter would correspond to a average difference in stellar population age of $\approx 0.5$~Gyr, using a simple solar metallicity single burst BC03 stellar population model. This may be evidence that these galaxies have had more complex star formation histories than the central cluster ellipticals, in agreement with the observed evolution of the morphology--density relation (e.g. Dressler et al. 1999; Postman et al. 2005), star formation (Poggianti et al. 2006), and observations of their dynamics at $z \approx 0.5$ (Moran et al. 2005, 2007).  They could have evolved from late--type galaxies infalling into the cluster and which have lost their disk through environmental effects. It is interesting that in our clusters at $z=1.26$, we observe a larger fraction of red S0/a, late--type red galaxies on the red sequence. This fraction is especially large in the highest redshift, low mass cluster of our sample, Lynx W. They might be the intermediate population still showing more extended disks before being transformed into S0s. In this case, their color transformation occurred before their morphological transformation, in agreement with previous results (e.g. Couch et al. 1998; Poggianti et al. 1999; Dressler et al. 1999; see however also McIntosh et al. 2004).

\item Faint elliptical and S0 galaxies show larger CMR scatters, e.g., on average ages younger by $\approx 0.5$~Gyr, using a simple solar metallicity single burst BC03 stellar population model. This is also evidence that faint objects go through more extended events of star formation as shown from the analysis of local samples (e.g. Thomas et al. 2005) and predictions from semi--analytical models  (e.g. De Lucia et al. 2006). This dependence on galaxy luminosity/mass and environment is also observed in local samples (Hogg et al. 2004; Bernardi et al. 2005; McIntosh et al. 2005; Gallazzi et al. 2006).

\item CMR parameters show negligeable evolution as a function of redshift. The Lynx cluster CMR zero points exhibit redder  $(U-B)_{z=0}$ colors (by $\approx 2 \sigma$) even when compared to lower redshift clusters of similar X--ray luminosity and velocity dispersion (see also Mei et al. 2006b). This might be due a different stellar formation history in these two clusters, and a possible higher spread in the CMR zero point at higher redshifts, which is predicted from recent semi--analytical models of galaxy formation (Menci et al. 2008). However, there is an uncertainty of few hundreds of a magnitude in the calibration of the $z_{850}$ filter that could also be the cause of this difference (Sirianni et al. 2005; Mei et al. 2006b). 

\item Massive clusters in our sample show no evolution in the fraction of early--type galaxies on the CMR. This implies that the increase in the late-type fraction
    observed at this redshift (Postman et al. 2005; Desai et
    al. 2007) might come from an increase in bluer, star forming
    galaxies.

\item The Lynx W cluster shows a early--type fraction ($0.47\pm0.19$) that is $\approx 2~\sigma$ lower than the average early--type fraction of the other clusters ($0.87\pm0.07$).  This is shown to be a real evolution with redshift in less luminous clusters in our sample. Most of the CMR late--type galaxies in this cluster are morphologically classified (Postman et al. 2005) as S0/a. These galaxies might still have more extended disks and might eventually evolve into S0 galaxies. Cassata et al. (2008) finds a similar decrease of the early--type fraction on the red sequence in the field, that is $\approx 50\%$ at z=2.  We cannot exclude, however, a larger presence of interlopers in this cluster.

\item We observe a faint, red late--type galaxy population only in the less dense cluster regions, and bright, red late--type galaxies only in dense regions. Most of the bright, red late--type population are S0/a red sequence galaxies observed in the Lynx superclusters. We would expect these galaxies to transform from S0/a to S0 galaxies.

\end{itemize}

\vskip 0.5 truecm

\begin{acknowledgements}
ACS was developed under NASA contract NAS 5-32865, and this research 
has been supported by NASA grant NAG5-7697 and 
by an equipment grant from  Sun Microsystems, Inc.  
The {Space Telescope Science
Institute} is operated by AURA Inc., under NASA contract NAS5-26555.
We are grateful to K.~Anderson, J.~McCann, S.~Busching,
 A.~Framarini, S.~Barkhouser,
and T.~Allen for their invaluable contributions to the ACS project at JHU. 
S.M. thanks Marco Sirianni for useful insights on color conversions between the Johnson UBV system and the ACS filters, Emmanuel Bertin for a discussion on the Sextractor magnitude calculation, and Ricardo Demarco, Marisa Girardi, and Tadayuki Kodama for fruitful discussions. We thank the anonymous referee for the very helpful comments.
\end{acknowledgements}

\newpage

\clearpage

\begin{table*}
\caption{ACS Intermediate Redshift Cluster Survey sample \label{sample}}
\vspace{0.25cm}
\resizebox{!}{2.9cm}{
\begin{tabular}{llccccccccccccccc}
\tableline \tableline\\
Cluster & $z$ & $\sigma_v$ &$L^{X_a}_{bol}$&$R_{200}^b$ &$M_{tot}^{X_c}$&X--ray &Vel. Disp. &$Age^d$ \\
&&(km/s)&($10^{44} h^{-2}_{70}$  erg/sec) &Mpc&$10^{14} M_{\odot}$&Ref.&Ref.&(Gyr)\\ \hline
MS 1054--0321& 0.831&$1156 \pm 82$& $28.48 \pm 2.96$&1.8 &$21.3\pm4.0$&1&2&3.5 \\
RXJ~0152.7--1357 & 0.834&$1203^{+96}_{-123}$&$18.40\pm0.78$&1.9 &$6.1\pm1.7$&1&3&3.5 \\
{\it RXJ~0152.7--1357 N}& 0.834&$919 \pm 168$&$10.67\pm0.67$&1.4  &$2.7\pm0.8$&1&4&3.5 \\
{\it RXJ~0152.7--1357 S}& 0.830&$737\pm126$&$7.73\pm0.40$&1.1   &$3.5\pm 1.5$&1&4&3.5\\
CL1604+4304&0.897&$703\pm110$& $1.43$ &1.1&&5&6&3.5 \\
CL1604+4321&0.924&$582\pm167$& $<0.78$&0.9&&5&6&3.5\\
RDCS~J0910+5422&1.106&$675\pm190$&$2.83\pm0.35$&0.9&$4.9\pm2.9$&1&7&3.1 \\
RDCS~J1252.9-2927&1.237&$747^{+74}_{-84}$&$5.99\pm1.10$&0.9&$1.6\pm0.4$&1&8 &2.7\\
RX~J0849+4452&1.261&$740^{+113}_{-134}$ &$2.83\pm0.17$&0.9&$2.9\pm1.5$&1&9&2.5\\
RX~J0848+4453& 1.270& $650 \pm 170$&$1.04 \pm0.73$&0.8 &$1.4\pm1.0$&1&10&2.5\\
\tableline \tableline\\
\end{tabular}}
\scriptsize{\\ 
$a$:  Bolometric luminosities derived within an over-density $\Delta_z =500$ for an Einstein--de Sitter universe.
Note that an X-ray luminosity with "$<$" is a upper limit.   \\
$b$: $R_{200}$ refers to the radius at which the 
cluster mean density is 200 times the critical density and is derived from the cluster velocity dispersion (Carlberg et al. 1997). \\
$c$: Total masses were estimated out to $R_{500}$ using a cluster $\beta$ model together with the measured emission--weighted X--ray  temperature (Ettori et al. 2004)\\ 
$d$: Minimum mean luminosity--weighted age as derived from the intrinsic scatter around the color--magnitude relation (Blakeslee et al. 2003a, 2006; Homeier et al. 2006; Mei et al. 2006a,b) \\ \\
References:  (1) Ettori et al. (2004); (2) Tran et al. (2007); (3) Girardi et al. (2005); (4) Demarco et al. (2005); (5) Kocevski et al. (2008);  (6) Gal et al. (2008); (7) Mei et al. (2006a); (8) Demarco et al. (2007);  (9) Jee et al. (2006); (10) Stanford et al. (2001)}
\end{table*}

\vspace{2cm}
\begin{table}
\caption{ACS Intermediate Redshift Cluster Survey sample -- CMR fits.   \label{fitcmr_mlm}}
\vspace{0.25cm}
\resizebox{!}{3.2cm}{
\begin{tabular}{llclccccccccccccc}
\tableline \tableline\\
Cluster&$ACS Color^a$&$ACS Mag^a$&$Type^b$&$N^c$&$c_0^d$&$Slope^d$&$Scatter^d$&$(U-B)_{z=0}^e$&$|\frac{\delta (U-B)_z}{\delta M_{B,z=0}}|^e$&$\sigma (U-B)_{z=0}^e$ \\
 &(mag)&(mag)&&&$(mag)$&&(mag)&(mag)&&(mag) \\
 \tableline\\
MS 1054--0321&($V_{606}-z_{850}$) &    $i_{775}$    &   E+S0 & 73 &       2.26  $\pm$       0.02 &     -0.052  $\pm$    0.029 &    0.124  $\pm$    0.015 &       0.37  $\pm$       0.01 &     -0.029  $\pm$    0.016 &    0.070  $\pm$    0.008     \\
 &   &      &         E & 42 &       2.24  $\pm$       0.02 &     -0.067  $\pm$    0.023 &    0.089  $\pm$    0.013 &       0.36  $\pm$       0.01 &     -0.038  $\pm$    0.013 &    0.050  $\pm$    0.007     \\

 RX~J0152.7--1357 &  ($r_{625}-z_{850}$)  &      $i_{775}$    &  E+S0 & 56 &       1.93  $\pm$       0.02 &     -0.040  $\pm$    0.017 &    0.079  $\pm$    0.008 &       0.38  $\pm$       0.01 &     -0.026  $\pm$    0.011 &    0.050  $\pm$    0.005     \\
  &        &       &         E & 36 &       1.94  $\pm$       0.02 &     -0.034  $\pm$    0.020 &    0.072  $\pm$    0.009 &       0.38  $\pm$       0.01 &     -0.022  $\pm$    0.013 &    0.046  $\pm$    0.006     \\

CL1604+4304 &   ($V_{606}-I_{814}$)  &         $I_{814}$    &      E+S0 & 39 &       1.78  $\pm$       0.01 &     -0.075  $\pm$    0.009 &    0.040  $\pm$    0.004 &       0.37  $\pm$       0.01 &     -0.058  $\pm$    0.007 &    0.031  $\pm$    0.003     \\
 &        &       &         E & 23 &       1.78  $\pm$       0.01 &     -0.074  $\pm$    0.012 &    0.042  $\pm$    0.006 &       0.37  $\pm$       0.01 &     -0.057  $\pm$    0.009 &    0.032  $\pm$    0.005     \\

 CL1604+4321&   ($V_{606}-I_{814}$)  &       $I_{814}$    &  E+S0 &    26 &       1.80  $\pm$       0.01 &     -0.042  $\pm$    0.020 &    0.057  $\pm$    0.008 &       0.35  $\pm$       0.01 &     -0.032  $\pm$    0.015 &    0.043  $\pm$    0.006     \\
 &       &      &         E & 19 &       1.81  $\pm$       0.01 &     -0.064  $\pm$    0.024 &    0.052  $\pm$    0.009 &       0.37  $\pm$       0.01 &     -0.049  $\pm$    0.018 &    0.039  $\pm$    0.007     \\

RDCS~J0910+5422 &   ($i_{775}-z_{850}$)  &         $z_{850}$    &    E+S0 & 30 &       1.01  $\pm$       0.01 &     -0.030  $\pm$    0.017 &    0.055  $\pm$    0.008 &       0.31  $\pm$       0.01 &     -0.033  $\pm$    0.019 &    0.060  $\pm$    0.009     \\
 &       &      &         E & 20 &       1.03  $\pm$       0.01 &     -0.032  $\pm$    0.014 &    0.044  $\pm$    0.010 &       0.33  $\pm$       0.01 &     -0.035  $\pm$    0.015 &    0.048  $\pm$    0.011     \\

RDCS~J1252.9-2927 &  ($i_{775}-z_{850}$)  &          $z_{850}$    &     E+S0 & 42 &       0.97  $\pm$       0.01 &     -0.034  $\pm$    0.017 &    0.066  $\pm$    0.013 &       0.36  $\pm$       0.02 &     -0.058  $\pm$    0.029 &    0.112  $\pm$    0.022     \\
 &       &       &         E & 25 &       0.97  $\pm$       0.01 &     -0.020  $\pm$    0.009 &    0.025  $\pm$    0.005 &       0.37  $\pm$       0.02 &     -0.034  $\pm$    0.015 &    0.042  $\pm$    0.008     \\

RX~J0849+4452 &   ($i_{775}-z_{850}$)  &      $z_{850}$    &                               E+S0 & 18 &       0.99  $\pm$       0.01 &     -0.021  $\pm$    0.021 &    0.039  $\pm$    0.008 &       0.44  $\pm$       0.02 &     -0.038  $\pm$    0.038 &    0.070  $\pm$    0.014     \\
 &          &&                             E & 10 &       0.99  $\pm$       0.01 &     -0.025  $\pm$    0.019 &    0.026  $\pm$    0.012 &       0.44  $\pm$       0.02 &     -0.045  $\pm$    0.034 &    0.047  $\pm$    0.022     \\
RX~J0848+4453  &   ($i_{775}-z_{850}$)  &             $z_{850}$    &                    E+S0 &  9 &       0.99  $\pm$       0.02 &     -0.055  $\pm$    0.018 &    0.027  $\pm$    0.015 &       0.46  $\pm$       0.04 &     -0.100  $\pm$    0.033 &    0.049  $\pm$    0.027     \\
 &          &&                           E &  6 &       0.98  $\pm$       0.02 &     -0.045  $\pm$    0.032 &    0.025  $\pm$    0.023 &       0.44  $\pm$       0.04 &     -0.082  $\pm$    0.058 &    0.045  $\pm$    0.042     \\

\tableline \tableline\\
\end{tabular}}
\scriptsize{\\ 
$a$: ACS color and magnitude used in this analysis. \\
$b$: Galaxy morphological type from Postman et al. (2005). \\
$c$: Number of galaxies used for the fit. \\
$d$: CMR fitted zero point $c_0$, slope and scatter in ACS colors and magnitude within $R_{200}$ and for the same range in $M_*$, e.g., at the same magnitude limit as the Postman et al. (2005) morphological classification. \\
$e$: CMR fitted zero point $(U-B)_{z=0}$, slope $|\frac{\delta (U-B)_{z=0}}{\delta M_{B,z=0}}|$, and scatter $\sigma (U-B)_{z=0}$ in the $(U-B)$ rest--frame. These are the zero--points, slopes and scatter labeled by footnote $d$ converted to the $(U-B)$ rest--frame, as detailed in Appendix~II.
}
\end{table}

\clearpage

\newpage

\centerline{\bf \large APPENDIX I: CMR Fit Parameters}

\vskip 0.5 truecm
\normalsize
\begin{table}[hb]
\begin{center}
\caption{Cluster MS 1054--0321 fitted CMR parameters.   \label{tcmr1}}
\vspace{0.25cm}
\resizebox{!}{4.cm}{
\begin{tabular}{llccclccccccccccc}
\tableline \tableline\\
$Sample^a$ &$Log_{10}(Density)^a$ & $R^a$ & $m^{lim}$&$M_{B,z=0}^{lim}$&$Type^b$&$N^c$&$c_0^d$&$Slope^d$&$Scatter^d$&$(U-B)_{z=0}^e$&$|\frac{\delta (U-B)_{z=0}}{\delta M_{B,z=0}}|^e$&$\sigma (U-B)_{z=0}^e$ \\
 &$Log_{10}(Gal/Mpc^2)$ & Mpc & mag&mag&&&$mag$&&mag&mag&&mag \\
\tableline\\
SRF&&&&&&&&\\
 & 1 &       1.80 &       22.7 &      -20.2 &      E+S0 & 61 &       2.25  $\pm$       0.02 &     -0.059  $\pm$    0.030 &    0.116  $\pm$    0.015 &       0.36  $\pm$       0.01 &     -0.033  $\pm$    0.017 &    0.065  $\pm$    0.008     \\
 & 1 &       1.80 &       22.7 &      -20.2 &         E & 37 &       2.24  $\pm$       0.02 &     -0.068  $\pm$    0.023 &    0.084  $\pm$    0.014 &       0.36  $\pm$       0.01 &     -0.038  $\pm$    0.013 &    0.047  $\pm$    0.008     \\
 & 2 &       1.80 &       22.7 &      -20.2 &      E+S0 & 50 &       2.25  $\pm$       0.02 &     -0.051  $\pm$    0.035 &    0.116  $\pm$    0.017 &       0.36  $\pm$       0.01 &     -0.029  $\pm$    0.020 &    0.065  $\pm$    0.010     \\
 & 2 &       1.80 &       22.7 &      -20.2 &         E & 31 &       2.25  $\pm$       0.02 &     -0.064  $\pm$    0.028 &    0.086  $\pm$    0.016 &       0.37  $\pm$       0.01 &     -0.036  $\pm$    0.016 &    0.048  $\pm$    0.009     \\
 & 1 &       0.90 &       22.7 &      -20.2 &      E+S0 & 53 &       2.25  $\pm$       0.02 &     -0.050  $\pm$    0.034 &    0.113  $\pm$    0.016 &       0.36  $\pm$       0.01 &     -0.028  $\pm$    0.019 &    0.063  $\pm$    0.009     \\
 & 1 &       0.90 &       22.7 &      -20.2 &         E & 34 &       2.24  $\pm$       0.02 &     -0.068  $\pm$    0.025 &    0.084  $\pm$    0.016 &       0.36  $\pm$       0.01 &     -0.038  $\pm$    0.014 &    0.047  $\pm$    0.009     \\
 & 2 &       0.90 &       22.7 &      -20.2 &      E+S0 & 49 &       2.26  $\pm$       0.02 &     -0.045  $\pm$    0.035 &    0.116  $\pm$    0.017 &       0.36  $\pm$       0.01 &     -0.025  $\pm$    0.020 &    0.065  $\pm$    0.010     \\
 & 2 &       0.90 &       22.7 &      -20.2 &         E & 30 &       2.25  $\pm$       0.02 &     -0.062  $\pm$    0.029 &    0.086  $\pm$    0.017 &       0.37  $\pm$       0.01 &     -0.035  $\pm$    0.016 &    0.048  $\pm$    0.010     \\
 & 1 &       1.00 &       22.7 &      -20.2 &      E+S0 & 56 &       2.25  $\pm$       0.02 &     -0.055  $\pm$    0.031 &    0.113  $\pm$    0.016 &       0.36  $\pm$       0.01 &     -0.031  $\pm$    0.017 &    0.063  $\pm$    0.009     \\
 & 1 &       1.00 &       22.7 &      -20.2 &         E & 37 &       2.24  $\pm$       0.02 &     -0.067  $\pm$    0.024 &    0.085  $\pm$    0.014 &       0.36  $\pm$       0.01 &     -0.038  $\pm$    0.013 &    0.048  $\pm$    0.008     \\
 & 2 &       1.00 &       22.7 &      -20.2 &      E+S0 & 50 &       2.25  $\pm$       0.02 &     -0.051  $\pm$    0.035 &    0.116  $\pm$    0.017 &       0.36  $\pm$       0.01 &     -0.029  $\pm$    0.020 &    0.065  $\pm$    0.010     \\
 & 2 &       1.00 &       22.7 &      -20.2 &         E & 31 &       2.25  $\pm$       0.02 &     -0.064  $\pm$    0.028 &    0.086  $\pm$    0.016 &       0.37  $\pm$       0.01 &     -0.036  $\pm$    0.016 &    0.048  $\pm$    0.009     \\
\tableline \\
SRMS&&&&&&$$&&\\
 & 1 &       1.80 &       23.0 &      -19.9 &      E+S0 & 73 &       2.26  $\pm$       0.02 &     -0.052  $\pm$    0.029 &    0.124  $\pm$    0.015 &       0.37  $\pm$       0.01 &     -0.029  $\pm$    0.016 &    0.070  $\pm$    0.008     \\
 & 1 &       1.80 &       23.0 &      -19.9 &         E & 42 &       2.24  $\pm$       0.02 &     -0.067  $\pm$    0.023 &    0.089  $\pm$    0.013 &       0.36  $\pm$       0.01 &     -0.038  $\pm$    0.013 &    0.050  $\pm$    0.007     \\
 & 2 &       1.80 &       23.0 &      -19.9 &      E+S0 & 61 &       2.27  $\pm$       0.02 &     -0.035  $\pm$    0.027 &    0.118  $\pm$    0.014 &       0.36  $\pm$       0.01 &     -0.020  $\pm$    0.015 &    0.066  $\pm$    0.008     \\
 & 2 &       1.80 &       23.0 &      -19.9 &         E & 36 &       2.25  $\pm$       0.02 &     -0.063  $\pm$    0.025 &    0.090  $\pm$    0.016 &       0.37  $\pm$       0.01 &     -0.035  $\pm$    0.014 &    0.050  $\pm$    0.009     \\
 & 1 &       0.90 &       23.0 &      -19.9 &      E+S0 & 64 &       2.27  $\pm$       0.02 &     -0.035  $\pm$    0.027 &    0.115  $\pm$    0.015 &       0.36  $\pm$       0.01 &     -0.020  $\pm$    0.015 &    0.065  $\pm$    0.008     \\
 & 1 &       0.90 &       23.0 &      -19.9 &         E & 39 &       2.24  $\pm$       0.02 &     -0.067  $\pm$    0.024 &    0.088  $\pm$    0.015 &       0.36  $\pm$       0.01 &     -0.038  $\pm$    0.013 &    0.049  $\pm$    0.008     \\
 & 2 &       0.90 &       23.0 &      -19.9 &      E+S0 & 60 &       2.27  $\pm$       0.02 &     -0.032  $\pm$    0.028 &    0.117  $\pm$    0.015 &       0.36  $\pm$       0.01 &     -0.018  $\pm$    0.016 &    0.066  $\pm$    0.008     \\
 & 2 &       0.90 &       23.0 &      -19.9 &         E & 35 &       2.25  $\pm$       0.02 &     -0.062  $\pm$    0.025 &    0.090  $\pm$    0.016 &       0.37  $\pm$       0.01 &     -0.035  $\pm$    0.014 &    0.050  $\pm$    0.009     \\
 & 1 &       1.00 &       23.0 &      -19.9 &      E+S0 & 67 &       2.26  $\pm$       0.02 &     -0.038  $\pm$    0.026 &    0.115  $\pm$    0.014 &       0.36  $\pm$       0.01 &     -0.021  $\pm$    0.015 &    0.065  $\pm$    0.008     \\
 & 1 &       1.00 &       23.0 &      -19.9 &         E & 42 &       2.24  $\pm$       0.02 &     -0.067  $\pm$    0.024 &    0.089  $\pm$    0.013 &       0.36  $\pm$       0.01 &     -0.038  $\pm$    0.013 &    0.050  $\pm$    0.007     \\
 & 2 &       1.00 &       23.0 &      -19.9 &      E+S0 & 61 &       2.27  $\pm$       0.02 &     -0.035  $\pm$    0.026 &    0.118  $\pm$    0.014 &       0.36  $\pm$       0.01 &     -0.020  $\pm$    0.015 &    0.066  $\pm$    0.008     \\
 & 2 &       1.00 &       23.0 &      -19.9 &         E & 36 &       2.25  $\pm$       0.02 &     -0.063  $\pm$    0.025 &    0.090  $\pm$    0.015 &       0.37  $\pm$       0.01 &     -0.035  $\pm$    0.014 &    0.050  $\pm$    0.008     \\
\tableline \tableline\\
\end{tabular}
}
\end{center}
\scriptsize{
$a$: Galaxy sample used in this analysis. The abbreviation $SFR$ stands for Same Reference Frame and $SRMS$ for Same Range in terms of $M_*$, as per Postman et al. (2005) and as used in the CMR paper series. Galaxies where selected according to neighboring galaxy density $Log_{10}(Density)$, distance from the cluster center $R$, and ACS magnitude (see Table~\ref{fitcmr_mlm}) limit  $m^{lim}$, which corresponds to a rest--frame magnitude limit $M_{B,z=0}^{lim}$.\\
$b$: Galaxy morphological type from Postman et al. (2005). \\
$c$: Number of galaxies used for the fit. \\
$d$: CMR fitted zero point $c_0$, slope and scatter in ACS colors and magnitude  (see Table~\ref{fitcmr_mlm}). \\
$e$: CMR fitted zero point $(U-B)_{z=0}$, slope $|\frac{\delta (U-B)_{z=0}}{\delta M_{B,z=0}}|$, and scatter $\sigma (U-B)_{z=0}$ in the $(U-B)$ rest--frame.  These are the zero--points, slopes and scatter corresponding to footnote $d$ converted to the $(U-B)$ rest--frame, as detailed in Appendix~II.
} 
\end{table}
\begin{table*}[ht]
\begin{center}
\caption{Cluster RX~J0152.7--1357 fitted CMR parameters.  \label{tcmr2}}
\vspace{0.25cm}
\resizebox{!}{4.cm}{
\begin{tabular}{llccclccccccccccc}
\tableline \tableline\\
$Sample^a$ &$Log_{10}(Density)^a$ & $R^a$ & $m^{lim}$&$M_{B,z=0}^{lim}$&$Type^b$&$N^c$&$c_0^d$&$Slope^d$&$Scatter^d$&$(U-B)_{z=0}^e$&$|\frac{\delta (U-B)_{z=0}}{\delta M_{B,z=0}|}^e$&$\sigma (U-B)_{z=0}^e$ \\
 &$Log_{10}(Gal/Mpc^2)$ & Mpc & mag&mag&&&$mag$&&mag&mag&&mag \\
\tableline\\
SRF&&&&&&&&\\
 & 1 &       1.90 &       22.8 &      -20.2 &      E+S0 & 50 &       1.95  $\pm$       0.02 &     -0.021  $\pm$    0.018 &    0.076  $\pm$    0.009 &       0.38  $\pm$       0.01 &     -0.013  $\pm$    0.012 &    0.049  $\pm$    0.006     \\
 & 1 &       1.90 &       22.8 &      -20.2 &         E & 33 &       1.96  $\pm$       0.02 &     -0.011  $\pm$    0.020 &    0.066  $\pm$    0.009 &       0.38  $\pm$       0.01 &     -0.007  $\pm$    0.013 &    0.042  $\pm$    0.006     \\
 & 2 &       1.90 &       22.8 &      -20.2 &      E+S0 & 38 &       1.96  $\pm$       0.02 &     -0.024  $\pm$    0.023 &    0.081  $\pm$    0.012 &       0.39  $\pm$       0.01 &     -0.015  $\pm$    0.015 &    0.052  $\pm$    0.008     \\
 & 2 &       1.90 &       22.8 &      -20.2 &         E & 24 &       1.98  $\pm$       0.03 &     -0.011  $\pm$    0.025 &    0.065  $\pm$    0.011 &       0.39  $\pm$       0.02 &     -0.007  $\pm$    0.016 &    0.042  $\pm$    0.007     \\
 & 1 &       0.95 &       22.8 &      -20.2 &      E+S0 & 37 &       1.95  $\pm$       0.02 &     -0.029  $\pm$    0.022 &    0.080  $\pm$    0.012 &       0.38  $\pm$       0.01 &     -0.019  $\pm$    0.014 &    0.051  $\pm$    0.008     \\
 & 1 &       0.95 &       22.8 &      -20.2 &         E & 25 &       1.96  $\pm$       0.03 &     -0.021  $\pm$    0.026 &    0.071  $\pm$    0.010 &       0.38  $\pm$       0.02 &     -0.013  $\pm$    0.017 &    0.045  $\pm$    0.006     \\
 & 2 &       0.95 &       22.8 &      -20.2 &      E+S0 & 34 &       1.96  $\pm$       0.03 &     -0.028  $\pm$    0.025 &    0.080  $\pm$    0.013 &       0.39  $\pm$       0.02 &     -0.018  $\pm$    0.016 &    0.051  $\pm$    0.008     \\
 & 2 &       0.95 &       22.8 &      -20.2 &         E & 22 &       1.97  $\pm$       0.03 &     -0.015  $\pm$    0.029 &    0.068  $\pm$    0.012 &       0.39  $\pm$       0.02 &     -0.010  $\pm$    0.019 &    0.043  $\pm$    0.008     \\
 & 1 &       1.00 &       22.8 &      -20.2 &      E+S0 & 41 &       1.95  $\pm$       0.02 &     -0.026  $\pm$    0.021 &    0.080  $\pm$    0.011 &       0.38  $\pm$       0.01 &     -0.017  $\pm$    0.013 &    0.051  $\pm$    0.007     \\
 & 1 &       1.00 &       22.8 &      -20.2 &         E & 27 &       1.96  $\pm$       0.02 &     -0.018  $\pm$    0.023 &    0.072  $\pm$    0.010 &       0.38  $\pm$       0.01 &     -0.012  $\pm$    0.015 &    0.046  $\pm$    0.006     \\
 & 2 &       1.00 &       22.8 &      -20.2 &      E+S0 & 37 &       1.96  $\pm$       0.02 &     -0.029  $\pm$    0.023 &    0.080  $\pm$    0.012 &       0.39  $\pm$       0.01 &     -0.019  $\pm$    0.015 &    0.051  $\pm$    0.008     \\
 & 2 &       1.00 &       22.8 &      -20.2 &         E & 23 &       1.97  $\pm$       0.03 &     -0.016  $\pm$    0.026 &    0.067  $\pm$    0.012 &       0.39  $\pm$       0.02 &     -0.010  $\pm$    0.017 &    0.043  $\pm$    0.008     \\
\tableline \\
SRMS&&&&&&$$&&\\
 & 1 &       1.90 &       23.0 &      -20.0 &      E+S0 & 56 &       1.93  $\pm$       0.02 &     -0.040  $\pm$    0.017 &    0.079  $\pm$    0.008 &       0.38  $\pm$       0.01 &     -0.026  $\pm$    0.011 &    0.050  $\pm$    0.005     \\
 & 1 &       1.90 &       23.0 &      -20.0 &         E & 36 &       1.94  $\pm$       0.02 &     -0.034  $\pm$    0.020 &    0.072  $\pm$    0.009 &       0.38  $\pm$       0.01 &     -0.022  $\pm$    0.013 &    0.046  $\pm$    0.006     \\
 & 2 &       1.90 &       23.0 &      -20.0 &      E+S0 & 40 &       1.95  $\pm$       0.02 &     -0.033  $\pm$    0.020 &    0.081  $\pm$    0.011 &       0.38  $\pm$       0.01 &     -0.021  $\pm$    0.013 &    0.052  $\pm$    0.007     \\
 & 2 &       1.90 &       23.0 &      -20.0 &         E & 25 &       1.97  $\pm$       0.02 &     -0.021  $\pm$    0.023 &    0.066  $\pm$    0.010 &       0.39  $\pm$       0.01 &     -0.013  $\pm$    0.015 &    0.042  $\pm$    0.006     \\
 & 1 &       0.95 &       23.0 &      -20.0 &      E+S0 & 38 &       1.95  $\pm$       0.02 &     -0.032  $\pm$    0.021 &    0.079  $\pm$    0.011 &       0.38  $\pm$       0.01 &     -0.020  $\pm$    0.013 &    0.050  $\pm$    0.007     \\
 & 1 &       0.95 &       23.0 &      -20.0 &         E & 26 &       1.95  $\pm$       0.02 &     -0.028  $\pm$    0.024 &    0.070  $\pm$    0.010 &       0.38  $\pm$       0.01 &     -0.018  $\pm$    0.015 &    0.045  $\pm$    0.006     \\
 & 2 &       0.95 &       23.0 &      -20.0 &      E+S0 & 35 &       1.96  $\pm$       0.02 &     -0.032  $\pm$    0.023 &    0.079  $\pm$    0.012 &       0.39  $\pm$       0.01 &     -0.020  $\pm$    0.015 &    0.050  $\pm$    0.008     \\
 & 2 &       0.95 &       23.0 &      -20.0 &         E & 23 &       1.97  $\pm$       0.03 &     -0.022  $\pm$    0.024 &    0.068  $\pm$    0.011 &       0.39  $\pm$       0.02 &     -0.014  $\pm$    0.015 &    0.043  $\pm$    0.007     \\
 & 1 &       1.00 &       23.0 &      -20.0 &      E+S0 & 42 &       1.95  $\pm$       0.02 &     -0.029  $\pm$    0.020 &    0.079  $\pm$    0.011 &       0.38  $\pm$       0.01 &     -0.019  $\pm$    0.013 &    0.050  $\pm$    0.007     \\
 & 1 &       1.00 &       23.0 &      -20.0 &         E & 28 &       1.95  $\pm$       0.02 &     -0.024  $\pm$    0.020 &    0.071  $\pm$    0.009 &       0.38  $\pm$       0.01 &     -0.015  $\pm$    0.013 &    0.045  $\pm$    0.006     \\
 & 2 &       1.00 &       23.0 &      -20.0 &      E+S0 & 38 &       1.95  $\pm$       0.02 &     -0.033  $\pm$    0.021 &    0.079  $\pm$    0.011 &       0.38  $\pm$       0.01 &     -0.021  $\pm$    0.013 &    0.050  $\pm$    0.007     \\
 & 2 &       1.00 &       23.0 &      -20.0 &         E & 24 &       1.96  $\pm$       0.02 &     -0.024  $\pm$    0.023 &    0.067  $\pm$    0.010 &       0.39  $\pm$       0.01 &     -0.015  $\pm$    0.015 &    0.043  $\pm$    0.006     \\
\tableline \tableline\\
\end{tabular}
}
\end{center}
\footnotesize{Footnotes $a$,$b$,$c$,$d$, and $e$ as in Table~\ref{tcmr1}}
\end{table*}
\begin{table*}[hb]
\begin{center}
\caption{Cluster CL1604+4304 fitted CMR parameters.  \label{tcmr3}}
\vspace{0.25cm}
\resizebox{!}{4.2cm}{
\begin{tabular}{llccclccccccccccc}
\tableline \tableline\\
$Sample^a$ &$Log_{10}(Density)^a$ & $R^a$ & $m^{lim}$&$M_{B,z=0}^{lim}$&$Type^b$&$N^c$&$c_0^d$&$Slope^d$&$Scatter^d$&$(U-B)_{z=0}^e$&$|\frac{\delta (U-B)_{z=0}}{\delta M_{B,z=0}}|^e$&$\sigma (U-B)_{z=0}^e$ \\
 &$Log_{10}(Gal/Mpc^2)$ & Mpc & mag&mag&&&$mag$&&mag&mag&&mag \\
\tableline\\
SRF&&&&&&&&\\
 & 1 &       1.10 &       22.8 &      -20.2 &      E+S0 & 22 &       1.79  $\pm$       0.01 &     -0.056  $\pm$    0.015 &    0.037  $\pm$    0.005 &       0.37  $\pm$       0.01 &     -0.043  $\pm$    0.012 &    0.029  $\pm$    0.004     \\
 & 1 &       1.10 &       22.8 &      -20.2 &         E & 13 &       1.80  $\pm$       0.01 &     -0.041  $\pm$    0.018 &    0.032  $\pm$    0.008 &       0.37  $\pm$       0.01 &     -0.032  $\pm$    0.014 &    0.025  $\pm$    0.006     \\
 & 2 &       1.10 &       22.8 &      -20.2 &      E+S0 & 13 &       1.79  $\pm$       0.01 &     -0.060  $\pm$    0.020 &    0.035  $\pm$    0.008 &       0.37  $\pm$       0.01 &     -0.046  $\pm$    0.015 &    0.027  $\pm$    0.006     \\
 & 2 &       1.10 &       22.8 &      -20.2 &         E &  8 &       1.80  $\pm$       0.01 &     -0.034  $\pm$    0.011 &    0.015  $\pm$    0.006 &       0.36  $\pm$       0.01 &     -0.026  $\pm$    0.008 &    0.012  $\pm$    0.005     \\
 & 1 &       0.55 &       22.8 &      -20.2 &      E+S0 & 16 &       1.79  $\pm$       0.02 &     -0.055  $\pm$    0.019 &    0.043  $\pm$    0.006 &       0.37  $\pm$       0.02 &     -0.042  $\pm$    0.015 &    0.033  $\pm$    0.005     \\
 & 1 &       0.55 &       22.8 &      -20.2 &         E &  9 &       1.81  $\pm$       0.01 &     -0.039  $\pm$    0.024 &    0.039  $\pm$    0.012 &       0.37  $\pm$       0.01 &     -0.030  $\pm$    0.019 &    0.030  $\pm$    0.009     \\
 & 2 &       0.55 &       22.8 &      -20.2 &      E+S0 &  9 &       1.80  $\pm$       0.02 &     -0.076  $\pm$    0.036 &    0.043  $\pm$    0.012 &       0.39  $\pm$       0.02 &     -0.059  $\pm$    0.028 &    0.033  $\pm$    0.009     \\
 & 2 &       0.55 &       22.8 &      -20.2 &         E &  5 &       1.82  $\pm$       0.01 &     -0.028  $\pm$    0.004 &    0.013  $\pm$    0.008 &       0.37  $\pm$       0.01 &     -0.022  $\pm$    0.003 &    0.010  $\pm$    0.006     \\
 & 1 &       1.00 &       22.8 &      -20.2 &      E+S0 & 22 &       1.79  $\pm$       0.01 &     -0.056  $\pm$    0.015 &    0.037  $\pm$    0.005 &       0.37  $\pm$       0.01 &     -0.043  $\pm$    0.012 &    0.029  $\pm$    0.004     \\
 & 1 &       1.00 &       22.8 &      -20.2 &         E & 13 &       1.80  $\pm$       0.01 &     -0.041  $\pm$    0.018 &    0.032  $\pm$    0.008 &       0.37  $\pm$       0.01 &     -0.032  $\pm$    0.014 &    0.025  $\pm$    0.006     \\
 & 2 &       1.00 &       22.8 &      -20.2 &      E+S0 & 13 &       1.79  $\pm$       0.01 &     -0.060  $\pm$    0.020 &    0.034  $\pm$    0.008 &       0.37  $\pm$       0.01 &     -0.046  $\pm$    0.015 &    0.026  $\pm$    0.006     \\
 & 2 &       1.00 &       22.8 &      -20.2 &         E &  8 &       1.80  $\pm$       0.01 &     -0.034  $\pm$    0.011 &    0.015  $\pm$    0.006 &       0.36  $\pm$       0.01 &     -0.026  $\pm$    0.008 &    0.012  $\pm$    0.005     \\
\tableline \\
SRMS&&&&&&$$&&\\
 & 1 &       1.10 &       24.0 &      -19.0 &      E+S0 & 39 &       1.78  $\pm$       0.01 &     -0.075  $\pm$    0.009 &    0.040  $\pm$    0.004 &       0.37  $\pm$       0.01 &     -0.058  $\pm$    0.007 &    0.031  $\pm$    0.003     \\
 & 1 &       1.10 &       24.0 &      -19.0 &         E & 23 &       1.78  $\pm$       0.01 &     -0.074  $\pm$    0.012 &    0.042  $\pm$    0.006 &       0.37  $\pm$       0.01 &     -0.057  $\pm$    0.009 &    0.032  $\pm$    0.005     \\
 & 2 &       1.10 &       24.0 &      -19.0 &      E+S0 & 20 &       1.78  $\pm$       0.01 &     -0.076  $\pm$    0.016 &    0.043  $\pm$    0.006 &       0.37  $\pm$       0.01 &     -0.059  $\pm$    0.012 &    0.033  $\pm$    0.005     \\
 & 2 &       1.10 &       24.0 &      -19.0 &         E & 11 &       1.80  $\pm$       0.02 &     -0.077  $\pm$    0.052 &    0.046  $\pm$    0.020 &       0.39  $\pm$       0.02 &     -0.059  $\pm$    0.040 &    0.035  $\pm$    0.015     \\
 & 1 &       0.55 &       24.0 &      -19.0 &      E+S0 & 29 &       1.78  $\pm$       0.01 &     -0.072  $\pm$    0.010 &    0.040  $\pm$    0.005 &       0.37  $\pm$       0.01 &     -0.056  $\pm$    0.008 &    0.031  $\pm$    0.004     \\
 & 1 &       0.55 &       24.0 &      -19.0 &         E & 17 &       1.78  $\pm$       0.02 &     -0.072  $\pm$    0.013 &    0.046  $\pm$    0.006 &       0.37  $\pm$       0.02 &     -0.056  $\pm$    0.010 &    0.035  $\pm$    0.005     \\
 & 2 &       0.55 &       24.0 &      -19.0 &      E+S0 & 14 &       1.80  $\pm$       0.02 &     -0.070  $\pm$    0.017 &    0.038  $\pm$    0.010 &       0.38  $\pm$       0.02 &     -0.054  $\pm$    0.013 &    0.029  $\pm$    0.008     \\
 & 2 &       0.55 &       24.0 &      -19.0 &         E &  7 &       1.82  $\pm$       0.01 &     -0.060  $\pm$    0.037 &    0.041  $\pm$    0.029 &       0.39  $\pm$       0.01 &     -0.046  $\pm$    0.029 &    0.032  $\pm$    0.022     \\
 & 1 &       1.00 &       24.0 &      -19.0 &      E+S0 & 39 &       1.78  $\pm$       0.01 &     -0.075  $\pm$    0.009 &    0.040  $\pm$    0.004 &       0.37  $\pm$       0.01 &     -0.058  $\pm$    0.007 &    0.031  $\pm$    0.003     \\
 & 1 &       1.00 &       24.0 &      -19.0 &         E & 23 &       1.78  $\pm$       0.01 &     -0.074  $\pm$    0.012 &    0.043  $\pm$    0.006 &       0.37  $\pm$       0.01 &     -0.057  $\pm$    0.009 &    0.033  $\pm$    0.005     \\
 & 2 &       1.00 &       24.0 &      -19.0 &      E+S0 & 20 &       1.78  $\pm$       0.01 &     -0.076  $\pm$    0.016 &    0.043  $\pm$    0.006 &       0.37  $\pm$       0.01 &     -0.059  $\pm$    0.012 &    0.033  $\pm$    0.005     \\
 & 2 &       1.00 &       24.0 &      -19.0 &         E & 11 &       1.80  $\pm$       0.02 &     -0.076  $\pm$    0.049 &    0.046  $\pm$    0.019 &       0.39  $\pm$       0.02 &     -0.059  $\pm$    0.038 &    0.035  $\pm$    0.015     \\

\tableline \tableline\\
\end{tabular}
}
\end{center}
\footnotesize{Footnotes $a$,$b$,$c$,$d$, and $e$ as in Table~\ref{tcmr1}}
\end{table*}
\begin{table*}
\begin{center}
\caption{Cluster CL1604+4321 fitted CMR parameters. \label{tcmr4}}
\vspace{0.25cm}
\resizebox{!}{4.22cm}{
\begin{tabular}{llccclccccccccccc}
\tableline \tableline\\
$Sample^a$ &$Log_{10}(Density)^a$ & $R^a$ & $m^{lim}$&$M_{B,z=0}^{lim}$&$Type^b$&$N^c$&$c_0^d$&$Slope^d$&$Scatter^d$&$(U-B)_{z=0}^e$&$|\frac{\delta (U-B)_{z=0}}{\delta M_{B,z=0}}|^e$&$\sigma (U-B)_{z=0}^e$ \\
 &$Log_{10}(Gal/Mpc^2)$ & Mpc & mag&mag&&&$mag$&&mag&mag&&mag \\
\tableline\\
SRF&&&&&&&&\\
 & 1 &       0.90 &       23.0 &      -20.2 &      E+S0 & 15 &       1.79  $\pm$       0.02 &     -0.061  $\pm$    0.043 &    0.048  $\pm$    0.011 &       0.35  $\pm$       0.02 &     -0.046  $\pm$    0.033 &    0.036  $\pm$    0.008     \\
 & 1 &       0.90 &       23.0 &      -20.2 &         E & 11 &       1.80  $\pm$       0.02 &     -0.090  $\pm$    0.049 &    0.040  $\pm$    0.013 &       0.38  $\pm$       0.02 &     -0.068  $\pm$    0.037 &    0.030  $\pm$    0.010     \\
 & 2 &       0.90 &       23.0 &      -20.2 &      E+S0 & 10 &       1.80  $\pm$       0.04 &     -0.085  $\pm$    0.071 &    0.043  $\pm$    0.018 &       0.37  $\pm$       0.03 &     -0.065  $\pm$    0.054 &    0.033  $\pm$    0.014     \\
 & 2 &       0.90 &       23.0 &      -20.2 &         E &  7 &       1.83  $\pm$       0.01 &     -0.053  $\pm$    0.047 &    0.031  $\pm$    0.019 &       0.38  $\pm$       0.01 &     -0.040  $\pm$    0.036 &    0.024  $\pm$    0.014     \\
 & 1 &       0.45 &       23.0 &      -20.2 &      E+S0 & 11 &       1.79  $\pm$       0.02 &     -0.055  $\pm$    0.049 &    0.038  $\pm$    0.013 &       0.35  $\pm$       0.02 &     -0.042  $\pm$    0.037 &    0.029  $\pm$    0.010     \\
 & 1 &       0.45 &       23.0 &      -20.2 &         E &  7 &       1.80  $\pm$       0.02 &     -0.044  $\pm$    0.058 &    0.022  $\pm$    0.011 &       0.35  $\pm$       0.02 &     -0.033  $\pm$    0.044 &    0.017  $\pm$    0.008     \\
 & 2 &       0.45 &       23.0 &      -20.2 &      E+S0 &  8 &       1.78  $\pm$       0.04 &     -0.099  $\pm$    0.082 &    0.035  $\pm$    0.020 &       0.36  $\pm$       0.03 &     -0.075  $\pm$    0.062 &    0.027  $\pm$    0.015     \\
 & 2 &       0.45 &       23.0 &      -20.2 &         E &  5 &       1.82  $\pm$       0.01 &     -0.034  $\pm$    0.021 &    0.016  $\pm$    0.009 &       0.36  $\pm$       0.01 &     -0.026  $\pm$    0.016 &    0.012  $\pm$    0.007     \\
 & 1 &       1.00 &       23.0 &      -20.2 &      E+S0 & 15 &       1.79  $\pm$       0.02 &     -0.061  $\pm$    0.041 &    0.048  $\pm$    0.011 &       0.35  $\pm$       0.02 &     -0.046  $\pm$    0.031 &    0.036  $\pm$    0.008     \\
 & 1 &       1.00 &       23.0 &      -20.2 &         E & 11 &       1.80  $\pm$       0.02 &     -0.091  $\pm$    0.049 &    0.040  $\pm$    0.012 &       0.38  $\pm$       0.02 &     -0.069  $\pm$    0.037 &    0.030  $\pm$    0.009     \\
 & 2 &       1.00 &       23.0 &      -20.2 &      E+S0 & 10 &       1.80  $\pm$       0.04 &     -0.083  $\pm$    0.067 &    0.044  $\pm$    0.017 &       0.37  $\pm$       0.03 &     -0.063  $\pm$    0.051 &    0.033  $\pm$    0.013     \\
 & 2 &       1.00 &       23.0 &      -20.2 &         E &  7 &       1.83  $\pm$       0.01 &     -0.053  $\pm$    0.047 &    0.031  $\pm$    0.019 &       0.38  $\pm$       0.01 &     -0.040  $\pm$    0.036 &    0.024  $\pm$    0.014     \\
\tableline \\
SRMS&&&&&&$$&&\\
 & 1 &       0.90 &       24.0 &      -19.2 &      E+S0 & 26 &       1.80  $\pm$       0.01 &     -0.042  $\pm$    0.020 &    0.057  $\pm$    0.008 &       0.35  $\pm$       0.01 &     -0.032  $\pm$    0.015 &    0.043  $\pm$    0.006     \\
 & 1 &       0.90 &       24.0 &      -19.2 &         E & 19 &       1.81  $\pm$       0.01 &     -0.064  $\pm$    0.024 &    0.052  $\pm$    0.009 &       0.37  $\pm$       0.01 &     -0.049  $\pm$    0.018 &    0.039  $\pm$    0.007     \\
 & 2 &       0.90 &       24.0 &      -19.2 &      E+S0 & 17 &       1.82  $\pm$       0.01 &     -0.045  $\pm$    0.021 &    0.048  $\pm$    0.010 &       0.37  $\pm$       0.01 &     -0.034  $\pm$    0.016 &    0.036  $\pm$    0.008     \\
 & 2 &       0.90 &       24.0 &      -19.2 &         E & 12 &       1.83  $\pm$       0.01 &     -0.052  $\pm$    0.019 &    0.029  $\pm$    0.010 &       0.38  $\pm$       0.01 &     -0.039  $\pm$    0.014 &    0.022  $\pm$    0.008     \\
 & 1 &       0.45 &       24.0 &      -19.2 &      E+S0 & 15 &       1.79  $\pm$       0.01 &     -0.053  $\pm$    0.020 &    0.038  $\pm$    0.008 &       0.35  $\pm$       0.01 &     -0.040  $\pm$    0.015 &    0.029  $\pm$    0.006     \\
 & 1 &       0.45 &       24.0 &      -19.2 &         E & 11 &       1.80  $\pm$       0.01 &     -0.062  $\pm$    0.023 &    0.029  $\pm$    0.009 &       0.36  $\pm$       0.01 &     -0.047  $\pm$    0.017 &    0.022  $\pm$    0.007     \\
 & 2 &       0.45 &       24.0 &      -19.2 &      E+S0 & 11 &       1.80  $\pm$       0.01 &     -0.053  $\pm$    0.016 &    0.034  $\pm$    0.014 &       0.36  $\pm$       0.01 &     -0.040  $\pm$    0.012 &    0.026  $\pm$    0.011     \\
 & 2 &       0.45 &       24.0 &      -19.2 &         E &  8 &       1.81  $\pm$       0.01 &     -0.054  $\pm$    0.016 &    0.018  $\pm$    0.009 &       0.36  $\pm$       0.01 &     -0.041  $\pm$    0.012 &    0.014  $\pm$    0.007     \\
 & 1 &       1.00 &       24.0 &      -19.2 &      E+S0 & 26 &       1.80  $\pm$       0.01 &     -0.042  $\pm$    0.020 &    0.057  $\pm$    0.008 &       0.35  $\pm$       0.01 &     -0.032  $\pm$    0.015 &    0.043  $\pm$    0.006     \\
 & 1 &       1.00 &       24.0 &      -19.2 &         E & 19 &       1.81  $\pm$       0.01 &     -0.064  $\pm$    0.024 &    0.052  $\pm$    0.009 &       0.37  $\pm$       0.01 &     -0.049  $\pm$    0.018 &    0.039  $\pm$    0.007     \\
 & 2 &       1.00 &       24.0 &      -19.2 &      E+S0 & 17 &       1.82  $\pm$       0.01 &     -0.045  $\pm$    0.021 &    0.048  $\pm$    0.011 &       0.37  $\pm$       0.01 &     -0.034  $\pm$    0.016 &    0.036  $\pm$    0.008     \\
 & 2 &       1.00 &       24.0 &      -19.2 &         E & 12 &       1.83  $\pm$       0.01 &     -0.052  $\pm$    0.020 &    0.029  $\pm$    0.010 &       0.38  $\pm$       0.01 &     -0.039  $\pm$    0.015 &    0.022  $\pm$    0.008     \\

\tableline \tableline\\
\end{tabular}
}
\end{center}
\footnotesize{Footnotes $a$,$b$,$c$,$d$, and $e$ as in Table~\ref{tcmr1}}
\end{table*}
\begin{table*}
\begin{center}
\caption{Table CMR RDCS~J0910+5422 \label{tcmr5}}
\vspace{0.25cm}
\resizebox{!}{4.2cm}{
\begin{tabular}{llccclccccccccccc}
\tableline \tableline\\
$Sample^a$ &$Log_{10}(Density)^a$ & $R^a$ & $m^{lim}$&$M_{B,z=0}^{lim}$&$Type^b$&$N^c$&$c_0^d$&$Slope^d$&$Scatter^d$&$(U-B)_{z=0}^e$&$|\frac{\delta (U-B)_{z=0}}{\delta M_{B,z=0}}|^e$&$\sigma (U-B)_{z=0}^e$ \\
 &$Log_{10}(Gal/Mpc^2)$ & Mpc & mag&mag&&&$mag$&&mag&mag&&mag \\
\tableline\\
SRF&&&&&&&&\\
 & 1 &       0.90 &       23.2 &      -20.2 &      E+S0 & 20 &       1.00  $\pm$       0.02 &     -0.060  $\pm$    0.040 &    0.061  $\pm$    0.009 &       0.31  $\pm$       0.02 &     -0.066  $\pm$    0.044 &    0.067  $\pm$    0.010     \\
 & 1 &       0.90 &       23.2 &      -20.2 &         E & 14 &       1.03  $\pm$       0.01 &     -0.049  $\pm$    0.045 &    0.049  $\pm$    0.013 &       0.34  $\pm$       0.01 &     -0.054  $\pm$    0.049 &    0.054  $\pm$    0.014     \\
 & 2 &       0.90 &       23.2 &      -20.2 &      E+S0 & 12 &       0.98  $\pm$       0.02 &     -0.111  $\pm$    0.055 &    0.048  $\pm$    0.012 &       0.31  $\pm$       0.02 &     -0.122  $\pm$    0.060 &    0.053  $\pm$    0.013     \\
 & 2 &       0.90 &       23.2 &      -20.2 &         E &  7 &       1.01  $\pm$       0.02 &     -0.095  $\pm$    0.042 &    0.044  $\pm$    0.016 &       0.34  $\pm$       0.02 &     -0.104  $\pm$    0.046 &    0.048  $\pm$    0.018     \\
 & 1 &       0.45 &       23.2 &      -20.2 &      E+S0 & 18 &       1.00  $\pm$       0.02 &     -0.045  $\pm$    0.053 &    0.062  $\pm$    0.010 &       0.30  $\pm$       0.02 &     -0.049  $\pm$    0.058 &    0.068  $\pm$    0.011     \\
 & 1 &       0.45 &       23.2 &      -20.2 &         E & 13 &       1.03  $\pm$       0.02 &     -0.033  $\pm$    0.043 &    0.047  $\pm$    0.014 &       0.33  $\pm$       0.02 &     -0.036  $\pm$    0.047 &    0.052  $\pm$    0.015     \\
 & 2 &       0.45 &       23.2 &      -20.2 &      E+S0 & 12 &       0.98  $\pm$       0.02 &     -0.111  $\pm$    0.056 &    0.048  $\pm$    0.012 &       0.31  $\pm$       0.02 &     -0.122  $\pm$    0.061 &    0.053  $\pm$    0.013     \\
 & 2 &       0.45 &       23.2 &      -20.2 &         E &  7 &       1.00  $\pm$       0.02 &     -0.095  $\pm$    0.045 &    0.044  $\pm$    0.017 &       0.33  $\pm$       0.02 &     -0.104  $\pm$    0.049 &    0.048  $\pm$    0.019     \\
 & 1 &       1.00 &       23.2 &      -20.2 &      E+S0 & 21 &       1.00  $\pm$       0.01 &     -0.060  $\pm$    0.034 &    0.059  $\pm$    0.010 &       0.31  $\pm$       0.01 &     -0.066  $\pm$    0.037 &    0.065  $\pm$    0.011     \\
 & 1 &       1.00 &       23.2 &      -20.2 &         E & 14 &       1.03  $\pm$       0.01 &     -0.049  $\pm$    0.047 &    0.049  $\pm$    0.013 &       0.34  $\pm$       0.01 &     -0.054  $\pm$    0.052 &    0.054  $\pm$    0.014     \\
 & 2 &       1.00 &       23.2 &      -20.2 &      E+S0 & 12 &       0.98  $\pm$       0.02 &     -0.110  $\pm$    0.055 &    0.048  $\pm$    0.012 &       0.31  $\pm$       0.02 &     -0.121  $\pm$    0.060 &    0.053  $\pm$    0.013     \\
 & 2 &       1.00 &       23.2 &      -20.2 &         E &  7 &       1.00  $\pm$       0.02 &     -0.096  $\pm$    0.045 &    0.043  $\pm$    0.016 &       0.33  $\pm$       0.02 &     -0.105  $\pm$    0.049 &    0.047  $\pm$    0.018     \\
\tableline \\
SRMS&&&&&&$$&&\\
 & 1 &       0.90 &       24.0 &      -19.4 &      E+S0 & 30 &       1.01  $\pm$       0.01 &     -0.030  $\pm$    0.017 &    0.055  $\pm$    0.008 &       0.31  $\pm$       0.01 &     -0.033  $\pm$    0.019 &    0.060  $\pm$    0.009     \\
 & 1 &       0.90 &       24.0 &      -19.4 &         E & 20 &       1.03  $\pm$       0.01 &     -0.032  $\pm$    0.014 &    0.044  $\pm$    0.010 &       0.33  $\pm$       0.01 &     -0.035  $\pm$    0.015 &    0.048  $\pm$    0.011     \\
 & 2 &       0.90 &       24.0 &      -19.4 &      E+S0 & 20 &       0.99  $\pm$       0.02 &     -0.025  $\pm$    0.022 &    0.052  $\pm$    0.010 &       0.28  $\pm$       0.02 &     -0.027  $\pm$    0.024 &    0.057  $\pm$    0.011     \\
 & 2 &       0.90 &       24.0 &      -19.4 &         E & 11 &       1.02  $\pm$       0.02 &     -0.042  $\pm$    0.028 &    0.044  $\pm$    0.014 &       0.32  $\pm$       0.02 &     -0.046  $\pm$    0.031 &    0.048  $\pm$    0.015     \\
 & 1 &       0.45 &       24.0 &      -19.4 &      E+S0 & 28 &       1.01  $\pm$       0.01 &     -0.024  $\pm$    0.018 &    0.054  $\pm$    0.009 &       0.30  $\pm$       0.01 &     -0.026  $\pm$    0.020 &    0.059  $\pm$    0.010     \\
 & 1 &       0.45 &       24.0 &      -19.4 &         E & 19 &       1.03  $\pm$       0.01 &     -0.034  $\pm$    0.014 &    0.042  $\pm$    0.011 &       0.33  $\pm$       0.01 &     -0.037  $\pm$    0.015 &    0.046  $\pm$    0.012     \\
 & 2 &       0.45 &       24.0 &      -19.4 &      E+S0 & 20 &       0.99  $\pm$       0.02 &     -0.025  $\pm$    0.022 &    0.052  $\pm$    0.010 &       0.28  $\pm$       0.02 &     -0.027  $\pm$    0.024 &    0.057  $\pm$    0.011     \\
 & 2 &       0.45 &       24.0 &      -19.4 &         E & 11 &       1.02  $\pm$       0.02 &     -0.043  $\pm$    0.027 &    0.044  $\pm$    0.014 &       0.32  $\pm$       0.02 &     -0.047  $\pm$    0.030 &    0.048  $\pm$    0.015     \\
 & 1 &       1.00 &       24.0 &      -19.4 &      E+S0 & 31 &       1.01  $\pm$       0.01 &     -0.032  $\pm$    0.016 &    0.054  $\pm$    0.008 &       0.31  $\pm$       0.01 &     -0.035  $\pm$    0.018 &    0.059  $\pm$    0.009     \\
 & 1 &       1.00 &       24.0 &      -19.4 &         E & 20 &       1.03  $\pm$       0.01 &     -0.032  $\pm$    0.015 &    0.044  $\pm$    0.010 &       0.33  $\pm$       0.01 &     -0.035  $\pm$    0.016 &    0.048  $\pm$    0.011     \\
 & 2 &       1.00 &       24.0 &      -19.4 &      E+S0 & 20 &       0.99  $\pm$       0.02 &     -0.025  $\pm$    0.022 &    0.052  $\pm$    0.010 &       0.28  $\pm$       0.02 &     -0.027  $\pm$    0.024 &    0.057  $\pm$    0.011     \\
 & 2 &       1.00 &       24.0 &      -19.4 &         E & 11 &       1.02  $\pm$       0.02 &     -0.043  $\pm$    0.028 &    0.044  $\pm$    0.014 &       0.32  $\pm$       0.02 &     -0.047  $\pm$    0.031 &    0.048  $\pm$    0.015     \\

\tableline \tableline\\
\end{tabular}
}
\end{center}
\footnotesize{Footnotes $a$,$b$,$c$,$d$, and $e$ as in Table~\ref{tcmr1}}
\end{table*}
\begin{table*}
\begin{center}
\caption{Table CMR  RDCS~J1252.9-2927\label{tcmr6}}
\vspace{0.25cm}
\resizebox{!}{4.2cm}{
\begin{tabular}{llccclccccccccccc}
\tableline \tableline\\
$Sample^a$ &$Log_{10}(Density)^a$ & $R^a$ & $m^{lim}$&$M_{B,z=0}^{lim}$&$Type^b$&$N^c$&$c_0^d$&$Slope^d$&$Scatter^d$&$(U-B)_{z=0}^e$&$|\frac{\delta (U-B)_{z=0}}{\delta M_{B,z=0}}|^e$&$\sigma (U-B)_{z=0}^e$ \\
 &$Log_{10}(Gal/Mpc^2)$ & Mpc & mag&mag&&&$mag$&&mag&mag&&mag \\
\tableline\\
SRF&&&&&&&&\\
 & 1 &       0.90 &       23.8 &      -20.2 &      E+S0 & 37 &       0.97  $\pm$       0.01 &     -0.035  $\pm$    0.015 &    0.057  $\pm$    0.014 &       0.36  $\pm$       0.02 &     -0.059  $\pm$    0.025 &    0.096  $\pm$    0.024     \\
 & 1 &       0.90 &       23.8 &      -20.2 &         E & 25 &       0.97  $\pm$       0.01 &     -0.020  $\pm$    0.009 &    0.025  $\pm$    0.005 &       0.37  $\pm$       0.02 &     -0.034  $\pm$    0.015 &    0.042  $\pm$    0.008     \\
 & 2 &       0.90 &       23.8 &      -20.2 &      E+S0 & 36 &       0.97  $\pm$       0.01 &     -0.037  $\pm$    0.015 &    0.056  $\pm$    0.014 &       0.36  $\pm$       0.02 &     -0.063  $\pm$    0.025 &    0.095  $\pm$    0.024     \\
 & 2 &       0.90 &       23.8 &      -20.2 &         E & 24 &       0.97  $\pm$       0.01 &     -0.022  $\pm$    0.009 &    0.024  $\pm$    0.005 &       0.37  $\pm$       0.02 &     -0.037  $\pm$    0.015 &    0.041  $\pm$    0.008     \\
 & 1 &       0.45 &       23.8 &      -20.2 &      E+S0 & 25 &       0.97  $\pm$       0.01 &     -0.030  $\pm$    0.011 &    0.051  $\pm$    0.020 &       0.36  $\pm$       0.02 &     -0.051  $\pm$    0.019 &    0.086  $\pm$    0.034     \\
 & 1 &       0.45 &       23.8 &      -20.2 &         E & 16 &       0.97  $\pm$       0.01 &     -0.019  $\pm$    0.009 &    0.020  $\pm$    0.005 &       0.37  $\pm$       0.02 &     -0.032  $\pm$    0.015 &    0.034  $\pm$    0.008     \\
 & 2 &       0.45 &       23.8 &      -20.2 &      E+S0 & 25 &       0.97  $\pm$       0.01 &     -0.030  $\pm$    0.012 &    0.051  $\pm$    0.020 &       0.36  $\pm$       0.02 &     -0.051  $\pm$    0.020 &    0.086  $\pm$    0.034     \\
 & 2 &       0.45 &       23.8 &      -20.2 &         E & 16 &       0.97  $\pm$       0.01 &     -0.019  $\pm$    0.009 &    0.020  $\pm$    0.005 &       0.37  $\pm$       0.02 &     -0.032  $\pm$    0.015 &    0.034  $\pm$    0.008     \\
 & 1 &       1.00 &       23.8 &      -20.2 &      E+S0 & 39 &       0.97  $\pm$       0.01 &     -0.033  $\pm$    0.014 &    0.056  $\pm$    0.014 &       0.36  $\pm$       0.02 &     -0.056  $\pm$    0.024 &    0.095  $\pm$    0.024     \\
 & 1 &       1.00 &       23.8 &      -20.2 &         E & 25 &       0.97  $\pm$       0.01 &     -0.020  $\pm$    0.009 &    0.025  $\pm$    0.005 &       0.37  $\pm$       0.02 &     -0.034  $\pm$    0.015 &    0.042  $\pm$    0.008     \\
 & 2 &       1.00 &       23.8 &      -20.2 &      E+S0 & 37 &       0.97  $\pm$       0.01 &     -0.036  $\pm$    0.016 &    0.056  $\pm$    0.014 &       0.36  $\pm$       0.02 &     -0.061  $\pm$    0.027 &    0.095  $\pm$    0.024     \\
 & 2 &       1.00 &       23.8 &      -20.2 &         E & 24 &       0.97  $\pm$       0.01 &     -0.022  $\pm$    0.009 &    0.024  $\pm$    0.005 &       0.37  $\pm$       0.02 &     -0.037  $\pm$    0.015 &    0.041  $\pm$    0.008     \\
\tableline \\
SRMS&&&&&&$$&&\\
 & 1 &       0.90 &       24.0 &      -20.0 &      E+S0 & 42 &       0.97  $\pm$       0.01 &     -0.034  $\pm$    0.017 &    0.066  $\pm$    0.013 &       0.36  $\pm$       0.02 &     -0.058  $\pm$    0.029 &    0.112  $\pm$    0.022     \\
 & 1 &       0.90 &       24.0 &      -20.0 &         E & 25 &       0.97  $\pm$       0.01 &     -0.020  $\pm$    0.009 &    0.025  $\pm$    0.005 &       0.37  $\pm$       0.02 &     -0.034  $\pm$    0.015 &    0.042  $\pm$    0.008     \\
 & 2 &       0.90 &       24.0 &      -20.0 &      E+S0 & 41 &       0.97  $\pm$       0.01 &     -0.036  $\pm$    0.018 &    0.065  $\pm$    0.014 &       0.36  $\pm$       0.02 &     -0.061  $\pm$    0.030 &    0.110  $\pm$    0.024     \\
 & 2 &       0.90 &       24.0 &      -20.0 &         E & 24 &       0.97  $\pm$       0.01 &     -0.022  $\pm$    0.009 &    0.024  $\pm$    0.005 &       0.37  $\pm$       0.02 &     -0.037  $\pm$    0.015 &    0.041  $\pm$    0.008     \\
 & 1 &       0.45 &       24.0 &      -20.0 &      E+S0 & 28 &       0.97  $\pm$       0.01 &     -0.033  $\pm$    0.015 &    0.066  $\pm$    0.020 &       0.36  $\pm$       0.02 &     -0.056  $\pm$    0.025 &    0.112  $\pm$    0.034     \\
 & 1 &       0.45 &       24.0 &      -20.0 &         E & 16 &       0.97  $\pm$       0.01 &     -0.019  $\pm$    0.009 &    0.020  $\pm$    0.005 &       0.37  $\pm$       0.02 &     -0.032  $\pm$    0.015 &    0.034  $\pm$    0.008     \\
 & 2 &       0.45 &       24.0 &      -20.0 &      E+S0 & 28 &       0.97  $\pm$       0.01 &     -0.032  $\pm$    0.014 &    0.066  $\pm$    0.019 &       0.36  $\pm$       0.02 &     -0.054  $\pm$    0.024 &    0.112  $\pm$    0.032     \\
 & 2 &       0.45 &       24.0 &      -20.0 &         E & 16 &       0.97  $\pm$       0.01 &     -0.019  $\pm$    0.009 &    0.020  $\pm$    0.005 &       0.37  $\pm$       0.02 &     -0.032  $\pm$    0.015 &    0.034  $\pm$    0.008     \\
 & 1 &       1.00 &       24.0 &      -20.0 &      E+S0 & 44 &       0.97  $\pm$       0.01 &     -0.034  $\pm$    0.016 &    0.064  $\pm$    0.013 &       0.36  $\pm$       0.02 &     -0.058  $\pm$    0.027 &    0.108  $\pm$    0.022     \\
 & 1 &       1.00 &       24.0 &      -20.0 &         E & 25 &       0.97  $\pm$       0.01 &     -0.019  $\pm$    0.009 &    0.025  $\pm$    0.005 &       0.37  $\pm$       0.02 &     -0.032  $\pm$    0.015 &    0.042  $\pm$    0.008     \\
 & 2 &       1.00 &       24.0 &      -20.0 &      E+S0 & 42 &       0.97  $\pm$       0.01 &     -0.036  $\pm$    0.016 &    0.064  $\pm$    0.013 &       0.36  $\pm$       0.02 &     -0.061  $\pm$    0.027 &    0.108  $\pm$    0.022     \\
 & 2 &       1.00 &       24.0 &      -20.0 &         E & 24 &       0.97  $\pm$       0.01 &     -0.022  $\pm$    0.009 &    0.024  $\pm$    0.006 &       0.37  $\pm$       0.02 &     -0.037  $\pm$    0.015 &    0.041  $\pm$    0.010     \\
\tableline \tableline\\
\end{tabular}
}
\end{center}
\footnotesize{Footnotes $a$,$b$,$c$,$d$, and $e$ as in Table~\ref{tcmr1}}
\end{table*}
\clearpage     
\begin{table*}
\begin{center}
\caption{Cluster RX~J0849+4452 fitted CMR parameters. \label{tcmr7}}
\vspace{0.25cm}
\resizebox{!}{2.5cm}{
\begin{tabular}{llccclccccccccccc}
\tableline \tableline\\
$Sample^a$ &$Log_{10}(Density)^a$ & $R^a$ & $m^{lim}$&$M_{B,z=0}^{lim}$&$Type^b$&$N^c$&$c_0^d$&$Slope^d$&$Scatter^d$&$(U-B)_{z=0}^e$&$|\frac{\delta (U-B)_{z=0}}{\delta M_{B,z=0}}|^e$&$\sigma (U-B)_{z=0}^e$ \\
 &$Log_{10}(Gal/Mpc^2)$ & Mpc & mag&mag&&&$mag$&&mag&mag&&mag \\
\tableline\\
SRF/SRMS&&&&&&&&\\
 &       1 &        0.9 &      24.00 &     -20.20 &                          E+S0 & 18 &       0.99  $\pm$       0.01 &     -0.021  $\pm$    0.021 &    0.039  $\pm$    0.008 &       0.43  $\pm$       0.02 &     -0.038  $\pm$    0.038 &    0.070  $\pm$    0.014     \\
 &       1 &        0.9 &      24.00 &     -20.20 &                             E & 10 &       0.99  $\pm$       0.01 &     -0.025  $\pm$    0.019 &    0.026  $\pm$    0.012 &       0.43  $\pm$       0.02 &     -0.045  $\pm$    0.034 &    0.047  $\pm$    0.022     \\
 &       2 &        0.9 &      24.00 &     -20.20 &                          E+S0 & 18 &       0.99  $\pm$       0.01 &     -0.021  $\pm$    0.021 &    0.039  $\pm$    0.008 &       0.43  $\pm$       0.02 &     -0.038  $\pm$    0.038 &    0.070  $\pm$    0.014     \\
 &       2 &        0.9 &      24.00 &     -20.20 &                             E & 10 &       0.99  $\pm$       0.01 &     -0.025  $\pm$    0.019 &    0.025  $\pm$    0.012 &       0.43  $\pm$       0.02 &     -0.045  $\pm$    0.034 &    0.045  $\pm$    0.022     \\
 &       1 &        0.4 &      24.00 &     -20.20 &                          E+S0 & 13 &       0.99  $\pm$       0.01 &     -0.036  $\pm$    0.015 &    0.020  $\pm$    0.007 &       0.42  $\pm$       0.02 &     -0.065  $\pm$    0.027 &    0.036  $\pm$    0.013     \\
 &       1 &        0.4 &      24.00 &     -20.20 &                             E &  7 &       0.99  $\pm$       0.01 &     -0.027  $\pm$    0.014 &    0.011  $\pm$    0.008 &       0.43  $\pm$       0.02 &     -0.049  $\pm$    0.025 &    0.020  $\pm$    0.014     \\
 &       2 &        0.4 &      24.00 &     -20.20 &                          E+S0 & 13 &       0.99  $\pm$       0.01 &     -0.037  $\pm$    0.015 &    0.020  $\pm$    0.007 &       0.42  $\pm$       0.02 &     -0.066  $\pm$    0.027 &    0.036  $\pm$    0.013     \\
 &       2 &        0.4 &      24.00 &     -20.20 &                             E &  7 &       0.99  $\pm$       0.01 &     -0.027  $\pm$    0.013 &    0.011  $\pm$    0.008 &       0.43  $\pm$       0.02 &     -0.049  $\pm$    0.023 &    0.020  $\pm$    0.014     \\
 &       1 &        1.0 &      24.00 &     -20.20 &                          E+S0 & 18 &       0.99  $\pm$       0.01 &     -0.021  $\pm$    0.021 &    0.039  $\pm$    0.008 &       0.43  $\pm$       0.02 &     -0.038  $\pm$    0.038 &    0.070  $\pm$    0.014     \\
 &       1 &        1.0 &      24.00 &     -20.20 &                             E & 10 &       0.99  $\pm$       0.01 &     -0.024  $\pm$    0.020 &    0.026  $\pm$    0.011 &       0.43  $\pm$       0.02 &     -0.043  $\pm$    0.036 &    0.047  $\pm$    0.020     \\
 &       2 &        1.0 &      24.00 &     -20.20 &                          E+S0 & 18 &       0.99  $\pm$       0.01 &     -0.021  $\pm$    0.022 &    0.039  $\pm$    0.008 &       0.43  $\pm$       0.02 &     -0.038  $\pm$    0.040 &    0.070  $\pm$    0.014     \\
 &       2 &        1.0 &      24.00 &     -20.20 &                             E & 10 &       0.99  $\pm$       0.01 &     -0.025  $\pm$    0.020 &    0.025  $\pm$    0.011 &       0.43  $\pm$       0.02 &     -0.045  $\pm$    0.036 &    0.045  $\pm$    0.020     \\
\tableline \tableline\\
\end{tabular}
}
\end{center}
\footnotesize{Footnotes $a$,$b$,$c$,$d$, and $e$ as in Table~\ref{tcmr1}}
\end{table*}
\begin{table*}
\begin{center}
\caption{Cluster RX~J0848+4453 fitted CMR parameters. \label{tcmr8}}
\vspace{0.25cm}
\resizebox{!}{2.5cm}{
\begin{tabular}{llccclccccccccccc}
\tableline \tableline\\
$Sample^a$ &$Log_{10}(Density)^a$ & $R^a$ & $m^{lim}$&$M_{B,z=0}^{lim}$&$Type^b$&$N^c$&$c_0^d$&$Slope^d$&$Scatter^d$&$(U-B)_{z=0}^e$&$|\frac{\delta (U-B)_{z=0}}{\delta M_{B,z=0}}|^e$&$\sigma (U-B)_{z=0}^e$ \\
 &$Log_{10}(Gal/Mpc^2)$ & Mpc & mag&mag&&&$mag$&&mag&mag&&mag \\
\tableline\\
SRF/SRMS&&&&&&&&\\
 &       1 &        0.9 &      24.00 &     -20.20 &                          E+S0 &  9 &       0.99  $\pm$       0.02 &     -0.055  $\pm$    0.018 &    0.027  $\pm$    0.015 &       0.43  $\pm$       0.04 &     -0.100  $\pm$    0.033 &    0.049  $\pm$    0.027     \\
 &       1 &        0.9 &      24.00 &     -20.20 &                             E &  6 &       0.98  $\pm$       0.02 &     -0.045  $\pm$    0.032 &    0.025  $\pm$    0.023 &       0.42  $\pm$       0.04 &     -0.082  $\pm$    0.058 &    0.045  $\pm$    0.042     \\
 &       2 &        0.9 &      24.00 &     -20.20 &                          E+S0 &  9 &       0.99  $\pm$       0.02 &     -0.055  $\pm$    0.018 &    0.027  $\pm$    0.015 &       0.43  $\pm$       0.04 &     -0.100  $\pm$    0.033 &    0.049  $\pm$    0.027     \\
 &       2 &        0.9 &      24.00 &     -20.20 &                             E &  6 &       0.98  $\pm$       0.02 &     -0.047  $\pm$    0.032 &    0.025  $\pm$    0.023 &       0.42  $\pm$       0.04 &     -0.085  $\pm$    0.058 &    0.045  $\pm$    0.042     \\
 &       1 &        0.4 &      24.00 &     -20.20 &                          E+S0 &  7 &       0.99  $\pm$       0.02 &     -0.052  $\pm$    0.021 &    0.031  $\pm$    0.019 &       0.43  $\pm$       0.04 &     -0.094  $\pm$    0.038 &    0.056  $\pm$    0.035     \\
 &       1 &        0.4 &      24.00 &     -20.20 &                             E &  5 &       0.97  $\pm$       0.01 &     -0.055  $\pm$    0.024 &    0.018  $\pm$    0.028 &       0.39  $\pm$       0.02 &     -0.100  $\pm$    0.044 &    0.033  $\pm$    0.051     \\
 &       2 &        0.4 &      24.00 &     -20.20 &                          E+S0 &  7 &       0.99  $\pm$       0.02 &     -0.053  $\pm$    0.021 &    0.031  $\pm$    0.019 &       0.43  $\pm$       0.04 &     -0.096  $\pm$    0.038 &    0.056  $\pm$    0.035     \\
 &       2 &        0.4 &      24.00 &     -20.20 &                             E &  5 &       0.97  $\pm$       0.01 &     -0.054  $\pm$    0.024 &    0.017  $\pm$    0.028 &       0.40  $\pm$       0.02 &     -0.098  $\pm$    0.044 &    0.031  $\pm$    0.051     \\
 &       1 &        1.0 &      24.00 &     -20.20 &                          E+S0 &  9 &       0.99  $\pm$       0.02 &     -0.055  $\pm$    0.018 &    0.027  $\pm$    0.015 &       0.43  $\pm$       0.04 &     -0.100  $\pm$    0.033 &    0.049  $\pm$    0.027     \\
 &       1 &        1.0 &      24.00 &     -20.20 &                             E &  6 &       0.98  $\pm$       0.02 &     -0.045  $\pm$    0.032 &    0.024  $\pm$    0.023 &       0.42  $\pm$       0.04 &     -0.082  $\pm$    0.058 &    0.044  $\pm$    0.042     \\
 &       2 &        1.0 &      24.00 &     -20.20 &                          E+S0 &  9 &       0.99  $\pm$       0.02 &     -0.055  $\pm$    0.017 &    0.027  $\pm$    0.015 &       0.43  $\pm$       0.04 &     -0.100  $\pm$    0.031 &    0.049  $\pm$    0.027     \\
 &       2 &        1.0 &      24.00 &     -20.20 &                             E &  6 &       0.98  $\pm$       0.02 &     -0.046  $\pm$    0.033 &    0.025  $\pm$    0.023 &       0.42  $\pm$       0.04 &     -0.084  $\pm$    0.060 &    0.045  $\pm$    0.042     \\
\tableline \tableline\\
\end{tabular}
}
\end{center}
\footnotesize{Footnotes $a$,$b$,$c$,$d$, and $e$ as in Table~\ref{tcmr1}}
\end{table*}


\vskip 0.5cm
\large{\bf APPENDIX II: Conversion to rest--frame magnitudes and colors}
\vskip 0.5cm
\small
{Following the same approach adopted in our CMR paper series (see details in Blakeslee et al. 2006), ACS galaxy colors as observed at the cluster redshift were converted to $(U-B)$ rest--frame colors. Using Bruzual \& Charlot (2003) stellar population models, and passive evolution, we calculated colors using theoretical SEDs (Spectral Energy Distribution) and filter sensitivity curves. We then fitted $(U-B)_{z=0}$ as a function of ACS colors, for SEDs with a galaxy formation redshift between 1.8 and 7, corresponding to ages expected for the early--type population, and metallicities equal to about half (40\%) solar, solar and 2.5 solar:
\begin{equation}
(U-B)_{z=0} =  Zp + Slope  (ACS \ Color)
\end{equation}
where {\it Zp} is the linear fit zero point and {\it Slope} is the slope.

We used Johnson U and B filters and Vega magnitude for the $(U-B)_{z=0}$ rest--frame color and ACS filters and AB magnitudes as measured at the cluster redshift. Errors on the zero point and the slope were calculated by bootstrapping 1000 simulations.
The error on the fit zero point can be reduced to a few thousandths of a magnitude if the fit to the ASC colors is shifted to the ACS color mean. Errors on slope, of primary interest to us in this paper, are the same in the two cases.
The ACS filter sensitivity curves used are the same as those used by Sirianni et al. (2005). For the Johnson $U$ and $B$ filters, we used sensitivity curves from Ma\'iz Apell\'aniz (2006). This author has pointed out that a number of sensitivity curves for the Johnson UBV system have been published in past years (Buser \& Kurucs 1978; Bessel 1990) that show important differences in the definition of the U filter sensitivity curve (see also Blakeslee et al. 2006, who present a discussion leading to a difference in $(U-B)_{z=0}$ of $\approx$ 10\%). The U filter sensitivity curves proposed by Bessel (1990) and by Buser \& Kurucs (1978) were shown to give inaccurate descriptions of the data, leading  Ma\'iz Apell\'aniz (2006) to derive a new U filter sensitivity function  better describing observed $(U-B)$ colors. For the B and V filters, the author concluded that Bessel (1990) provided an accurate description.  We use the Bessel (1990) B filter sensitivity curve.

Obtained zero point and slopes are given in Table~\ref{table_color}, and the fit in Figure~\ref{color}.
The linear relation between rest--frame and ACS cluster colors is tight at redshifts less than z$\approx$1.2, while it shows more dispersion at $z > 1.2$, i.e., for the farthest clusters observed. In fact, while at $z < 1.2$ our observations in the ACS bands match the $(U-B)_{z=0}$ rest--fame colors quite well, at $z > 1.1$ the (U-B) rest--frame magnitude would be lying more towards the infrared than the ACS ($i_{775}-z_{850}$) color. 

\begin{table*}[!ht]
\begin{center}
\caption{Conversion between the $(U-B)_{z=0}$ rest--frame color and ACS colors. \label{table_color}}
\resizebox{!}{2cm}{
\begin{tabular}{llccccccccccccccc}
\tableline \tableline\\
Color&z&Zp&$\sigma_{Zp}$&Slope&$\sigma_{Slope}$&$\sigma_{fit}$ \\ \\
\tableline \\
 ($V_{606}-z_{850}$)  &        0.83  &      -0.928  &       0.009  &       0.561  &       0.004  &      0.0060
                  \\
 ($r_{625}-z_{850}$)  &        0.83  &      -0.880  &       0.006  &       0.639  &       0.003  &      0.0024
                  \\
 ($V_{606}-I_{814}$)  &        0.90  &      -1.048  &       0.010  &       0.771  &       0.005  &      0.0004
                  \\
 ($i_{775}-z_{850}$)  &        1.11  &      -0.817  &       0.006  &       1.098  &       0.006  &      0.0051
                  \\
 ($i_{775}-z_{850}$)  &        1.24  &      -1.267  &       0.036  &       1.691  &       0.036  &      0.0154
                  \\
 ($i_{775}-z_{850}$)  &        1.26  &      -1.337  &       0.047  &       1.797  &       0.050  &      0.0107
                 \\
\tableline \tableline\\
\end{tabular}}
\end{center}
\end{table*}

\begin{figure}
\centerline{\includegraphics[scale=0.5,angle=180]{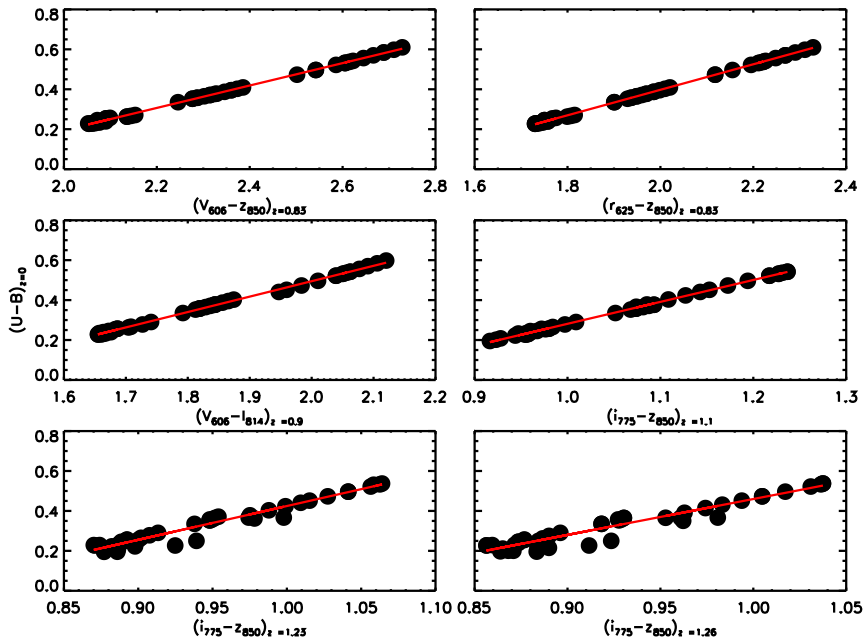}}
\caption {
Bruzual \& Charlot (2003) stellar population model predictions for $(U-B)_{z=0}$ color vs ACS colors at the cluster redshifts. The circles cover galaxy formation redshifts between 1.8 to 7, and metallicities of half solar, solar and 2.5 solar. The linear relation between rest--frame and ACS cluster colors is tight at redshift less than z$\approx$1.2, while it shows more dispersion at $z > 1.2$, for the farthest clusters observed. {\label{color}}}
\end{figure}

We also used the same models to convert ACS magnitudes and colors to $M_{B,z=0}$, by fitting the following relation:
\begin{equation}
M_{B,z=0} = (ACS \ mag) + Zp + Slope  (ACS \ Color)
\end{equation}
where {\it Zp} is the linear fit zero point and {\it Slope} is the slope. We used the same age and metallicity range as above. Results are given in Table~\ref{table_mag} and shown in Figure~\ref{magnitude}. To obtain absolute rest--frame magnitudes  $M_{B,z=0}$, we added $-5 log_{10}(\frac{Dist(pc)}{10pc})$ to the zero point fit, where $Dist$ is the distance in parsec of a hypothetical object at redshift $z$=0.02 (our rest--frame).
Higher order polynomial fits are not warranted by the data; a linear fit proved sufficient.

The conversion parameters we obtain depend on the models and the ranges of age and metallicity used. Within the uncertainties, they compare well with those used in Blakeslee et al. (2006) and Jee et al. (2007). Rest--frame absolute $M_{B,z=0}$ in Mei et al. (2006a,b) were expressed in AB magnitude, and were calculated using Bessel (1990) U and B filter responses.

 Some of the clusters in this sample (e.g., RX~J0152.7-1357 [Demarco et al. 2005]) show a significant dispersion in redshift. We have checked that assuming a single redshift in this conversions does not change our CMR parameters. We converted the color of each single galaxy in RX~J0152.7-1357 to the rest--frame $(U-B)_{z=0}$ color and  absolute rest--frame $M_{B,z=0}$ magnitude, and our results do not change at the 0.001~mag level.

\begin{figure}
\centerline{\includegraphics[scale=0.5,angle=180]{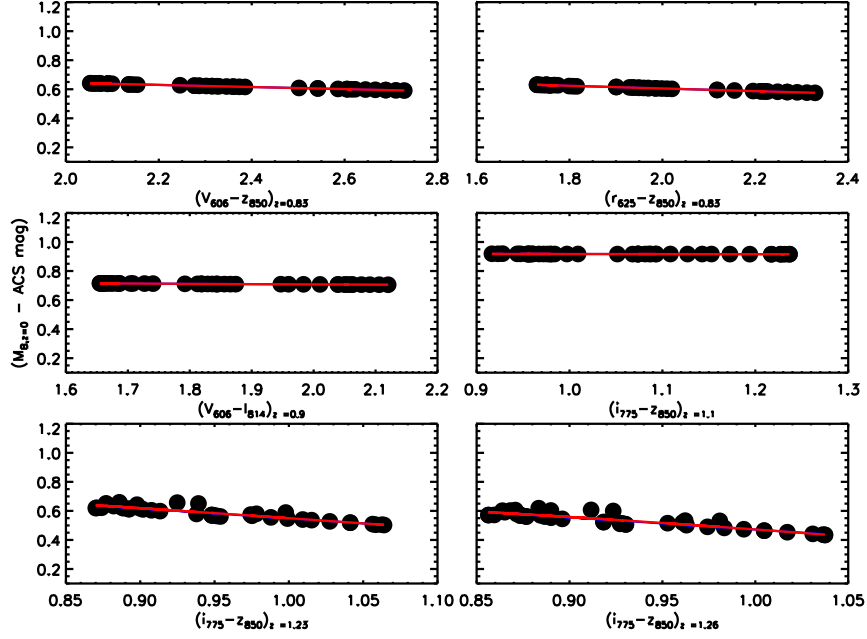}}
\caption {
Bruzual \& Charlot (2003) stellar population model predictions for $M_{B,z=0}$ vs ACS colors at the cluster redshifts. The circles cover galaxy formation redshifts between 1.8 to 7, and metallicities of half solar, solar and 2.5 solar. ACS colors and magnitudes are always expressed in AB magnitudes $M_{B,z=0}$ in Vega magnitude. The linear relation between rest--frame and ACS cluster colors is tight at redshifts less than z$\approx$1.2, while it shows more dispersion at $z > 1.2$, for the farthest clusters observed. {\label{magnitude}}}
\end{figure}

\newpage
\begin{table*}[!hb]
\begin{center}
\caption{Conversion between $M_{B,z=0}$ and ACS colors and magnitudes. To obtain the absolute rest--frame $M_{B,z=0}$ magnitude $-5 log_{10}(\frac{Dist(pc)}{10pc})$ should be added to the fit zero point, where $Dist$ is the distance in parsec of an hypothetical object at redshift $z$=0.02 (our rest--frame). \label{table_mag}}
\vspace{0.15cm}
\resizebox{!}{2cm}{
\begin{tabular}{llccccccccccccccc}
\tableline \tableline\\
Color&z&ACS mag&Zp&$\sigma_{Zp}$&Slope&$\sigma_{Slope}$&$\sigma_{fit}$ \\ \\
\tableline \\

 ($V_{606}-z_{850}$)  &        0.83  &      $i_{775}$  &       0.790  &       0.001  &      -0.073  &       0.001  &      0.0001
                  \\
 ($r_{625}-z_{850}$)  &        0.83  &      $i_{775}$  &       0.790  &       0.002  &      -0.092  &       0.001  &      0.0002
                  \\
 ($V_{606}-I_{814}$)  &        0.90  &      $I_{814}$  &       0.750  &       0.003  &      -0.021  &       0.002  &      0.0005
                  \\
 ($i_{775}-z_{850}$)  &        1.11  &      $z_{850}$  &       0.923  &       0.001  &      -0.006  &       0.001  &      0.0002
                  \\
 ($i_{775}-z_{850}$)  &        1.24  &      $z_{850}$  &       1.242  &       0.029  &      -0.693  &       0.029  &      0.0124
                  \\
 ($i_{775}-z_{850}$)  &        1.26  &      $z_{850}$  &       1.316  &       0.038  &      -0.845  &       0.039  &      0.0105
                  \\
\tableline \tableline\\
\end{tabular}}
\end{center}
\end{table*}
}

\end{document}